\documentclass[journal]{IEEEtran}
\usepackage{amsmath}
\usepackage{amsthm}
\usepackage{amssymb}
\usepackage{amsfonts}
\usepackage{cite}
\usepackage[dvipsnames]{xcolor}
\usepackage{subfig}
\usepackage{graphicx}
\usepackage{soul}
\usepackage{mathtools}
\usepackage{hyperref}
\usepackage[linesnumbered,ruled]{algorithm2e}
\usepackage{balance}

\ifodd 0
\newcommand{\EditRevision}[1]{\textcolor[rgb]{0,0,1}{#1}}
\newcommand{\CommentWong}[1]{\textcolor[rgb]{1,0,0}{[CW: #1]}}

\newcommand{\CommentRitesh}[1]{\textcolor[rgb]{0,0,1}{[Ritesh: #1]}}
\newcommand{\EditRitesh}[1]{\textcolor[rgb]{0,0,1}{#1}}
\else
\newcommand{\EditRevision}[1]{#1}
\newcommand{\CommentWong}[1]{}

\newcommand{\CommentRitesh}[1]{}
\newcommand{\EditRitesh}[1]{}
\fi

\def\E{\mathbb{E}}
\def\x{\boldsymbol{x}}
\def\y{\boldsymbol{y}}
\def\q{\boldsymbol{q}}
\def\v{\boldsymbol{v}}
\def\V{\boldsymbol{V}}
\def\w{\boldsymbol{w}}
\def\A{\boldsymbol{A}}
\def\a{\boldsymbol{a}}
\def\F{{\mathcal{F}}}

\def\var{\operatorname{Var}}

\def\F{\mathcal{F}}
\def\M{\mathcal{M}}
\def\Fviral{\F \text{ viral}}
\newcommand{\normaldensity}[3]{\mathcal{N}({#1}; {#2}, {#3})}

\def\pif{\pi_{\text{vf}}}

\def\pih{\pi_{\text{vf}}}
\def\pip{\pi_{\text{ind}}}

\def\lasso{\textsc{Lasso}}
\def\sqrtglasso{\textsc{Sqrt-Glasso}}
\def\glasso{\textsc{Glasso}}
\def\sqrtoglasso{\textsc{Sqrt-Oglasso}}
\def\comp{\textsc{Comp}}
\def\complasso{\textsc{Comp-Lasso}}
\def\compsqrtglasso{\textsc{Comp-Sqrt-Glasso}}
\def\compsqrtoglasso{\textsc{Comp-Sqrt-Oglasso}}
\def\mOne{\textbf{M1}}
\def\mTwo{\textbf{M2}}

\SetKwInput{KwInput}{Input}
\SetKwInput{KwOutput}{Output}

\newtheorem{definition}{Definition}

\newtheorem{lemma}{Lemma}

\DeclarePairedDelimiterX{\norm}[1]{\lVert}{\rVert}{#1}

\graphicspath{{figures/}}
\renewcommand{\baselinestretch}{1.0}

\begin{document}

\title{Contact Tracing Information Improves the Performance of Group Testing Algorithms}

\author{Ritesh~Goenka,
Shu-Jie~Cao,
Chau-Wai~Wong,~\IEEEmembership{Member,~IEEE,}
Ajit~Rajwade,~\IEEEmembership{Senior Member,~IEEE,}
Dror~Baron,~\IEEEmembership{Senior Member,~IEEE}%
\thanks{This work was supported in part by the Science and Engineering Research Board (SERB) Matrics Grant \#10013890, IITB-WRCB (Wadhwani Center for Bioengineering) Grant \#DONWR04-002, and DST-Rakshak grant \#DST0000-005, and in part by the National Science Foundation under Award ECCS-2030430. A subset of this work was presented at the 2021 IEEE International Conference on Acoustics, Speech and Signal Processing~\cite{Goenka2021}.}
}

\maketitle

\begin{abstract}
Group testing can help maintain a widespread testing program using fewer resources amid a pandemic.
In group testing, we are given $n$ samples, one per individual. These samples are arranged into $m < n$ pooled samples, where each pool is obtained by mixing a subset of the $n$ individual samples. 
Infected individuals are then identified using a group testing algorithm.
In this paper, we use side information (SI) collected from contact tracing (CT) within 
nonadaptive/single-stage group testing algorithms.
We generate CT SI data by incorporating characteristics of disease spread between individuals.
These data are fed into two signal and measurement models for group testing, 
and numerical results show that our algorithms provide improved sensitivity and specificity. We also show how to incorporate CT SI into the design of the pooling matrix. 
That said, our numerical results suggest that the utilization of SI in the pooling matrix design \EditRevision{based on minimization of a weighted coherence measure} 
does not yield significant performance gains beyond the incorporation of SI in the group testing algorithm.
\end{abstract}


\IEEEpeerreviewmaketitle

\section{Introduction}
\label{sec:intro}
Widespread testing has been promoted for combating the ongoing COVID-19 pandemic.
Samples are typically collected from nasal or oropharyngeal swabs, and then processed by a reverse transcription polymerase chain reaction (RT-PCR) machine.
However, widespread testing is hindered by supply chain constraints and long testing times. 

Pooled or {\em group testing} has been suggested for improving testing efficiency \cite{Dorfman1943,aldridge2019group,Hogan2020,Abdelhamid2020,Zhu2020,Yi_arxiv,Ghosh2021,zhu2020paris,Heiderzadeh2020,Nikolopoulos2020,Shental2020,lin2020comparisons,cohen2020multilevel,lin2020positively,Nikolopoulos2021b,ahn2021adaptive,arasli2021group}.
Group testing involves mixing a subset of $n$ individual samples into $m < n$ pools. The measurement process can be expressed as $\boldsymbol{y} = \mathfrak{N}(\boldsymbol{Ax})$,
where $\boldsymbol{x}$ is a vector that quantifies the health status of the $n$ individuals,
$\boldsymbol{A}$ is an $m \times n$ binary pooling matrix with $A_{ij} = 1$ if the $j$th individual contributes to the $i$th pool, else $A_{ij} = 0$,
$\boldsymbol{y}$ is a vector of $m$ noisy measurements or tests,
and $\mathfrak{N}$ represents a probabilistic noise model that relates the noiseless pooled results, $\boldsymbol{Ax}$, to 
$\boldsymbol{y}$. We consider two signal and noise models.

\noindent{}{\bf Model \textbf{M1}:} A {\em binary noise} model used by Zhu et al.~\cite{Zhu2020}, 
where $\boldsymbol{x}$ is a binary {vector},
$\boldsymbol{w} = \boldsymbol{Ax}$
is an auxiliary (noiseless) vector,
and the measurement $y_i\in\{0,1\}$ depends probabilistically on $w_i$,
where $\Pr(y_i=1|w_i=0)$ and $\Pr(y_i=0|w_i>0)$ are probabilities of erroneous tests.

\noindent{}{\bf Model \textbf{M2}:} A {\em multiplicative noise} model of the form 
$\boldsymbol{y} = \boldsymbol{Ax} \circ \boldsymbol{z}$ as used in Ghosh et al.~\cite{Ghosh2021},
where $\circ$ represents elementwise multiplication, $\boldsymbol{z}$ is a vector of $m$ noisy elements defined as
$z_i = (1+\EditRevision{q_a})^{\eta_i}$, $\EditRevision{q_a} \in (0, 1]$ is a known amplification factor for RT-PCR, 
$\eta_i \sim \mathcal{N}(0,\sigma^2)$, and $\sigma^2 \ll 1$ is a known parameter controlling the noise in RT-PCR. Under model \textbf{M2}, $\boldsymbol{x}$ and $\boldsymbol{y}$ represent viral loads in the 
$n$ individuals and $m$ pools, respectively. Assuming reasonably high viral loads in $\boldsymbol{x}$, 
Poisson effects in $\boldsymbol{y}$ can be ignored \cite{Ghosh2021}.

For both models, we wish to estimate $\boldsymbol{x}$ from $\boldsymbol{y}$ and $\boldsymbol{A}$.
We use single-stage {\em nonadaptive} algorithms as in \cite{Zhu2020, Ghosh2021}, rather than two-stage algorithms,
which employ a second stage of tests depending on results from the first stage, as in 
Heidarzadeh and Narayanan~\cite{Heiderzadeh2020} or the classical 
Dorfman approach~\cite{Dorfman1943}. 
The advantage of nonadaptive algorithms is that they reduce testing time, which is high for RT-PCR.

Algorithms that estimate $\boldsymbol{x}$ from $\boldsymbol{y}$ and $\boldsymbol{A}$~\cite{Ghosh2021, Shental2020} rely primarily on the {\em sparsity} of $\boldsymbol{x}$, which is a valid assumption for COVID-19 due to low prevalence rates \cite{Benatia2020}.
However, in addition to sparsity, the health 
status vector $\boldsymbol{x}$ contains plenty of 
statistical structure. For example,
Zhu et al.~\cite{Zhu2020} exploited probabilistic information such as the prevalence rate and structure in $\boldsymbol{x}$, and stated the potential benefits of using {\em side information} (SI).
Specific forms of SI include individuals' symptoms and family 
structure~\EditRevision{\cite{zhu2020paris, Goenka2021}}. \EditRevision{Family structure refers to information regarding which individuals belong to the same family. Besides its well-known conventional meaning, the concept of ``family'' could refer to a group of students sharing the same room in a hostel, security officers working regularly at the same checkpoint, or healthcare workers working in the same facility.}
Finally, Nikolopoulos et al. independently observed that community structure can improve the performance of group testing~\cite{Nikolopoulos2020,Nikolopoulos2021b}; 
these works focused on encoder design in conjunction with basic decoders. \EditRevision{On the other hand, we focus largely on innovative decoder designs and explore the effect of encoder designs.} 
{\em The overarching message of our paper is that SI can greatly improve group testing.}

\begin{figure*}[!t]
	\begin{tabular}{@{}cc@{}c@{}c@{}c@{}}
	\multicolumn{2}{l}{\underline{Sparsity:} \hspace{8mm} \underline{$2.12\%$}} & \underline{$3.98\%$} & \underline{$6.01\%$} & \underline{$8.86\%$}  \vspace{0mm} \\
  \rotatebox[origin=l]{90}{\hspace{14mm}\underline{\textbf{M1}}} & 
		\includegraphics[width=0.24\linewidth]{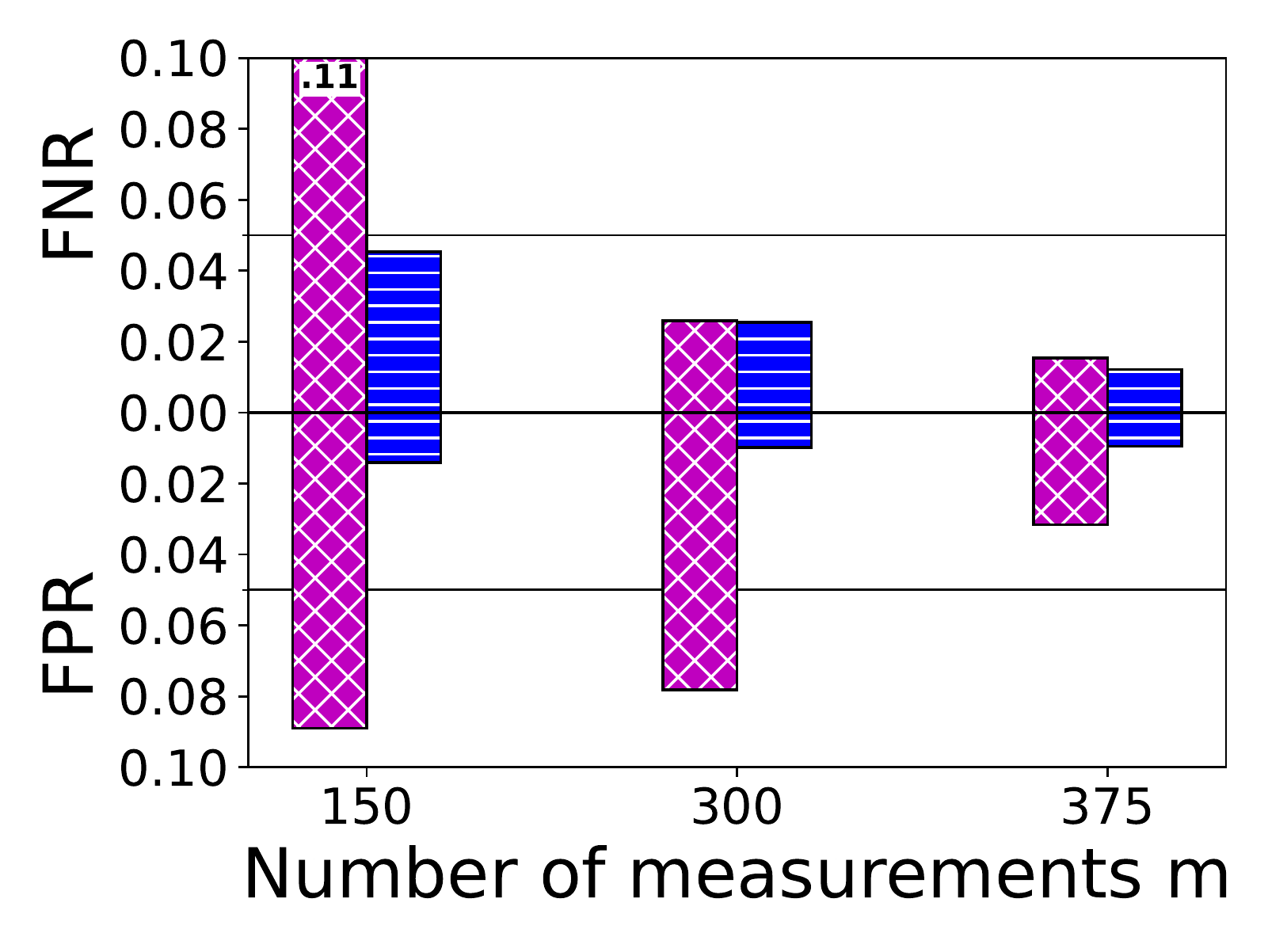} &
		\includegraphics[width=0.24\linewidth]{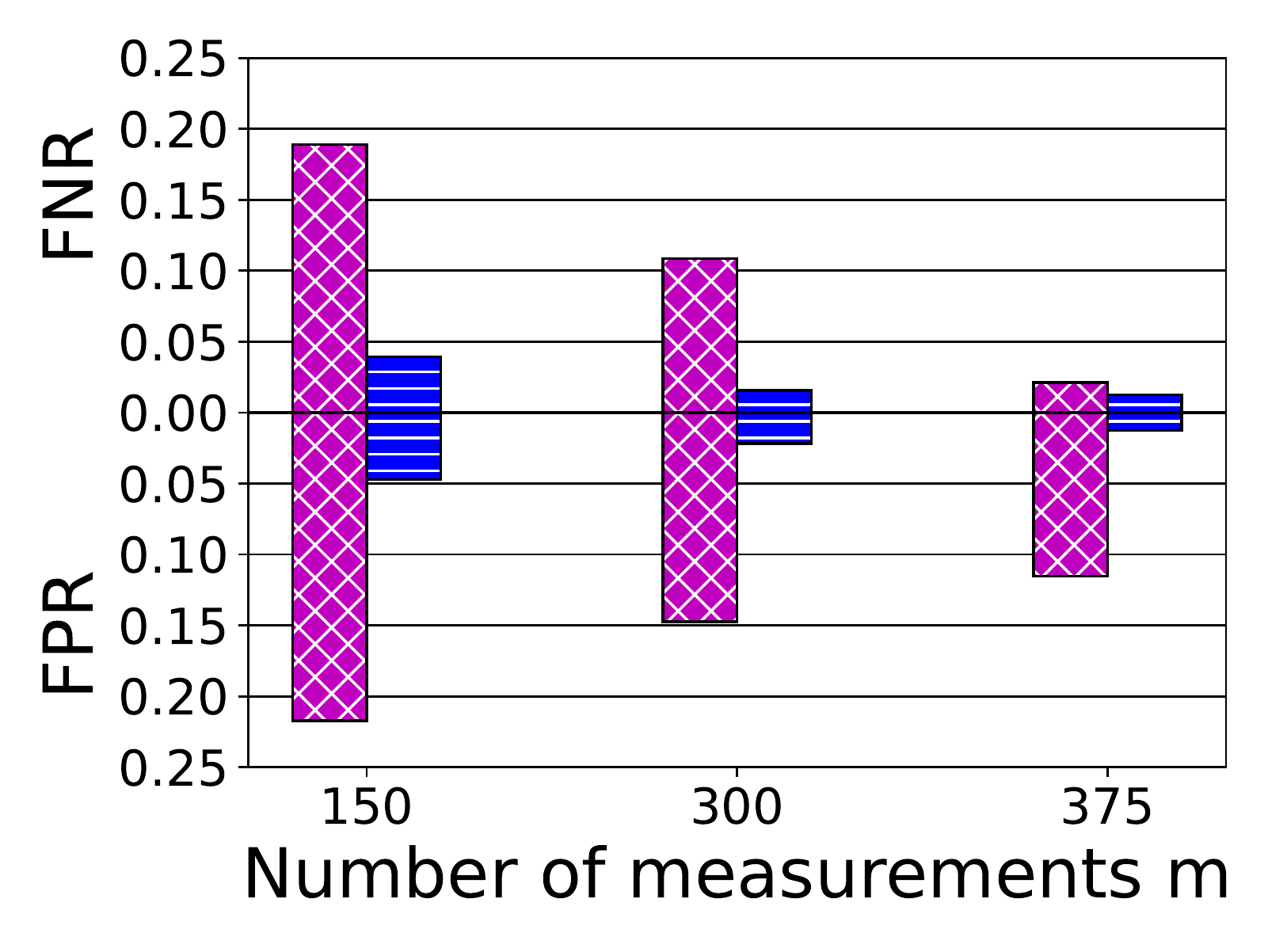} &
		\includegraphics[width=0.24\linewidth]{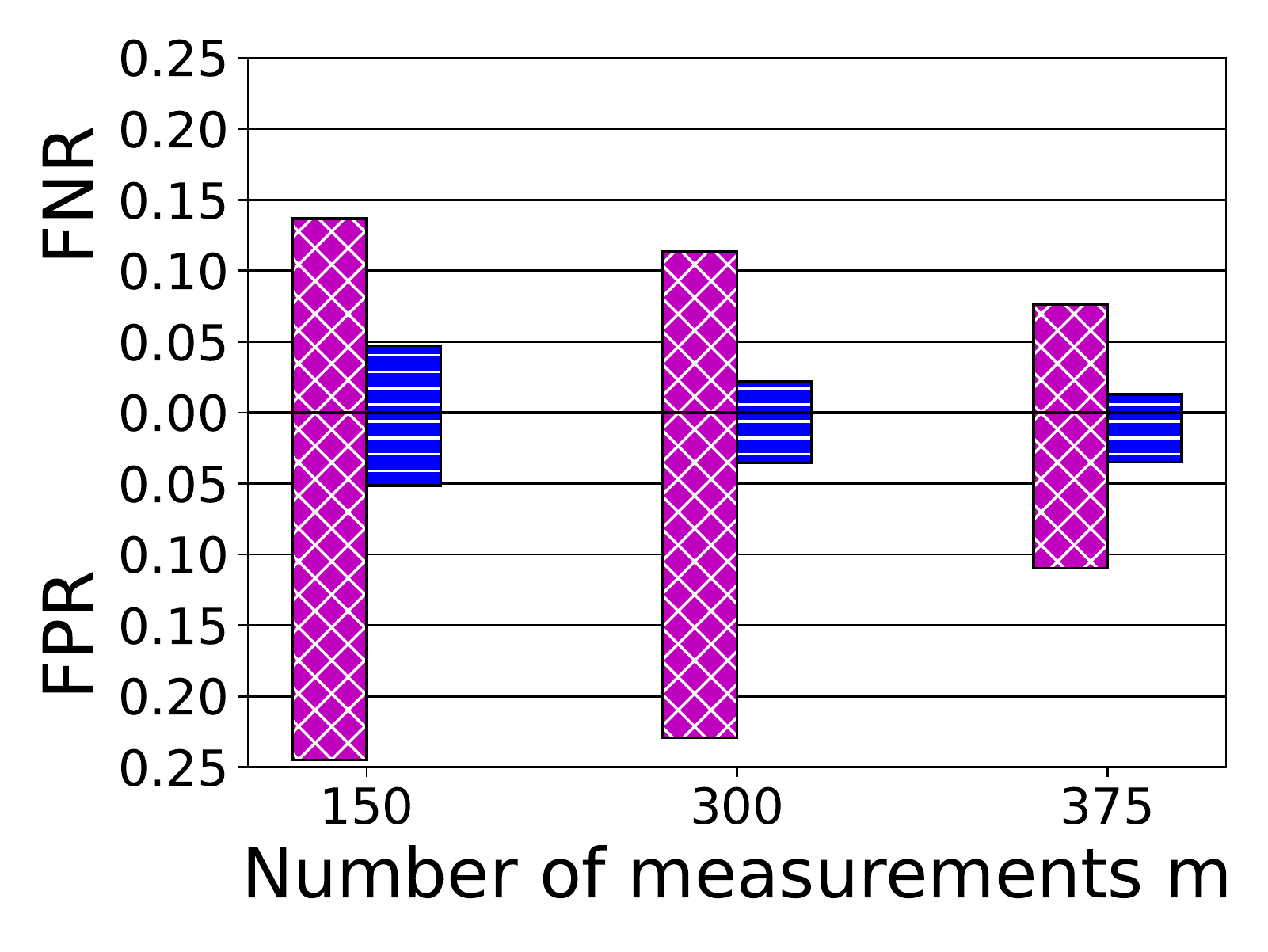} &
		\includegraphics[width=0.24\linewidth]{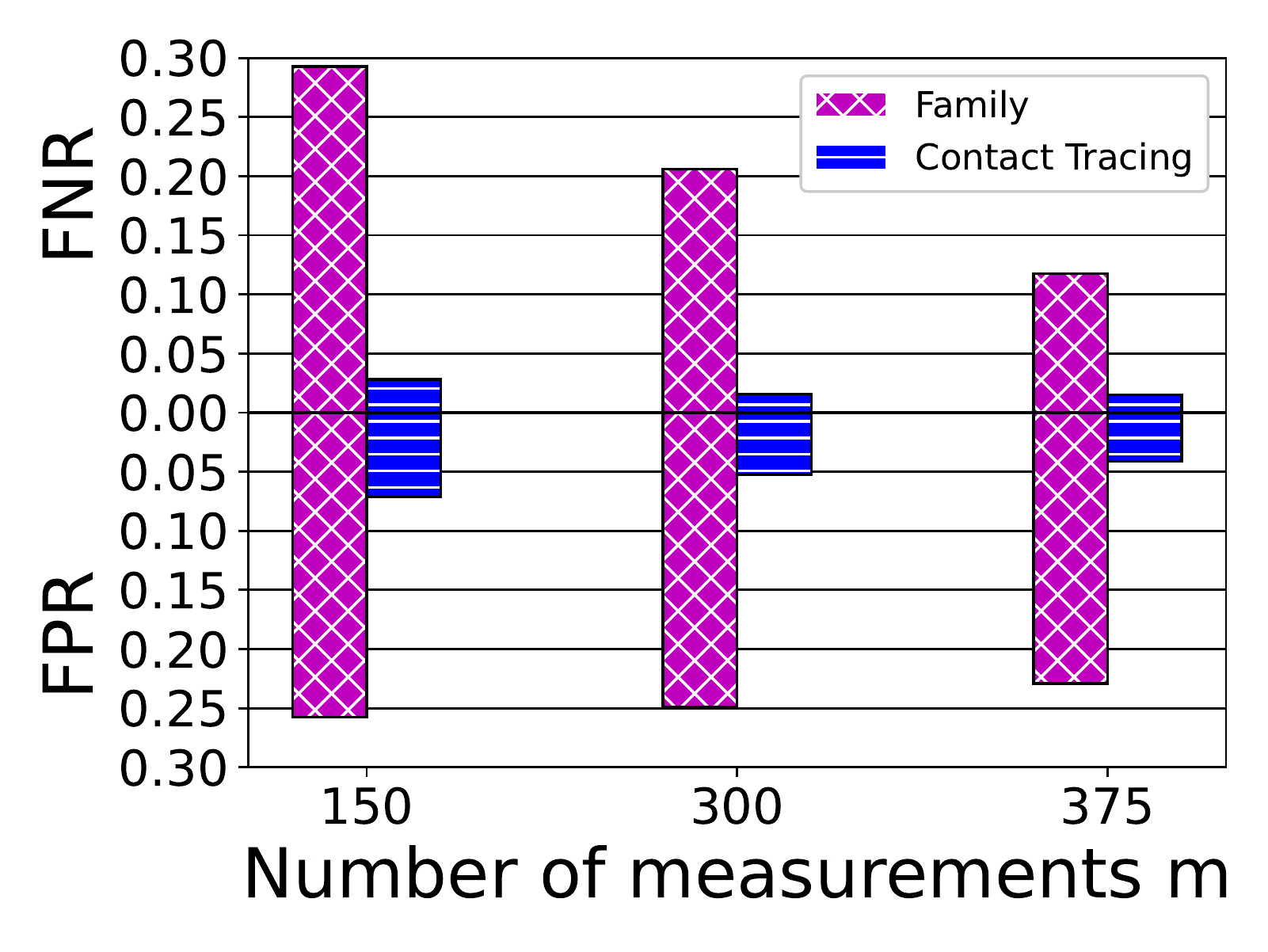} \vspace{-2mm} \\ 
		\rotatebox[origin=l]{90}{\hspace{14mm}\underline{\textbf{M2}}} & 
		\includegraphics[width=0.24\linewidth]{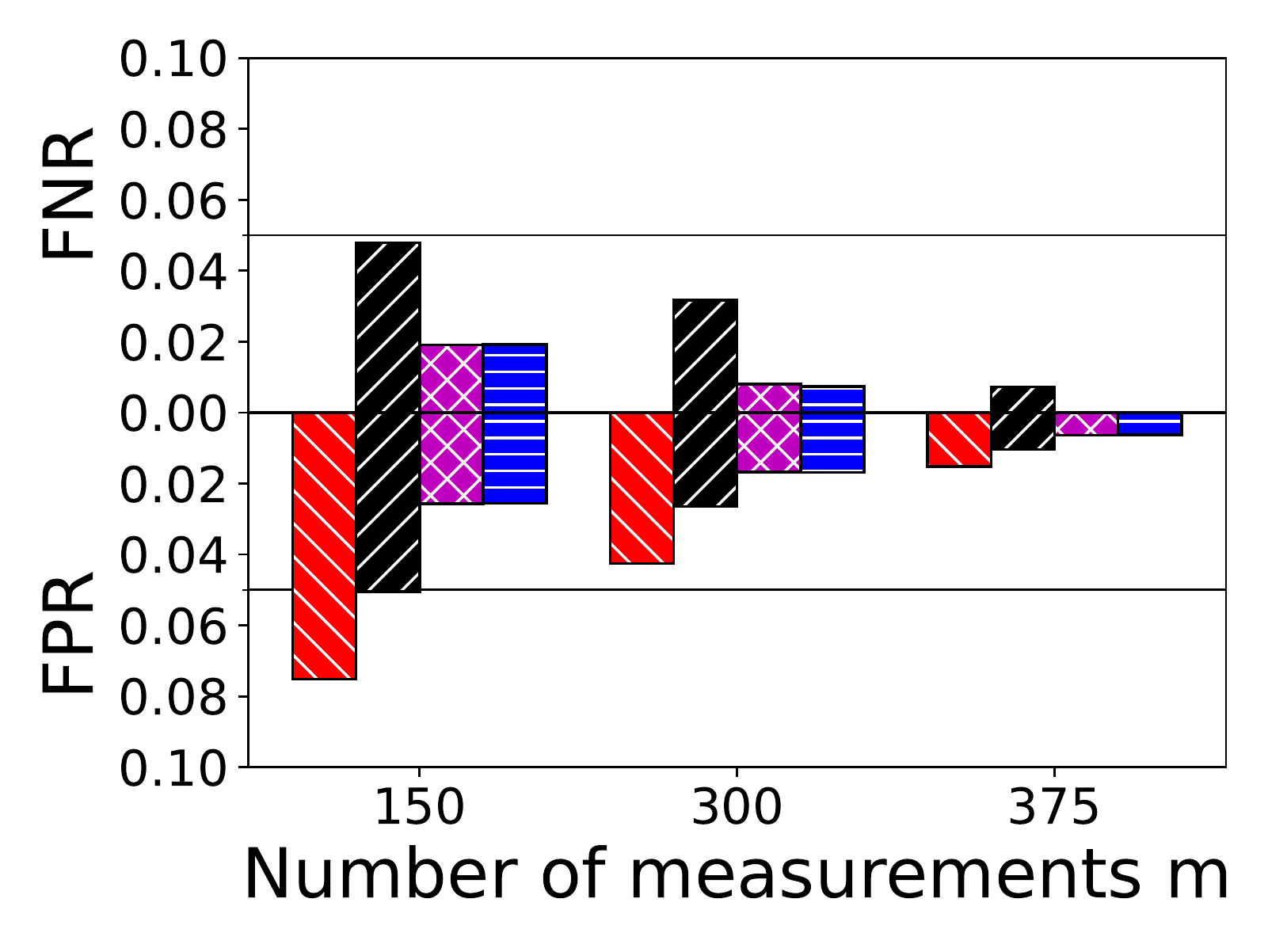} &
		\includegraphics[width=0.24\linewidth]{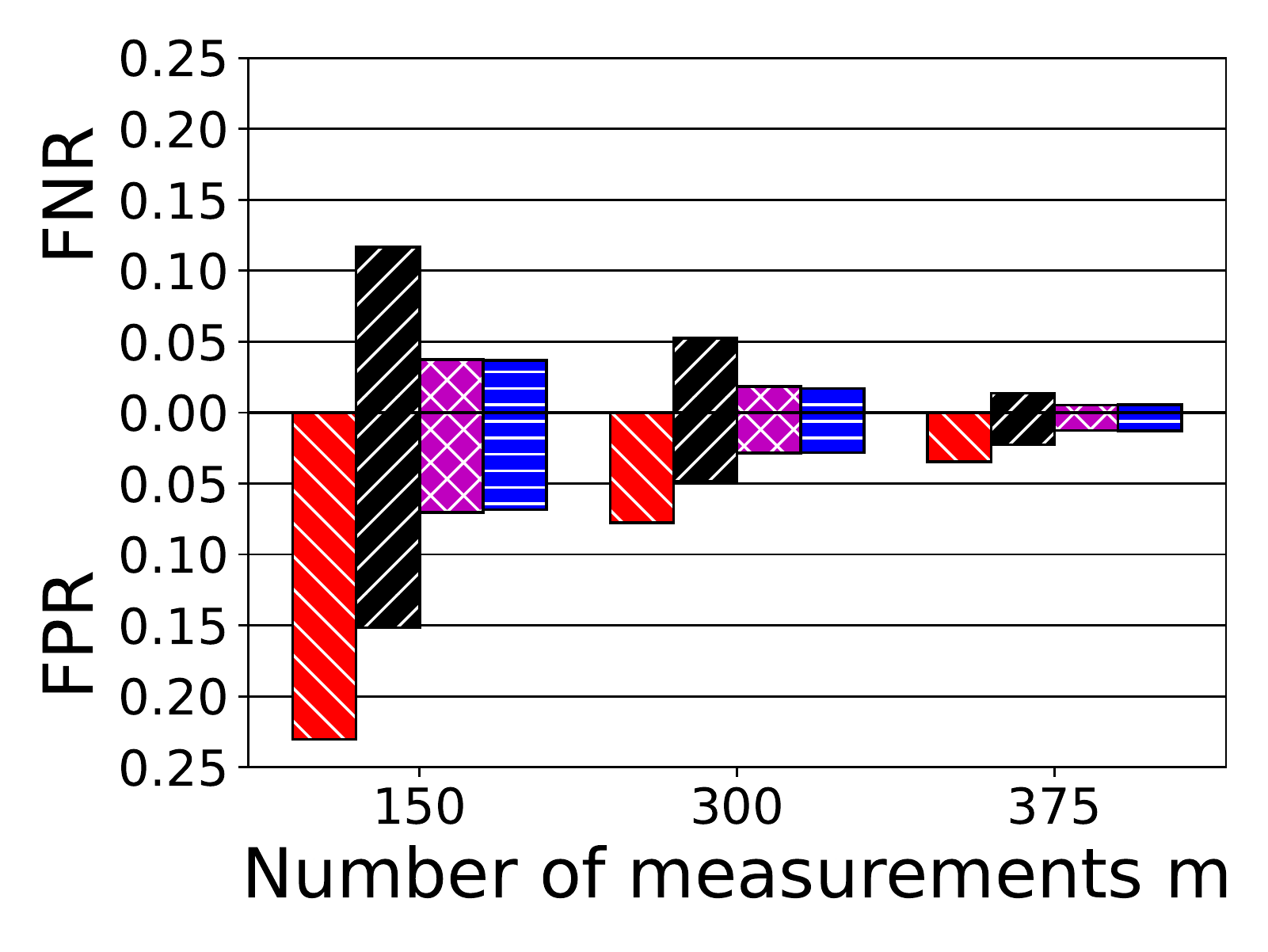} &
		\includegraphics[width=0.24\linewidth]{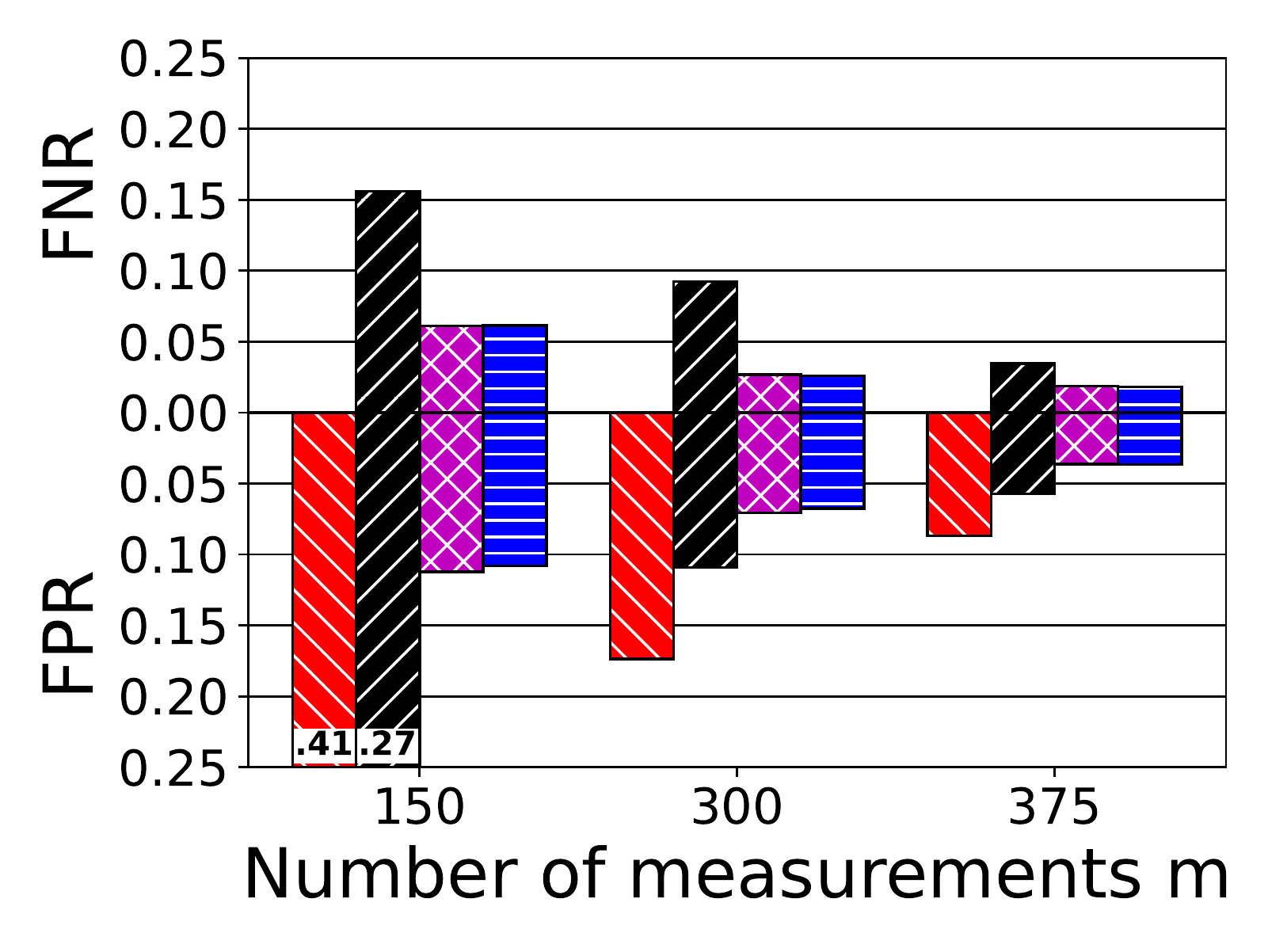} &
		\includegraphics[width=0.24\linewidth]{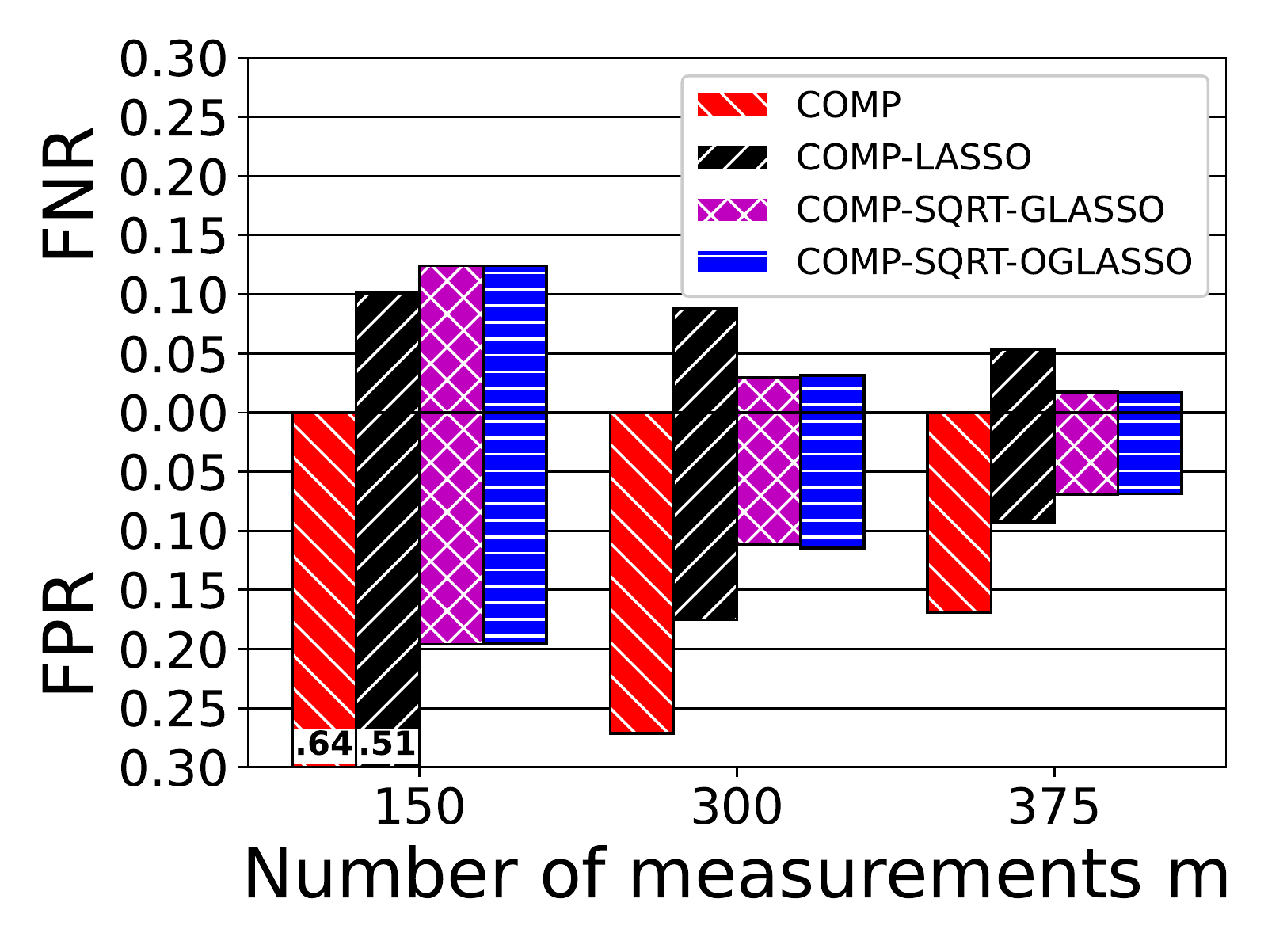}
	\end{tabular}
  \vspace{-3mm}
  \caption{Performance of the proposed group testing methods \textbf{M1} (top row) with binary noise and \textbf{M2} (bottom row) with multiplicative noise at four averaged sparsity levels and three measurement levels for a population of $n = 1000$ individuals. See Secs.~\ref{subsec:results_M1} and~\ref{subsec:results_M2} for more details.}
  \label{fig:res-m1-m2}
  \vspace{-4mm}
\end{figure*}

{\em Contact tracing} (CT) information has been widely collected and used for
controlling a pandemic~\cite{cdc_contact_tracing}. 
\EditRevision{Such information, including the duration of contact between pairs of individuals and measures of physical proximity, 
can be collected via some combinations of various modalities such as Bluetooth~\cite{Hekmati2020}, the global positioning system~\cite{Kleinman2020}, manual inquiries by social workers~\cite{Hohman2021, Ross2020}, and financial transaction data~\cite{CDC_CT,Kleinman2020}}. 
In this paper, we show how to estimate $\boldsymbol{x}$ while utilizing CT SI.

Our contributions are twofold.
First, in Sec.~\ref{sec:algos},
we show that CT SI used at the decoder can help algorithms such as generalized approximate message passing~(GAMP)~\cite{rangan2011generalized}
or least absolute shrinkage and selection operator (LASSO) variants \cite{Yuan2006,Jacob2009} better estimate $\boldsymbol{x}$ from $\boldsymbol{y}$ and $\boldsymbol{A}$. 
Our numerical results
are presented in Sec.~\ref{sec:results};
a typical result appears in Fig.~\ref{fig:res-m1-m2}.
Our work uses more SI than others~\cite{Nikolopoulos2020,Nikolopoulos2021b,ahn2021adaptive}
that only considered family/community
structure in binary group testing while designing the encoder. 

Our second contribution appears in Sec.~\ref{sec:matrix_des}, where we also incorporate SI at the encoder 
and demonstrate numerically that it provides modest performance benefits over the utilization of SI only within the decoder.

The rest of the paper is organized as follows.
In Sec.~\ref{sec:related_work}, we survey related work.
In Sec.~\ref{sec:data_gen}, we present a data generation model that 
allows us to simulate 
realistic datasets.
Our main contributions appear in Sec.~\ref{sec:algos},
where we propose two classes of group testing algorithms 
that epxloit CT SI, and in Sec.~\ref{sec:matrix_des}, 
where we provide techniques for pooling matrix design that incorporates SI.
Numerical results 
are presented in Sec.~\ref{sec:results}. 
We discuss in Sec.~\ref{sec:discussion} and conclude in Sec.~\ref{sec:conclusion}.

\section{Related Work}
\label{sec:related_work}

\subsection{Exploiting Side Information in Group Testing}

Side information (SI)
can be derived from travel history, medical history, 
and symptoms. 
These can be used to assign infection probabilities to individuals, who can be subsequently classified into low or high risk, and processed by risk-aware group testing algorithms.
A recent publication from the World Health Organization (WHO) \cite{Deckert2020} demonstrated a marked reduction in the number of tests required if the pooling was performed on subjects with similar infection probabilities, as opposed to subjects with widely varying infection probabilities. Moreover, these risk-specific algorithms can also incorporate strategies to minimize the number of tests. This type of hierarchical approach has been adopted in \cite{McMahan2012} in conjunction with the classical Dorfman pooling method \cite{Dorfman1943}. 
In a broadly similar approach in \cite{Bilder2010}, individuals are ordered as per their infection probabilities, and extra tests are assigned for individuals identified to be high risk.

There is relatively little prior art that makes use of SI about family structures 
or contact tracing information. One recent example is Zhu et al.~\cite{Zhu2020,zhu2020paris}, who used both individual infection probabilities derived from symptom SI and family information while performing the decoding. 
More recently, Nikolopolous et al. \cite{Nikolopoulos2021a, Nikolopoulos2021b} used SI about 
connected or overlapping communities to design good encoding matrices. The designed matrices boost the performance of basic decoders such as definite defectives (\textsc{Dd})~\cite{aldridge2019group}
or loopy belief propagation (LBP)~\cite{Baron2010}, in conjunction with a model of binary noise that is similar to \textbf{M1}. However, in contrast to our work, their emphasis is on the encoder side, without designing decoders that explicitly account for SI.
Moreover, we also have a model \textbf{M2} that allows estimation of the viral load in the infected subjects. 
Ahn et al.\cite{ahn2021adaptive} and Arasli and Ulukus~\cite{arasli2021group} considered correlations represented by arbitrary graphs drawn from a stochastic block model, albeit for adaptive group testing, unlike the non-adaptive approach in this paper. Their approaches generalized the i.i.d. assumption often used in group testing. Such correlations could be derived from family or contact tracing information. Graphs generated from this model contain node clusters with dense connections within a cluster and sparse connections between clusters. Such structures are typical of many social network graphs. The authors in \cite{lin2020positively,Lendle2012} provided an analytical proof of cost reduction in Dorfman pooling if there is a positive correlation between the different samples being pooled. Moreover, Lin et al.~\cite{lin2020comparisons} also presented a hierarchical agglomerative algorithm for pooled testing in a social graph. We note that \cite{ahn2021adaptive,arasli2021group,lin2020comparisons,Lendle2012} do not consider measurement noise. On the other hand, our work deals with nonadaptive algorithms and explores two different measurement noise models.

\subsection{Pooling Matrix Design}
\label{subsec:pooling_matrix_design}
\noindent{}{\bf Compressive recovery.} 
Decoding in group testing and compressive recovery are closely related,
as initially explored by Gilbert et al.~\cite{Gilbert2008}.
In both areas, designing group testing/sensing matrices is a key problem.
A popular quality measure for the sensing matrix $\A$ from the compressed sensing literature is the coherence~\cite{Davenport2012}, defined as 
$\mu(\A) \triangleq \max_{i \neq j, 1 \leq i,j \leq n} |\a_i \cdot \a_j| \, \big/ \, \|\a_i\|_2 \|\a_j\|_2$,
where 
$\a_i$ is the $i$th column vector of $\A$,
and $\a_i \cdot \a_j$ is the dot product of 
$\a_i$ and $\a_j$.
As the coherence is a worst-case measure, 
a large body of art \cite{Abolghasemi2010,Carin2011,Arguello2014} seeks to design $\A$ by minimizing the average squared coherence, 
$\mu_{\text{avg}}(\A) \triangleq \sum_{i \neq j, 1 \leq i,j \leq n} |\a_i \cdot \a_j|^2 \, \big/ \big( \|\a_i\|^2_2 \, \|\a_j\|^2_2 \big)$.

Since DeVore~\cite{Devore2007}, 
deterministic designs of binary sensing matrices with bounds on their coherence values have
been proposed~\cite{Naidu2016,Naidu2017,Li2014,lin2020comparisons}. Our  
previous work \cite{Goenka2021,Ghosh2021} used so-called Kirkman triple matrices, which are also deterministic in nature with many useful properties. However, all these methods construct matrices of \emph{specific} sizes. For example, 
DeVore~\cite{Devore2007} designed matrices of size \EditRevision{$q^2_p \times q^{l+1}_p$ for $1 < l < q_p$ where $q_p$ is a prime power} and $l$ is a natural number. In contrast, many group testing applications require matrices whose size does not adhere to  {\em arbitrary} constraints.

\begin{figure}[!t]
  \centering
  \centerline{\includegraphics[width=8cm]{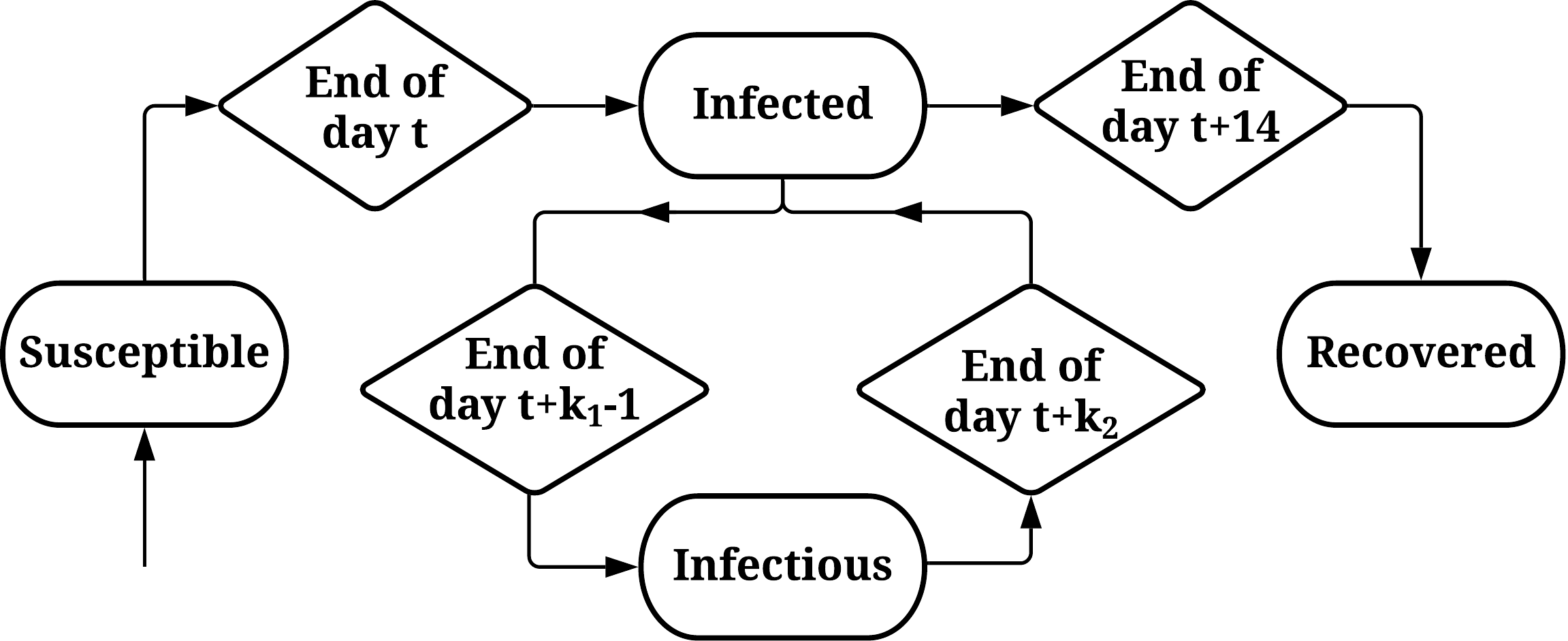}}
  \caption{\EditRevision{A diagram of state transitions among susceptible, infected (excluding infectious), infectious, and recovered for a node.} A node is infectious only between days $k_1$ and $k_2$ (both inclusive) after getting infected. In our work, we set $(k_1,k_2) = (3,7)$.}
  \vspace{-4mm}
\label{fig:state-transition-diagram}
\end{figure}

\section{Data Generation Model}
\label{sec:data_gen}
In this section, we present a generative infection model 
that incorporates CT SI, 
which we later use to prepare simulated data for algorithmic evaluation. We model a population of $n$ individuals using a dynamical or time-varying graphical model that contains nodes \EditRevision{$\{u_i\}_{i=1}^n$} and undirected edges $\big\{e_{ij}^{(t)}\big\}_{i,j=1}^n$.
On a given day $t$, an edge $e_{ij}^{(t)}$ between nodes $u_i$ and $u_j$ encodes CT SI 
$\big(\tau^{(t)}_{ij}, d^{(t)}_{ij}\big)$, which can be acquired via Bluetooth-based CT applications~\cite{Hekmati2020}. 
Here, $\tau^{(t)}_{ij}$ represents the contact duration and $d^{(t)}_{ij}$ represents a measure of the physical proximity between two individuals.
On day~$t$, a node can be in one of the following states: \emph{susceptible}, \emph{infected}, \emph{infectious}, and \emph{recovered}.
\EditRevision{Note that the infected state is defined in a narrow sense that excludes the infectious state, because states must be mutually exclusive.}
To keep the model simple, we assume that there are no reinfections, i.e., recovered is a terminal state, despite some reports of reinfection \cite{Haseltine2020}.
While our model is inspired by a classical compartmental 
model in epidemiology comprised
of susceptible, exposed, infectious, and recovered (SEIR) states considered for COVID-19~\cite{Carcione2020}, 
our state transitions explicitly use CT
SI and knowledge about the pandemic~\cite{WHOreport}.

We adopt a simplified infection dynamic wherein the
infectious period is preceded and followed by the infected state.
Our design parameters for the infection dynamics are based on a World Health Organization report on COVID-19~\cite{WHOreport}.
Specifically, a node $u_i$ remains infected but noninfectious for $k_1 = 3$ days.
On day $t+k_1$, the node becomes infectious and may transmit the disease to a susceptible neighboring node $u_j$ 
with probability $p_{i,j}^{(t+k_1)}$ whose construction is described below.
An infectious node can potentially transmit the infection until $k_2 = 7$ days after getting infected, and becomes noninfectious afterward. \EditRevision{Note that the above assumptions ensure that the transmission of infection on any given day is limited to one hop in the CT graph, i.e., it cannot be the case that individual $X$ infects individual $Y$, who in turn infects individual $Z$ on the same day.}
We also model a small fraction of stray infections that may occur, for example, due to sporadic contact with contaminated surfaces. 
Such infections only affect nodes in the susceptible state with a probability $p_1 = 2 \times 10^{-4}$
of our choice.
A state diagram appears in Fig.~\ref{fig:state-transition-diagram}. 
Regarding the viral load $x_i^{(t)}$ for node $i$ on day $t$, we assume $x_i^{(t)} = 0$ if the node is susceptible or recovered.
For an infected or infectious node, we make a simplified assumption that its viral load $x_i^{(t)} \sim \textrm{Uniform}(1, 2^{15})$,\footnote{
The cycle threshold for RT-PCR commonly ranges from $19$ to $34$ cycles \cite[Fig. 3]{Buchan2020}, where $34$ cycles corresponds to a low initial viral load of a few molecules, and each cycle roughly doubles the viral density.
Therefore, we estimate the largest possible viral load as $2^{34-19} = 2^{15}$.}
once drawn, remains constant throughout the combined $14$-day period of infection.

Next, we model the probability $p_{i,j}^{(t)}$ that the disease is transmitted from node $u_i$ to $u_j$ on day $t$.
We view infection times as a nonhomogeneous Poisson process with time-varying rate function $\lambda(t)$.
Consider a $\tau^{(t)}_{ij}$-hour contact on day $t$ when susceptible node $u_j$ is exposed to infectious node $u_i$. 
The average infection rate $\lambda_{ij}(t)$ for day $t$ is assumed to be proportional to both the viral load $x^{(t)}_i$ and the physical proximity $d^{(t)}_{ij}$, namely, $\lambda_{ij}(t) = \lambda_0 \, x^{(t)}_i \, d^{(t)}_{ij}$, where $\lambda_0$ is a tunable, baseline Poisson rate.
The probability that $u_j$ is infected by the end of contact period $\tau^{(t)}_{ij}$ is therefore $p^{(t)}_{i,j} = 1 - \exp \left( -\lambda_0 \, x^{(t)}_i \, d^{(t)}_{ij} \, \tau^{(t)}_{ij} \right)$ for $t \in [k_1, k_2] + t_i$.
From the standpoint of susceptible node $u_j$, all its neighbors $u_k$ that are infectious contribute to its probability of getting infected on day~$t$, namely, $1-\prod_{k} \big( 1 - p^{(t)}_{k,j} \big)$. \EditRevision{We remark that if an individual catches the infection on day $t$, then it will possibly be detected only in testing conducted on day $t+1$ since we assume that sample collection and pooled testing is performed at the beginning of the day.}

\begin{figure}[!t]
  \vspace{-5mm}
  \subfloat[width=0.48\linewidth][]{\includegraphics[width=4.3cm]{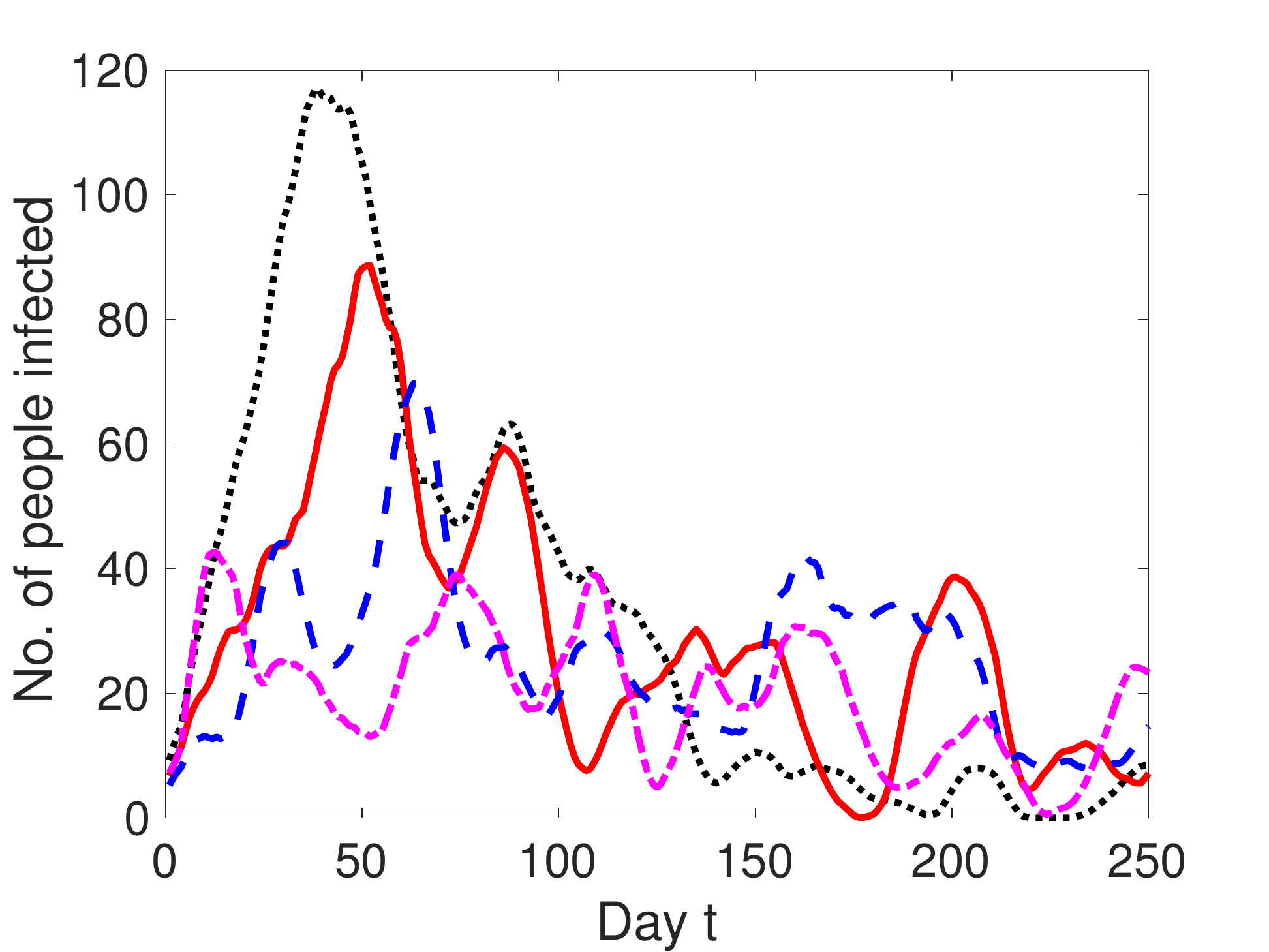}}
  \subfloat[width=0.48\linewidth][]{\includegraphics[width=4.3cm]{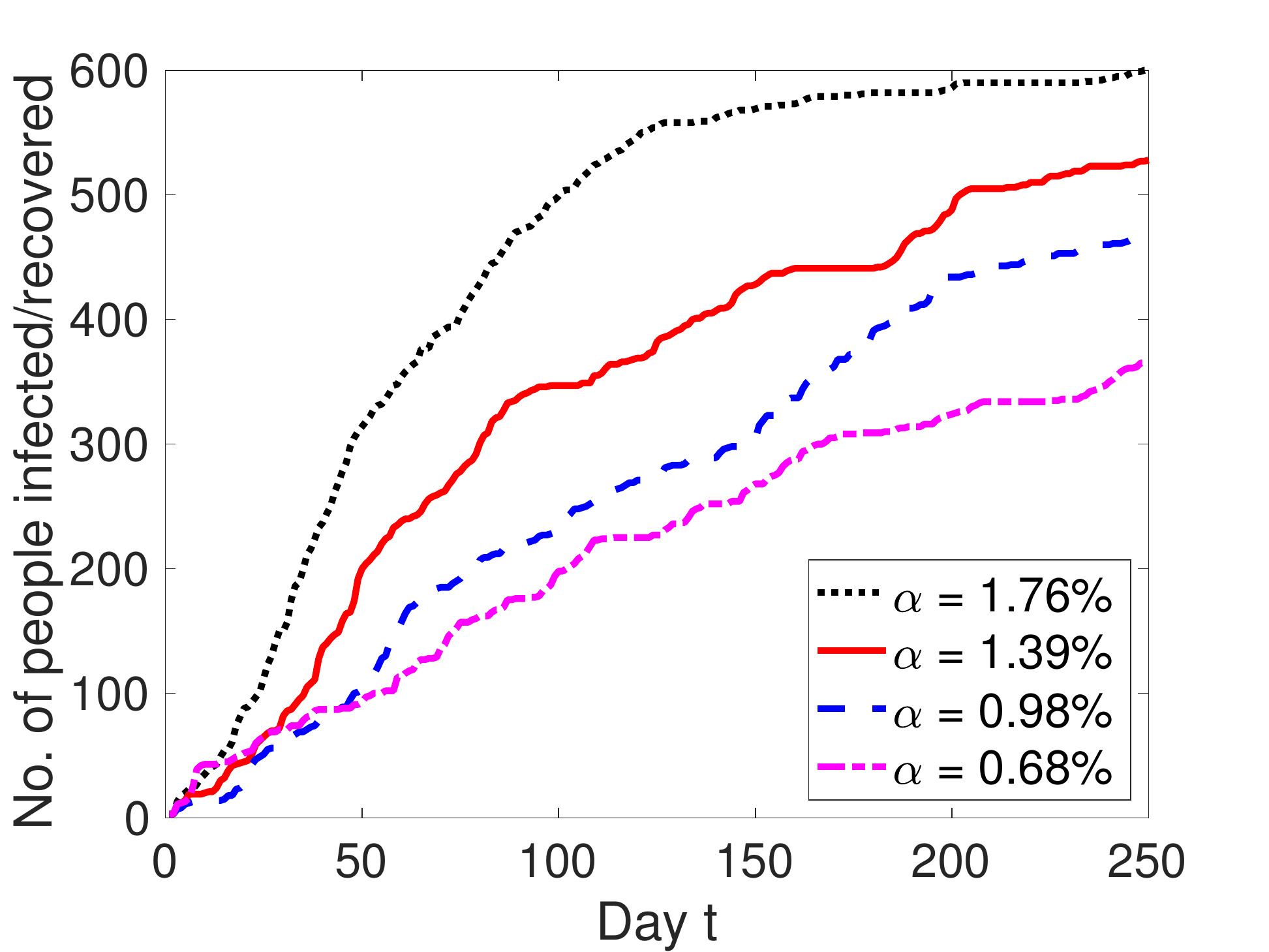}}
  \vspace{-2mm}
  \caption{(a) The number of active infections, and (b) cumulative infections at different inter-clique contact levels $\alpha$.
  We chose $50$-day windows for testing our proposed algorithms.}
  \vspace{-4mm}
\label{fig:infections-curves}
\end{figure}

While generating our data, we considered $n = 1,000$ 
nodes divided into cliques based on the distribution of family sizes in India \cite[pg. 18]{UNDoc}, for a duration of $t_{\text{max}} = 250$ days.
Fig.~\ref{fig:infections-curves} shows the number of active infections and the cumulative number of infections at the end of each day.
The clique structures were kept constant throughout the $t_{\text{max}}$ days, whereas inter-clique contacts corresponding to sporadic contacts between people were dynamically added and removed. 
We define the number of inter-clique contacts divided by the number of contacts for a given day as the inter-clique contact level, $\alpha$, where the contact is defined for a pair of individuals.
The varying $\alpha$ affects the sparsity of the underlying vector $\boldsymbol{x}$; it brings infections to new cliques/families.
Pooling of samples is performed at the beginning of each day from $t_{\text{peak}}-24$ to $t_{\text{peak}}+25$, 
where $t_{\text{peak}}$ is the day with the maximum number of active infections.

\EditRevision{The recent approach in \cite{LaGatta2021} fits a graph neural network to available data regarding COVID-19 infections, deaths and recoveries in a certain geographic location and obtains the temporally varying parameters of the differential equations associated with a standard SEIR model. Though there are in principle similarities to the approach presented here, the main difference is that we have used parameters based on recent documents published by the WHO \cite{Deckert2020}, whereas \cite{LaGatta2021} uses a data-driven approach to perform regression via a graph-based neural network. However, a neural network requires a large amount of data for training, which would not have been immediately available when the pandemic began. Our approach uses very basic knowledge regarding the spread of COVID-19.}

\section{Proposed Group Testing Algorithms}
\label{sec:algos}

This section describes two classes of group testing algorithms for reconstructing the health status vector $\x$ from the pooled tests, $\y$, and the pooling matrix, $\A$.

\subsection{Algorithms for Binary Noise}
\label{Sec:gamp_binary}
For model \textbf{M1}, Zhu et al.~\cite{Zhu2020} use
{\em generalized approximate message passing} (GAMP)~\cite{rangan2011generalized} for group testing estimation.
GAMP is comprised of two components.
The first component is comprised
of an input channel that relates a prior for $n$ individuals' 
health status, $\x=(x_i)_{i=1}^n$, 
to pseudo data, 
\begin{equation}
\v = \x + \q \in \mathbb{R}^n,
\end{equation}
where the $n$ coordinates of $\x$ are correlated, and
$\q$ is additive white Gaussian noise with $q_i \sim \mathcal{N}(0,\Delta)$.
\EditRevision{When the $i$th individual is in the infected/infectious state as defined in Sec.~\ref{sec:data_gen}, $x_i = 1$; otherwise, $x_i = 0$.
The pseudo data $\v$ is an internal variable of GAMP that can be considered as a corrupted version of the true unknown health-status vector $\x$. The pseudo data is later iteratively cleaned up to gradually reveal $\x$.
The unknown input $\x$ can be} estimated from pseudo data $\v$ using a denoising function or a denoiser:
\begin{equation}
\widehat{x}_{i}=g_{\text{in}} \left(\v \right) 
= \E \left[X_{i} \mid \V=\v \right] 
\label{input_d},
\end{equation}
where we use the convention that when both the capital and lower case versions of a symbol appear, the capital case is a random variable and the lower case is its realization, and
$\E \left[X_{i} | \v \right]$ represents $\E \left[X_{i} | \V=\v \right]$ when the context is clear.
The second component of GAMP is comprised of an output channel relating the auxiliary vector $\w$ to the noisy measurements $\y$ as reviewed in Sec.~\ref{sec:intro}.
We adopt the output channel {denoiser} of Zhu et al.~\cite{Zhu2020}, 
$h_{i}=g_{\text{out}}\left(y_{i}; k_{i}, \theta_{i}\right) = ( \E\left[W_{i} \mid y_{i}, k_{i},  \theta_{i}\right]-k_{i} ) / \theta_i$, 
where $\theta_{i}$ is the estimated variance of $h_i$, 
and $k_{i}$ is the mean of our estimate for $w_i$. 
Since $y_i$ depends probabilistically on $w_i$, we have 
$f \left(w_{i} \mid y_{i}, k_{i}, \theta_{i}\right) \propto \operatorname{Pr}\left(y_{i} \mid w_{i}\right) \, 
\exp \left[-\frac{\left(w_{i}-k_{i}\right)^{2}}{2 \theta_{i}}\right]$,
where \EditRevision{$f$ is a probability density function and} $W_i$ is approximated as Gaussian \EditRevision{per the central limit theorem when there are enough $1$'s in the $i$th row of pooling matrix $\A$.}
\EditRevision{Although GAMP's signal estimation capability is optimal in the asymptotic sense, it is practically useful in a finite parameter regime. A population $n$ greater than several hundred individuals, which is the case considered in this paper, allows GAMP to work reasonably well.}

The key to an efficient GAMP-based estimator  
is the input-channel denoiser.
While Zhu et al.~\cite{Zhu2020} considered Bernoulli elements in $\x$,
this paper accounts for probabilistic dependencies within $\x$.
Below, we provide details for the design of two denoisers.
Our first denoiser is based on a probabilistic model that
considers groups of people, such as family members.
Our second denoiser 
encodes the CT information into the prior for each individual's health status.

\subsubsection{Family Denoiser}
We formalize a family-based infection mechanism,
leading to a denoiser that improves the detection accuracy.
Define $\M_{\F}$ as the set of indices of all members of family $\F$.
We say that $\F$ is {\em viral} when there exists viral material in the family.
Next, define the infection probability of individual $i$ within a viral family 
$\F$ as $\pip = \Pr(X_i = 1 \mid \Fviral)$, for all $i \in \M_{\F}$, and $\pif = \Pr(\Fviral)$.
Note that the infection 
status of individuals in a viral family are conditionally 
independent and identically distributed (i.i.d.).

Under our definition, family $\F$ being viral need not be attributed to any individual $i \in \M_{\F}$.
After all, viral material can be on an infected pet or contaminated surface. 
For this model, once the family is viral, the virus
spreads independently with a fixed probability $\pip$.
Of course, our simplified model may not accurately reflect reality. 
That said, without a consensus in the literature on how coronavirus spreads, 
it is unrealistic to create a more accurate model.
On the other hand, our model is plausible, and we will see that it is mathematically tractable.
We further assume that individuals cannot be infected unless the family is viral,
i.e., $\Pr(X_i = 1 \mid \F \text{ not viral}) = 0$.
The family structure serves as SI and allows the group testing algorithm to impose the constraint that people living together have strongly correlated health status.

Next, we derive the denoiser \eqref{input_d} by incorporating the family-based infection mechanism.
Denote the pseudodata of the members of family $\F$ as $\v_{\F} = (v_i)_{i \in \M_{\F}}$, 
the family-based denoiser for $i$th individual can be decomposed as follows:
\begin{subequations}
\begin{align}
&g_{\text{in}}^\text{family}(\v_\F) \notag \\
=& \E \left[ X_{i} \mid \v_{\F} \right] 
= \Pr(X_i = 1 \mid \v_{\F}) \\ 
=& \Pr(X_i = 1, \Fviral \mid \v_{\F})  \\
=& \Pr(\Fviral \mid \v_{\F}) \Pr(X_i = 1 \mid \v_{\F},\Fviral),
\label{f_den}
\end{align}
\label{eq:denoiser-family}
\end{subequations}
where the first term of \eqref{f_den} is
\begin{align}
&\Pr(\Fviral \mid \v_{\F}) \nonumber \\
=& \, \frac{f(\v_{\F},\Fviral)}{f(\v_{\F},\Fviral) + f(\v_{\F},\F\text{ not viral}) }.
\label{p_infect}
\end{align}
The two quantities in \eqref{p_infect} can be further expanded as
\begin{subequations}
\begin{align}
&f(\v_{\F},\F\text{ not viral}) \\
= &(1-\pih) \, f(\v_{\F} \mid \F\text{ not viral}) \\
= &(1-\pih) \prod_{i\in \M_{\F}}
\normaldensity{v_i}{0}{\Delta}
\end{align}
\end{subequations}
and
\begin{subequations}
\begin{align}
&f(\v_{\F},\ \Fviral) = \pih \, f(\v_{\F} \mid \Fviral)\\
\begin{split}
  =\, &\pih\sum_{\x_k \in \Omega_{\F}} \prod_{i \in \M_{\F}} \\ 
  & \Big[ 
  f(v_{i} \mid X_{i}=x_{k,i})
  \Pr( X_i = x_{k,i} \mid \Fviral ) 
  \Big],
\end{split}
\label{sum over prod}
\normalsize
\end{align}
\end{subequations}
where $\normaldensity{x}{\mu}{\sigma^2} := \frac{1}{\sqrt{2 \pi \sigma^2}} \exp \left(  -\frac{(x-\mu)^2}{2 \sigma^2} \right)$, 
and $\Omega_{\F} = \{0...00,\  0...10,\ \dots,\ 1...11\}$ is a power set 
comprised of $2^{|\M_{\F}|}$ distinct infection patterns for family $\F$.
The second term of \eqref{f_den} can be simplified as follows:
\begin{subequations}
\begin{align}
\notag
&\Pr(X_i = 1 \mid \v_{\F},\Fviral) \\
= & \Pr(X_i = 1 \mid v_i,\Fviral)  \\
=&\Pr(X_i = 1, v_i \mid \Fviral) \, / \, \Pr(v_i \mid \Fviral) \\
=&\dfrac{\pip \, \normaldensity{v_i}{1}{\Delta}}
{\pip \, \normaldensity{v_i}{1}{\Delta} +  \left(1-\pip\right) \, \normaldensity{v_i}{0}{\Delta}} \\
=&  \left( 1 + \frac{1-\pip}{\pip} \cdot \frac{\normaldensity{v_i}{0}{\Delta}}{\normaldensity{v_i}{1}{\Delta}} \right)^{-1} \\
=& \left( 1 + \left( \pip^{-1} - 1 \right)
  \exp \Big[ \big(v_i-\tfrac{1}{2}\big) \big/ \Delta \Big] \right)^{-1}.
\end{align}
\end{subequations}

\subsubsection{Contact Tracing Denoiser}
While family structure SI characterizes part of the spread of the disease,
individual family members will presumably all come in close contact with each other;
hence CT SI will include cliques for these individuals. Additionally, CT SI
describes inter-family contacts. Therefore,
CT SI can characterize the spread of the disease more comprehensively than family SI.

Consider a hypothetical widespread testing program that relies on CT SI,
where all individuals are tested $8$ days before the testing program begins,
resulting in a good initial estimate of their ground-truth health status.
To exploit the CT SI, we encode it for each individual $i$ into the prior probability of infection, $\Pr(X_i=1)$,
and use the following scalar denoiser:
\begin{subequations}
\begin{align}
&g_{\text{in}}^\text{CT}(v_i) \nonumber \\
=& \E \left[ X_i \mid v_i \right] = \Pr \left( X_i = 1 \mid v_i \right) \\
=& f(v_i \mid X_i=1) \Pr(X_i=1) / f(v_i) \\
=& \left\{ 1 \!+\! \big[\Pr(X_i\!=\!1)^{-1} \! - \! 1\big]
  \exp \Big[ \big(v_i-\tfrac{1}{2}\big) \big/ \Delta \Big] \!\right\}^{-1}.
\end{align}
\label{eq:denoiser-ct}
\end{subequations}
\hspace{-0.8mm}Here, $\Pr(X_i\!=\!1)$ for day $k+1$ can be estimated by aggregating CT information of individual $i$ over a so-called \emph{SI period}\footnote{\EditRevision{Usage of the SI period from day $k-7$ to day $k$ for testing on day $k+1$ implicitly assumes that the test results are available within $24$ hours. However, in cases when it takes more than $24$ hours to generate test results, the SI period can be appropriately modified. For example, day $k-8$ to day $k-1$ for testing on day $k+1$ if it takes at most $48$ hours to generate test results.}} from day $k-7$ to day $k$ as follows
\begin{equation}
\widehat{\Pr}^{(k+1)}(X_{i}=1) = 1 - \prod_{d=k-7}^k{\prod_{j=1}^n{{ \!\big( 1-\widehat{p}^{(d)}_{i,j} \big) }}},
\label{eq:SI_period_aggre}
\end{equation}
where $\widehat{p}_{i,j}^{(d)}$ is the estimated probability of infection of individual $i$ due to 
contact with individual $j$.
This probability, $\widehat{p}_{i,j}^{(d)}$, 
may be determined by the CT information ($\tau_{ij}^{(d)}, d_{ij}^{(d)})$, as well as their infection status as follows:
\begin{equation}
\widehat{p}_{i,j}^{(d)}=\exp\left(-\big(\lambda \, \tau_{ij}^{(d)} \,  d_{ij}^{(d)} \, \Psi_{ij}^{(d)}+\epsilon\big)^{-1}\right),
\label{eq:est_pij_2nd}
\end{equation}
where $\Psi_{ij}^{(d)} =
1 - \widehat{\Pr}^{(d)}\!(X_i\!=\!0) \, \widehat{\Pr}^{(d)}\!(X_j\!=\!0){\color{orange}}$,
$\lambda$ is an unknown Poisson rate parameter,
and $\epsilon$ is used to avoid division by zero. 
We estimate $\lambda$ with maximum likelihood (ML) using the pseudodata of all individuals, i.e.,
\begin{equation}
\widehat{\lambda}^{\text{ML}} = \arg\max_{\lambda}\ \prod_{i=1}^n f(v_i|\lambda)
\label{eq:mle},
\end{equation}
where \(f(v_i|\lambda) = f(v_i|X_{i}=1) \,  \Pr(X_{i}=1|\lambda)+f(v_i|X_{i}=0) \, \Pr(X_{i}=0|\lambda)\).
Once $\widehat{\lambda}^{\text{ML}}$ is obtained, it is plugged into \eqref{eq:est_pij_2nd} for calculating the prior probability \EditRevision{in \eqref{eq:SI_period_aggre}}~\cite{dror_plugin}.
\EditRevision{As long as the pseudo data is not too noisy, the ML parameter will be close to the true one, and the estimated probability $\widehat{p}_{i,j}^{(d)}$ in \eqref{eq:est_pij_2nd} will be close to the true probability.}
A plug-in strategy can also used for our family denoiser $g_{\text{in}}^\text{family}(\v)$,  
where $\lambda = (\pih, \pip)$.

\subsection{Algorithms for Multiplicative Noise}
\label{subsec:alg_mult}
For model \textbf{M2}, recall that $\boldsymbol{x}$ and $\boldsymbol{y}$ represent viral loads of individual samples and pools,
respectively. The core algorithm presented in
\cite{Ghosh2021} uses the \lasso{} estimator \cite{THW2015}, 
\begin{equation}
\boldsymbol{\widehat{x}}^{\lasso{}} =\text{arg}\min_{\boldsymbol{x}} \|\boldsymbol{y}-\boldsymbol{Ax}\|^2_2 + \rho \|\boldsymbol{x}\|_1,
\end{equation}
where $\rho$ is a smoothness parameter. 
\lasso{} exploits the sparsity of $\boldsymbol{x}$ but uses no SI. Despite the multiplicative nature of the noise in $\y$, \lasso{} yields good estimation performance~\cite{Ghosh2021} for three measures: 
({\em i}) {\em relative root mean squared error} (RRMSE) $= \|\boldsymbol{x}-\boldsymbol{\widehat{x}}\|_2/ \|\boldsymbol{x}\|_2$;
({\em ii}) {\em false negative rate} (FNR) = $\#$incorrectly detected negatives$\big/ \#$true positives; and
({\em iii}) {\em false positive rate} (FPR) = $\#$incorrectly detected positives$\big/ \#$true negatives.
Note that FNR $=1-$ sensitivity and FPR $=1-$ specificity. 

In some cases, the $n$ individuals in $\boldsymbol{x}$ can be partitioned into $n_1 \ll n$ \emph{disjoint}
groups, for example, family members, 
who interact closely with each other and are 
likely to pass the virus between group members. 
This family-style structure leads to a situation where either group members are all uninfected, or a majority of group members are infected. Note that the family-style structure also includes groups of coworkers, 
students taking a course together, and people sharing accommodation. 
If reliable SI about how the $n$ individuals are partitioned into families is available, and only 
a small portion of families, $n_2 \ll n_1$, are infected, then \lasso{} can be replaced by {\em group \lasso{}} (\textsc{Glasso}) \cite{Yuan2006}. The \textsc{Glasso} is defined as:
\begin{equation}
\boldsymbol{\widehat{x}}^\glasso{} = \operatorname{arg}\min_{\boldsymbol{x}} \|\boldsymbol{y}-\boldsymbol{Ax}\|^2_2 + \rho \sum_{g=1}^{n_1} \|\boldsymbol{x}_g\|_2,
\end{equation}
where $\boldsymbol{x}_g$
is comprised of viral loads of people from the $g$th family. 
We observed through numerical experimentation that \sqrtglasso{}, which is an estimator with an $\ell_2$ data fidelity term \cite{Belloni2011} instead of a squared $\ell_2$ one, 
outperformed \textsc{Glasso}. The \sqrtglasso{} estimator is defined as:
\begin{equation}
\boldsymbol{\widehat{x}}^\sqrtglasso{} = \operatorname{arg}\min_{\boldsymbol{x}} \|\boldsymbol{y}-\boldsymbol{Ax}\|_2 + \rho \sum_{g=1}^{n_1} \|\boldsymbol{x}_g\|_2.
\label{eq:sqrt-glasso}
\end{equation}
In contrast, conventional \lasso{} without 
any notion of groups outperformed \textsc{Sqrt-Lasso} in our experiments.

In some cases, family SI is unavailable or does not precisely reflect the contact scenario. We may in turn use CT SI commonly available via Bluetooth~\cite{Hekmati2020}
to directly \emph{infer} group structure required by \eqref{eq:sqrt-glasso} using clique detection algorithms;
contacts between members of different families can also be considered to form small cliques.
In particular, we use the Bron--Kerbosch algorithm \cite{Bron1973} to find maximal cliques in the CT graph, and label each maximal clique as a group. Note that one could generate these groups differently~\cite[Sec.~7]{Jacob2009}, 
for example, decomposition into $k$-clique communities \cite{Palla2005}. Such a decomposition may partition the $n$ individuals into $n_3 \ll n$ groups that \emph{overlap} with each other, unlike the earlier case of disjoint families.
In a scenario with overlapping groups, we use the following {\em overlapping group square-root \lasso{}} (\sqrtoglasso{})
estimator~\cite{Jacob2009}:
\begin{equation}
\boldsymbol{\widehat{x}}^\sqrtoglasso{} = \operatorname{arg}\min_{\boldsymbol{x}} \|\boldsymbol{y}-\boldsymbol{Ax}\|_2 + \rho\, \Omega_{\text{overlap}}(\boldsymbol{x}),
\label{eq:oglasso1}
\end{equation}
where $\Omega_{\text{overlap}}(\boldsymbol{x}) = \inf_{\boldsymbol{v} \in \mathcal{V}_G, \sum_{g \in G} \boldsymbol{v}_g = \boldsymbol{x}}  \sum_{g \in G} \|\boldsymbol{v}_g\|_2$,
$G$ denotes a set of possibly overlapping groups each equal to a subset of $\{1,2,...,n\}$, 
$\boldsymbol{v} = (\boldsymbol{v}_g)_{g \in G}$, 
$g$ is a group belonging to $G$, $\boldsymbol{v}_g \in \mathbb{R}^n$ is a vector whose support is a subset of $g$ and $\mathcal{V}_G$ is the set of all such vectors $\boldsymbol{v}_g$. In case of overlapping groups, coordinate-descent algorithms are not guaranteed to converge to the optimal solution for the \textsc{Glasso} penalty \cite[Sec.~4.3.3]{THW2015}.
Furthermore, the \textsc{Oglasso} estimator typically has support that is a union of groups as desired in contrast to the \textsc{Glasso} estimator whose support is typically the complement of a union of groups \cite[Fig.~1 and Sec.~3]{Jacob2009}. 
The estimator in \eqref{eq:oglasso1} can also be equivalently expressed as:
\begin{equation}
\{\widehat{\boldsymbol{v}_g}\}_{g \in G} = \operatorname*{arg\,min}_{\{\boldsymbol{v}_g\}_{g \in G}} \norm[\big]{\boldsymbol{y} \! - \! \boldsymbol{A} \big( \! \sum_{g \in G}\boldsymbol{v}_g \! \big) }_2 + \rho \! \sum_{g \in G} \|\boldsymbol{v}_g\|_2,
\end{equation}
yielding the estimate $\boldsymbol{\widehat{x}}^\sqrtoglasso{} := \sum_{g \in G} \widehat{\boldsymbol{v}_g}$.

All three algorithms included a non-negativity constraint on $\boldsymbol{x}$, and were preceded by a step that executed {\em combinatorial orthogonal matching pursuit} (\comp{}). 
\comp{} declares all samples that contributed to a pool with zero viral loads to be negative.
This \comp{} preprocessing step reduces the problem size by eliminating elements from $\y$ and $\x$, thereby improving the performance and speed of all three algorithms.
Moreover, \textsc{Comp} preserves various properties of the pooling matrix $\A$ such as its restricted isometry constant (RIC) or restricted nullspace property (RSNP). All these benefits have been discussed in prior work \cite{Ghosh2021}. We refer to our algorithms as \complasso{}, \compsqrtglasso{}, and \compsqrtoglasso{}. 
After the \textsc{Comp} step, we sometimes obtain a positive pool with only one contributing sample because all other samples that contributed to the pool were eliminated. Such a sample is termed a ``sure positive.''
Such an approach is part of the \emph{definite defectives} (\textsc{Dd}) algorithm in the group testing literature~\cite[Sec. 2.4]{aldridge2019group}. However, \textsc{Dd} considers all samples that were neither eliminated by the \textsc{Comp} step nor declared sure positives, to be negative. This can give rise to many false negatives. In our approach, we instead incorporate the following post processing step: If a sample is declared a sure positive by \textsc{Dd} but negative by our algorithms such as \complasso{}, \compsqrtglasso{}, or \compsqrtoglasso{}, we declare it as positive.

\EditRevision{Compressed sensing estimators have been used in nonadaptive group testing earlier \cite{Ghosh2021, Gilbert2008}. However, to our best knowledge, there is no prior work that uses different forms of group sparsity for nonadaptive group testing. Furthermore, we clarify that unlike the GAMP algorithms from Sec.~\ref{Sec:gamp_binary}, the algorithms in this subsection do not make explicit use of any infection probabilities. They only make use of records about family memberships, and information regarding which pairs
of people were in contact with each other without using the time of contact $\tau^{(d)}_{ij}$ or proximity values $d^{(d)}_{ij}$. 
\EditRevision{We also remark that the algorithms for model \textbf{M2} do not utilize results of tests from previous days and consequently do not require any assumptions on the amount of time required to generate test results.}}

\section{Pooling Matrix Design}
\label{sec:matrix_des}

In the context of group testing applications such as pooled RT-PCR tests for COVID-19, it is imperative that the sensing matrix $\A$ obey some constraints.
First, we require the matrix to be binary. 
The binary constraint  simplifies the process of pipetting, 
which may be performed manually or by a liquid handling robot.
Second, we want the matrix to have fixed numbers of ones per row and per column, which are denoted by $r$ and $c$, respectively.
Using small $c$ and $r$ reduces problems such
as sample dilution (c.f., Aldridge et al.~\cite[Secs.~4.5 and 5.9]{aldridge2019group}). For this purpose, balanced binary matrices that satisfy these constraints (details below)
are appropriate.

In this section, we present a method to design compressed sensing matrices. Starting from randomly generated balanced binary matrices, 
which have fixed numbers of ones per column and row,
we present an efficient technique to iteratively modify the matrix by optimizing a quality measure of the matrix. Further, we discuss another quality measure that allows us to incorporate SI in the matrix design.

\subsection{Definition and Properties of Balanced Binary Matrices}

We begin by formally defining our constraints.
\begin{definition} \label{def:1}
    Let $m, n \in \mathbb{N}$. An $m \times n$ matrix $\A$ is said to be a \emph{balanced binary matrix} if it satisfies the following conditions:
    \begin{itemize}
        \item Each entry of $\A$ is either 0 or 1.
        \item Each row of $\A$ has $r$ ones, i.e. $\forall i \in [m], \sum_{j=1}^n A_{ij} = r$.
        \item Each column of $\A$ has $c$ ones, i.e. $\forall \! j \! \in \! [n], \sum_{i=1}^m \! A_{ij} \! = \! c$.
    \end{itemize}
\end{definition}
\noindent{}Here, $[k]$ denotes the set $\{1, 2, \dots, k\}$.
The quantities $m, n, r$, and $c$ must satisfy $mr = nc$, because the number of ones in the matrix appears in both sides of the equality. 
For $m, n, r, c \in \mathbb{N}$, denote the set of all balanced binary matrices 
that obey Definition~\ref{def:1} by $\mathfrak{M}(m,n;r,c)$.
We now describe the construction of a canonical matrix \cite[p. 165]{Bruladi}, which will be useful in our matrix design strategy.

\begin{definition}
    The \emph{canonical matrix} $\boldsymbol{\widetilde{A}} \in \mathfrak{M}(m,n;r,c)$ is the $m \times n$ binary matrix that has a $1$ at index $(i,j)$ if and only if $i-x \equiv 0$ mod $m$ has a solution satisfying $(j-1)c < x \le jc$.
\end{definition}

If $r$ divides $n$ (or, equivalently, $c$ divides $m$), 
then the rows and columns of the 
canonical
matrix can be reordered to obtain a matrix with a block-diagonal structure, where each block is a $c \times r$ matrix comprised entirely of ones.
Fig.~\ref{fig:canon-matrix} gives an example of a canonical matrix and its rearranged block diagonal form.

\begin{figure}[!t]
\centering
\subfloat[]{
\scalebox{0.83}{$\begin{bmatrix}
    1 & 0 & 0 & 1 & 0 & 0 & 1 & 0 & 0 \\
    1 & 0 & 0 & 1 & 0 & 0 & 1 & 0 & 0 \\
    0 & 1 & 0 & 0 & 1 & 0 & 0 & 1 & 0 \\
    0 & 1 & 0 & 0 & 1 & 0 & 0 & 1 & 0 \\
    0 & 0 & 1 & 0 & 0 & 1 & 0 & 0 & 1 \\
    0 & 0 & 1 & 0 & 0 & 1 & 0 & 0 & 1 \\
\end{bmatrix}$}}
\quad
\subfloat[]{
\scalebox{0.83}{$\begin{bmatrix}
    1 & 1 & 1 & 0 & 0 & 0 & 0 & 0 & 0 \\
    1 & 1 & 1 & 0 & 0 & 0 & 0 & 0 & 0 \\
    0 & 0 & 0 & 1 & 1 & 1 & 0 & 0 & 0 \\
    0 & 0 & 0 & 1 & 1 & 1 & 0 & 0 & 0 \\
    0 & 0 & 0 & 0 & 0 & 0 & 1 & 1 & 1 \\
    0 & 0 & 0 & 0 & 0 & 0 & 1 & 1 & 1 \\
\end{bmatrix}$}}
\vspace{-2mm}
\caption{Example of (a) a canonical matrix, and (b) its rearranged block diagonal form for $m = 6, n = 9, r = 3, c = 2$.}
\vspace{-4mm}
\label{fig:canon-matrix}
\end{figure}

We now define a quality measure on the set of $m \times n$ real-valued matrices.

\begin{definition}
    Let $\A$ be an $m \times n$ real-valued matrix. Then, its \emph{Gram matrix} $G(\A)$ is defined as $G(\A) = \A^T \A$.
\end{definition}
We define the quality measure $\psi: \mathbb{R}(m,n) \to \mathbb{R}$ as the squared Frobenius norm of the Gram matrix, i.e.,
\begin{equation}
    \psi(\A) = \|G(\A)\|^2_F := \sum_{i \in [m]} \sum_{j \in [n]} g_{i,j}^2,
    \label{eq:psiA}
\end{equation}
where $G(\A) = (g_{i,j})_{n \times n}$. Note that $\psi(\A)$ is equal (up to a scalar) to the average squared coherence $\mu_{\text{avg}}(\A)$ defined in Sec.~\ref{subsec:pooling_matrix_design}. We now provide lower and upper bounds on the value of $\psi(\A)$ for balanced binary matrices.
As a first step toward establishing these bounds, we state the following elementary result for
entries of $G(\boldsymbol{A})$. We omit the proof for brevity.

\begin{lemma}
\label{lem:basic}
    Let $\A \in \mathfrak{M}(m,n;r,c)$. Let $G(\A) = (g_{i,j})_{n \times n}$. Then,
    \begin{itemize}
        \item $g_{i,i} = c$ for each $i \in [n]$,
        \item $0 \le g_{i,j} \le c$ for each $i,j \in [n]$, and
        \item $ \displaystyle \sum_{i \in [n]} \sum_{j \in [n]} g_{i,j} = nrc.$
    \end{itemize}
\end{lemma}

The next lemma provides bounds on the values of the  
quality measure $\psi$ (\ref{eq:psiA}).

\begin{lemma}
\label{lem:bounds}
    Let $\A \in \mathfrak{M}(m,n;r,c)$. Then,
    \begin{equation}
        nc(r+c-1) \le \psi(\A) \le nrc^2.
    \end{equation}
    Further, the lower bound is achieved when each nonzero off-diagonal element of $G(\A)$ equals 1, which is possible only if $rc-c \le n-1$, 
    and the upper bound is achieved when each nonzero off-diagonal element of $G(\A)$ equals $c$.
\end{lemma}

\begin{proof}
    Using Lemma~\ref{lem:basic} in \eqref{eq:psiA}, it can be shown that
    \begin{align*}
        \psi(\A) = nc^2 + \sum_{i,j \in [n], i \ne j} g_{i,j}^2,
    \end{align*}
    with
    \begin{equation}
        \sum_{i,j \in [n], i \ne j} g_{i,j} = nrc - nc.
    \end{equation}
    To solve the above, we provide a solution to a more general problem, namely, given an arbitrary finite multiset $S$ comprised of positive integers, each less than or equal to $c$ with sum equal to $sc$ with $s \in \mathbb{N}$, find lower and upper bounds on $\varphi(S)$, where $\varphi(S)$ denotes the sum of squares of the integers in $S$. For any multiset $S$ satisfying these conditions, we define two transformations $f$ and $g$  as follows:
    \begin{itemize}
        \item $f(S)$ is obtained by replacing a maximum element $x$ of $S$ with $x$ copies of 1, and
        \item $g(S)$ is obtained by replacing the two smallest elements $a$ and $b$ of $S$ with
        \begin{itemize}
            \item the element $a+b$, if $a+b \le c$,
            \item the elements $c$ and $a+b-c$, otherwise.
        \end{itemize}
    \end{itemize}
    Using these definitions, it can be shown that:
    \begin{itemize}
        \item $\varphi(f(S)) \le \varphi(S) \le \varphi(g(S))$.
        \item $f^\ell(S) = S^-$ and $g^\ell(S) = S^+$ for a sufficiently large $\ell$ ($\ell \ge sc$), where $S^-$ is the multiset comprised of $sc$ copies of 1 and $S^+$ is the multiset comprised of $s$ copies of $c$. Further, we have $\varphi(S^-) = sc$ and $\varphi(S^+) = sc^2$.
        \item $\varphi(f(S)) < \varphi(S)$ if $S \ne S^-$, and $\varphi(S) < \varphi(g(S))$ if $S \ne S^+$.
    \end{itemize}
    Combining the first two properties, we obtain
    \begin{equation}
        sc \le \varphi(S) \le sc^2.
    \end{equation}
    Using the last property, we obtain that the lower bound is achieved if and only if $S = S^-$, and the upper bound is achieved if and only if $S = S^+$. Using these results in our original problem with $s = nr-n$ yields the desired result.
\end{proof}

\subsection{\EditRevision{Motivation for the Use of Balanced Binary Matrices}}
\label{subsec:justify-balanced-mat}
\EditRevision{The motivation for using
\emph{binary} matrices in RT-PCR pooling was discussed
in\cite{Ghosh2021}. The motivation for using \emph{balanced} matrices is to ensure that each pool contains contributions from the same number of samples. This makes the pooling process easier, especially if it is to be done manually. Also, as the sample volume for RT-PCR is fixed, a larger size for some pools would lead to sample dilution in those pools. Balanced binary matrices also ensure that no sample contributes to too many pools, which would have led to sample depletion. We refer the reader to 
Sec.~\ref{subsec:unbalanced_mat_result}
for some empirical performance comparisons between balanced binary matrices and some classes of unbalanced binary matrices. 
}

\subsection{Search Algorithm for Balanced Binary Matrix}
\label{sec:algo}

We begin by defining the \emph{interchange operation}~\cite[p. 170]{Bruladi}, which allows our algorithm to generate one balanced binary matrix from another.

\begin{definition}\label{def:interchange}
    For a binary matrix $\A$, an \emph{interchange} operation is defined as the replacement of a submatrix
    \begin{equation}
    \label{eqn:sub-matrix}
        \begin{pmatrix}
            1 & 0\\
            0 & 1
        \end{pmatrix}
        \textrm{ with }
        \begin{pmatrix}
            0 & 1\\
            1 & 0
        \end{pmatrix},
    \end{equation}
    or vice versa.
\end{definition}

The set $\mathfrak{M}(m,n;r,c)$ is closed under the interchange operation. Based on this operation, we define a relation $R$ on the set $\mathfrak{M}(m,n;r,c) \times \mathfrak{M}(m,n;r,c)$ as follows: $(\A,\boldsymbol{B}) \in R$ if and only if $\A$ can be obtained from $\boldsymbol{B}$ by performing a single interchange operation. This relation induces an undirected graph $G(m,n;r,c)$ with $\mathfrak{M}(m,n;r,c)$ as the set of vertices. It has been proved that the graph $G(m,n;r,c)$ is connected and its diameter (maximum distance between any two graph vertices) is less than or equal to $mr$ (or $nc$) \cite[Theorem 1]{Walkup}. 

\begin{algorithm}[!t]
\DontPrintSemicolon
    \KwInput{$m,n,r,c \in \mathbb{N}$ satisfying $r \le n, c \le m$ and $mr = nc$, $N \in \mathbb{N}$}
    \KwOutput{A matrix $\A$ optimized w.r.t. the metric $\psi$} 
    $\boldsymbol{A} = \boldsymbol{\widetilde{A}}$ (Initialize $\A$ with the canonical matrix)\;
    $\boldsymbol{G} = \A^T \A$; $V = \psi(\A)$\;
    \For{$i = 0$ \KwTo $N$}{
      Find a sub-matrix of $\A$ as in \eqref{eqn:sub-matrix}\;
      Perform an interchange operation on $\A$ to obtain another balanced binary matrix $\boldsymbol{A'}$\;
      Update the Gram matrix $\boldsymbol{G}$ to obtain $\boldsymbol{G'} := G(\boldsymbol{A'})$\;
      Compute $\Delta_{\psi} := \psi(\boldsymbol{A'}) - \psi(\boldsymbol{A})$\;
      \If{$\Delta_{\psi} < 0$}{
        $V = V + \Delta_{\psi}$\;
        $\boldsymbol{A} = \boldsymbol{A'}$; $\boldsymbol{G} = \boldsymbol{G'}$\;
      }
      \If{$V == nc(r+c-1)$}{\textbf{break }}
    }
\caption{Balanced Binary Pooling Matrix Design}
\label{algo:psi-opt}
\end{algorithm}

\begin{figure*}[!t]
    \centering
    \includegraphics[scale=0.62]{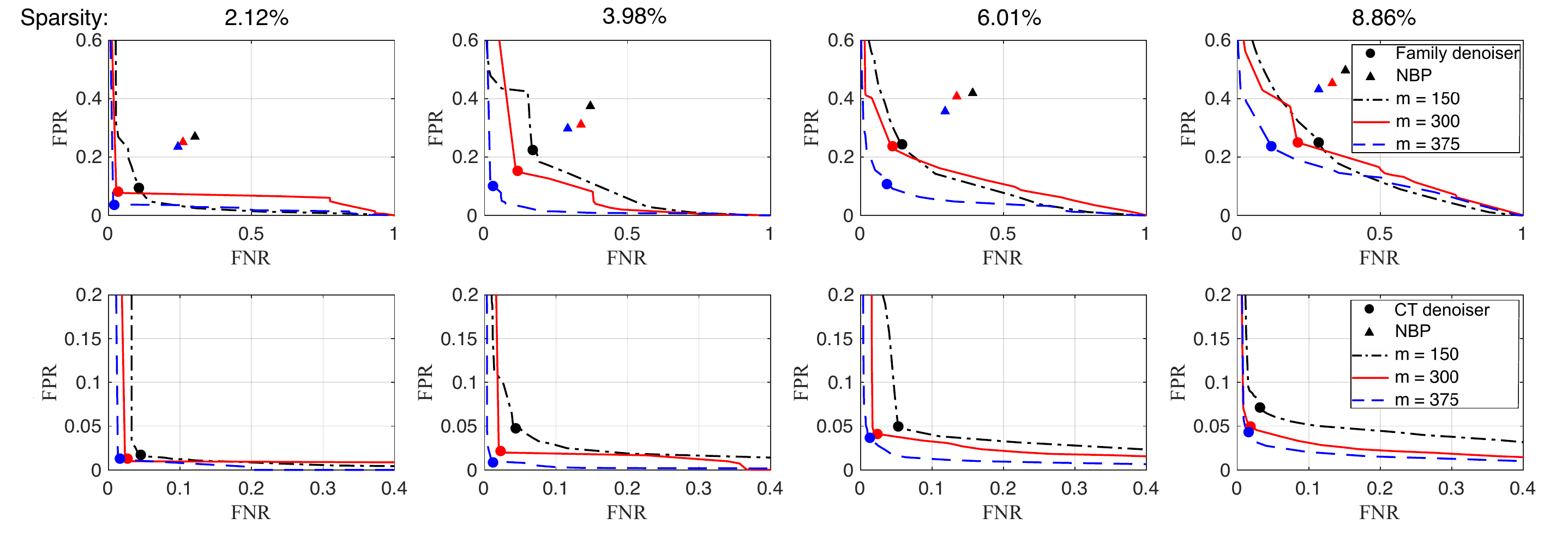}
    \vspace{-6mm}
    \caption{Performance of \textbf{M1} in terms of ROC when family denoiser (top row) and CT denoiser (bottom row) are used.
    Columns correspond to averaged sparsity levels ranging from $2.12\%$ to $8.86\%$.
    Within each plot, the performance under three measurement levels for a population of $n = 1000$ individuals is compared. 
    The \EditRevision{circular} dot on each curve corresponds to an operating point that minimizes the sum of FPR and FNR.
    The CT denoiser significantly outperforms the family denoiser with error rates mostly below $0.05$.
    The estimation problem is more challenging when fewer measurements are used at a higher sparsity level.
    \EditRevision{The results of a baseline algorithm (NBP~\cite{bickson2011fault}) that does not exploit SI are shown in triangular markers. The baseline performs significantly worse than both GAMP-based algorithms since its operating points are far away from the ROC curves for the family denoiser.} }
    \label{fig:roc_M1}
    \vspace{-4mm}
\end{figure*}

We now present Algorithm~\ref{algo:psi-opt}\footnote{\EditRevision{A MATLAB implementation of Algorithm~\ref{algo:psi-opt} can be found at
\href{https://github.com/Riteshgoenka/Group-Testing/blob/main/Pooling\%20Matrix\%20Design/Balanced\%20Binary\%20Matrices/Generation/psi_optimize_balanced.m}{https://github.com/Riteshgoenka/Group-Testing}.
}} for optimizing a pooling matrix with respect to (w.r.t.) the quality measure $\psi$. In Step 6 
of Algorithm~\ref{algo:psi-opt}, we compute $G'$ through an update instead of recomputing it. Further, we compute the new value of metric $\psi(\A')$ by computing the change in metric $\psi(\A') - \psi(\A)$. Finally, if $m \le n$, running the algorithm with $m,n$ interchanged and $r,c$ interchanged is faster. In such cases, the output matrix can be transposed to obtain the required matrix.

We define a matrix $\A \in \mathfrak{M}(m,n;r,c)$ to be $\psi$-optimal if it achieves the lower bound in Lemma~\ref{lem:bounds}. This lower bound is achieved when each nonzero off-diagonal element of the Gram matrix is equal to one. Note that this is equivalent to the dot product between any two distinct columns of $\A$ not exceeding one, which is further equivalent to the dot product between any two distinct rows of $\A$ not exceeding one. The upper bound is achieved in Lemma~\ref{lem:bounds} when each nonzero off-diagonal element of $G(\A)$ is equal to $c$. This is the case when the matrix has a block diagonal structure similar to the one that
the canonical matrix assumes with its rows/columns re-ordered when $c$ divides $m$. \EditRevision{We observe (but have not theoretically proved) that for many values of $m,n,r,c$ (we typically use $c = 3$) of our interest, Algorithm~\ref{algo:psi-opt} produces a $\psi$-optimal matrix. We always set $N = nrc$ in Algorithm~\ref{algo:psi-opt}. In TABLE~\ref{tab:mat-des}, we report the number of iterations and CPU runtime required to design $\psi$-optimal matrices for some $m,n,r,c$ values.}
\begin{table}[h!]
\centering
\caption{\EditRevision{Number of iterations required to terminate, $M$, and CPU runtime in seconds, $t_r$, for some designed $\psi$-optimal matrices. We set $N = nrc$ in Algorithm~\ref{algo:psi-opt} when designing these matrices.}
}
\begin{tabular}{ccccccc}
\hline
$m$ & $n$ & $r$ & $c$ & $N$ & $M$ & $t_r$\\
\hline
150 & 1000 & 20 & 3 & 60000 & 8368 & 0.60\\
300 & 1000 & 10 & 3 & 30000 & 3892 & 0.37\\
375 & 1000 & 8 & 3 & 24000 & 4430 & 0.36\\
1000 & 10000 & 40 & 4 & 1600000 & 82036 & 13.29\\
2500 & 10000 & 40 & 10 & 4000000 & 557706 & 518.56\\
1200 & 100000 & 250 & 3 & 75000000 & 1152886 & 1318.33\\
\hline
\end{tabular}
\label{tab:mat-des}
\end{table}

\subsection{Incorporating SI in Pooling Matrix Design}
We now describe how to incorporate SI in the design of pooling matrices. 
Let $G$ be the contact graph---an undirected graph whose vertex set is the population $\{1,2, \dots, n\}$ and an edge from $i$ to $j$ denotes a contact between individuals $i$ and $j$ with its weight $p_{ij}$ equal to the probability of transmission of the disease. Then, one possible intuitive cost function can be defined as:
\begin{equation*}
    \phi(\A) = \sum_{i,j \in [n]} b_{ij} \, (\a_i \cdot \a_j)^2,
    \label{eq:phi_A}
\end{equation*}
where for any $(i,j)$, we define $b_{ij} = 0$ if $p_{i,j} = 0$, and $b_{ij} = 1$ otherwise. 
It is evident that $\phi(\A)$ penalizes the squared dot product of those pairs of columns that correspond to individuals that can potentially transmit the disease between themselves. Moreover, $\phi(\A)$ ignores the dot product between individuals that are not in contact with each other. Thus, $\phi(\A)$ can be interpreted 
as a weighted form of $\psi(\A)$. We could have also designed a cost function that uses the actual values of $p_{i,j}$ instead of $b_{ij}$, but
did not do so to avoid excessive dependence on the precise values of $p_{i,j}$. Our algorithm for encoder design has two stages. In the first stage, we design a $\psi$-optimal matrix using Algorithm~\ref{algo:psi-opt}. In the second stage, we optimize the obtained matrix w.r.t. the quality measure $\phi(\A)$. 
Recall that Algorithm~\ref{algo:psi-opt} traverses the space of balanced binary matrices using the interchange operation. In the second stage, we adopt a similar approach where the algorithm traverses the space of balanced binary matrices by interchanging two columns of the matrix. This ensures that the value of $\psi(\A)$ is preserved while the matrix is optimized w.r.t. the quality measure $\phi(\A)$. \EditRevision{Note that interchanging the columns of the matrix $\boldsymbol{A}$ while optimizing over $\phi$ does not affect the $b_{ij}$ values. This is because $b_{ij}$ values are related to the probability that individuals $i$ and $j$ infect each other, and swapping columns of the $\boldsymbol{A}$ matrix does not alter these probability values.}
\EditRevision{We have also observed (but have not theoretically proved) that} for $m, n, r, c$ values of interest and the sparse contact tracing graphs described in Sec.~\ref{sec:data_gen}, our approach produces a $\psi,\phi$-optimal matrix, i.e., a matrix $\A$ satisfying $\psi(\A) = nc(r+c-1)$ and $\phi(\A) = 0$.

\section{Numerical Results}
\label{sec:results}

\subsection{Experimental Conditions} 
The data was simulated based on the data generation model described in Sec.~\ref{sec:data_gen},
and group testing inference was performed using the algorithms proposed in Sec.~\ref{sec:algos}. 
We generated datasets with $n=1000$ individuals using four levels of cross-clique contacts, leading to four averaged sparsity levels, $2.12\%$, $3.98\%$, $6.01\%$, and $8.86\%$, for $\boldsymbol{x}$.
At each sparsity level, we perform pooling experiments using Kirkman triple matrices 
as proposed in \cite{Ghosh2021} using $m \in \{150, 300, 375\}$.
Measurement vectors $\y$ for \textbf{M1} were generated using probabilities for erroneous binary tests, $\Pr(y_i=1|w_i=0)=0.001$ and $\Pr(y_i=0|w_i>0)=0.02$, per Hanel and Thurner~\cite{hanel2020boosting}.
Vectors $\y$ for \textbf{M2} were generated by setting the parameter reflecting the strength of noise in RT-PCR to $\sigma^2 = 0.01$. The parameter \EditRevision{$q_a$} for RT-PCR was set to $0.95$. Fig.~\ref{fig:res-m1-m2} shows the performance of the proposed algorithms in terms of FNR and FPR
averaged across the inference results obtained for the time window of $50$ days described in Sec.~\ref{sec:data_gen}. For \textbf{M2}, we report error rates by thresholding the estimated viral load; samples with estimated viral load below $\tau_n \triangleq 0.2$ were considered negative. The error rates do not change much when $\tau_n$ varies between $0$ and $1$. 
The parameter $\rho$ in \complasso{}, \compsqrtglasso{}, and \compsqrtoglasso{} was selected using cross-validation, as described in \cite{Zhang2014}.

\subsection{Main Numerical Results for M1} 
\label{subsec:results_M1}

We tested model~\textbf{M1} using both the family denoiser~\eqref{eq:denoiser-family} and CT denoiser~\eqref{eq:denoiser-ct}.
The top row of Fig.~\ref{fig:res-m1-m2} summarizes the performance of \textbf{M1}, and Fig.~\ref{fig:roc_M1} presents complete 
receiver operating characteristic (ROC)
curves.
Fig.~\ref{fig:res-m1-m2} reveals that the CT denoiser outperforms the family denoiser in all settings.
Both algorithms yield lower (better) FNR and FPR as the number of measurements, $m$, increases.
Moreover, the CT denoiser's error rates are below $0.05$, except for the challenging cases where the sparsity level is $8.86\%$ and $m\in\{150,300\}$.

Fig.~\ref{fig:roc_M1} illustrates the performance of family and CT denoisers for \textbf{M1} at different measurement and sparsity levels.
The \EditRevision{circular} dot on each curve is the operating point that minimizes the total error rate, i.e., the sum of FPR and FNR, which correspond to the concise results of Fig.~\ref{fig:res-m1-m2}.
The closer a dot is to the origin of the FPR--FNR plane, the better the performance it reflects.
Comparing the ROC curves in the top and bottom rows, we note that the CT denoiser significantly outperforms the family denoiser at all sparsity levels.
The CT denoiser, with most of its FNR and FPR $< 5\%$, can achieve as low as $15\%$ of the total error rate of the family denoiser.
Across different sparsity levels, the algorithm performs less accurately as the sparsity level increases. 
In each plot, lower measurement rates make it more challenging for the group testing algorithm.

We also examine the stability of the thresholds corresponding to the operating points we selected to report results in Fig.~\ref{fig:res-m1-m2}. 
Our empirical results reveal that at a given sparsity level, the variation of the threshold due to different design matrices or denoisers is less than $0.003$.
As the sparsity level increases from $2.12\%$ to $8.86\%$, the threshold only drops from $0.160$ to $0.137$.
Hence, the threshold for minimizing the total error rate is insensitive to the testing conditions.

\EditRevision{We compare our proposed group testing algorithms to a baseline nonparametric belief propagation~(NBP) algorithm~\cite{bickson2011fault} that does not exploit side information.
We attach an additional output channel denoiser to the NBP algorithm to process the RT-PCR noise.
We evaluate the performance in terms of FPR--FNR pairs and plot them using triangular markers in Fig.~\ref{fig:roc_M1}. Since the operating points are far from the ROC curves for GAMP using the family denoiser, we conclude that our proposed group testing algorithms that exploit SI significantly outperform the baseline that does not use SI.}

\subsection{Additional Experiments for M1} 
\label{subsec:results_M1_additional}
\subsubsection{Using Prior Knowledge of Infection Status}
We now examine the advantage that prior knowledge of the population's infection status in the startup phase
provides our algorithm for the \textbf{M1} model.
As defined in \eqref{eq:SI_period_aggre}--\eqref{eq:est_pij_2nd}, 
we iteratively use the updated probability of infection, $\widehat{\Pr}(X_i=1)$, 
estimated from an SI period of 8 immediately preceding days. 
For days $k < 8$, we had to use the ground-truth infection status of 
each individual in the startup phase to generate the results reported in Sec.~\ref{subsec:results_M1}.
However, ground-truth infection data from 
the startup phase may provide 
our approach an unfair advantage over the 
algorithms proposed for \textbf{M2}. Below, we investigate whether this advantage is significant.

\begin{figure}[!t]
  \centering
  \vspace{-4mm}
  \includegraphics[width=\linewidth]{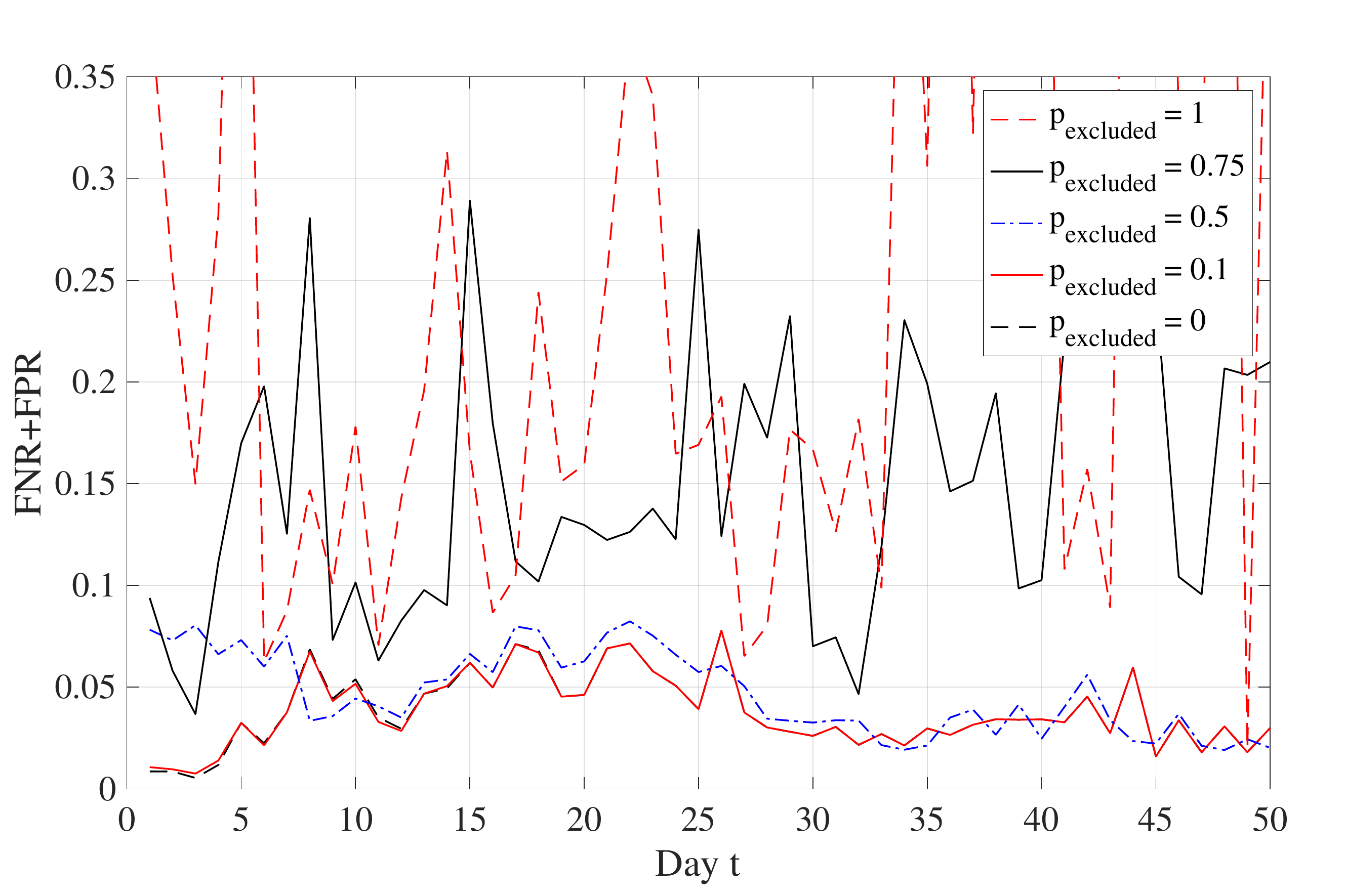}
  \vspace{-2mm}
  \caption{Performance of \textbf{M1} when a proportion, $p_{\text{excluded}}$, of the population's health states in the startup phase is unknown.
  The curves reveal that in the absence of up to $50\%$ prior knowledge of the infection status of the population, 
  the accuracy of \textbf{M1} is close to that when complete startup information is available.
  }
  \label{fig:edge_due_to_prior}
  \vspace{-4mm}
\end{figure}

We examine how varying the amount of startup
information impacts our algorithm's quality.
Specifically, we randomly replace a portion, $p_{\text{excluded}} \in \{0, 0.1, 0.5, 0.75, 1\}$, 
of the population's infection status by an estimated probability of infection, e.g., $5\%$, for a setup that has a true averaged sparsity level of $7.2\%$.
Using a probability instead of a binary value, $0$ or $1$, gives the algorithm soft probabilistic information instead of hard ground-truth style information.
Fig.~\ref{fig:edge_due_to_prior} shows that even with 50\% prior knowledge of the infection status of individuals, 
our detection accuracy for \textbf{M1} is close to that when using complete prior information after ramping up for eight days.
The averages of the total error rates across time for increasing $p_{\text{excluded}}$ are $0.038$, $0.039$, $0.046$, $0.148$, and $0.407$, respectively. 
We also tried to replace the startup infection status with an estimated probability of infection of $10\%$, but only observed negligible performance differences.
The results show that
the CT algorithm is robust to the absence of up to $50\%$ of startup infection information.

\subsubsection{Duration of Startup SI Period}
There is a trade-off between the accuracy of our algorithm and the startup SI infection status information that needs to be pre-collected before the initialization of the testing algorithm.
We investigated the impact of the duration of the startup SI on estimation performance by testing three startup SI durations, namely, 4, 8, and 12 days.
Our experimental results (omitted for brevity)
show that the estimation accuracy is somewhat insensitive to the duration of the SI period.
Hence, for the experiments conducted for this paper, we chose eight days as the SI period.

\subsection{Main Numerical Results for M2}
\label{subsec:results_M2}

\begin{figure*}[!t]
    \begin{minipage}[b]{.24\linewidth}
      \centering
      \centerline{\includegraphics[width=4.35cm]{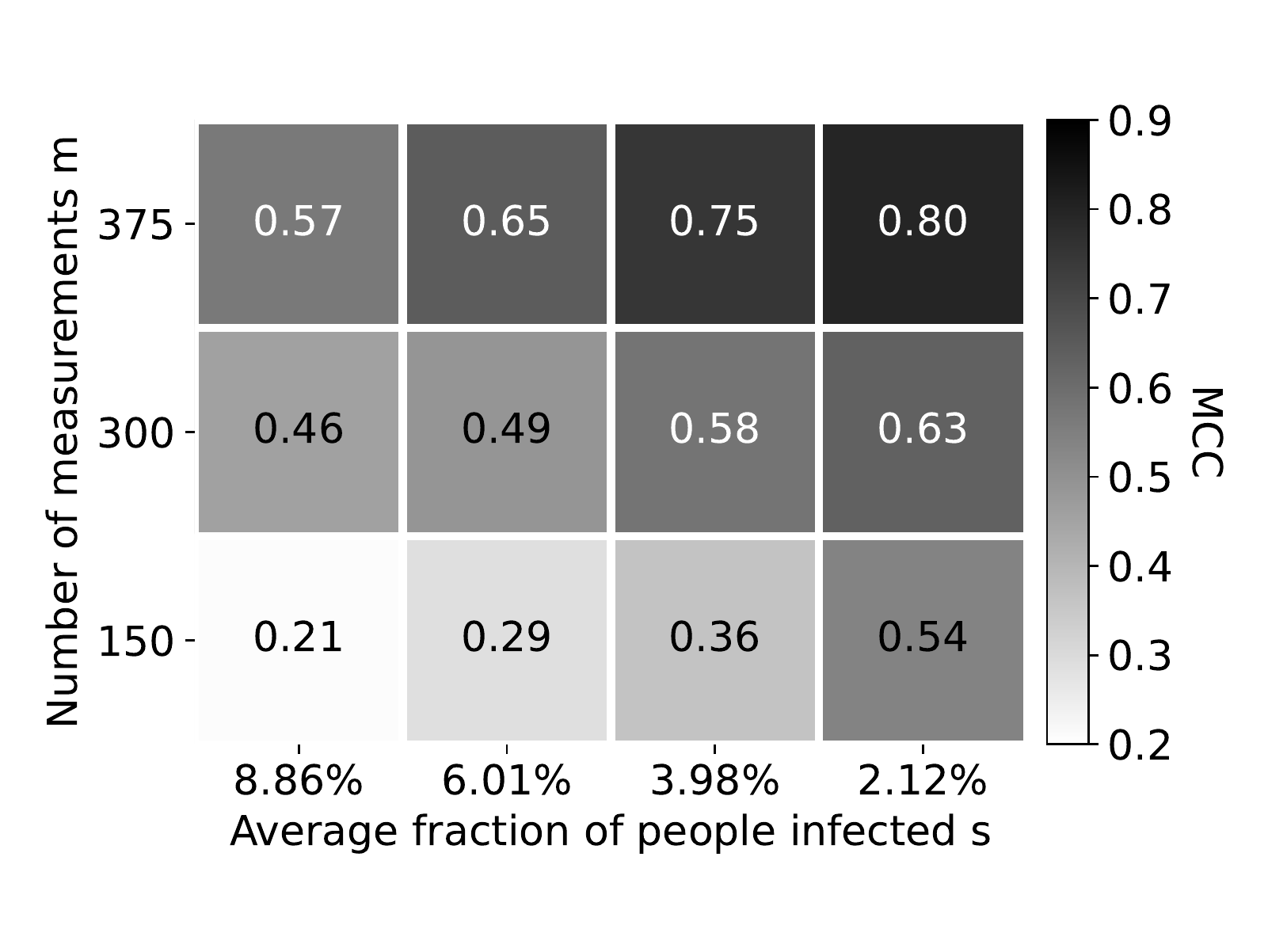}}
    \end{minipage}
    \hfill
    \begin{minipage}[b]{0.24\linewidth}
      \centering
      \centerline{\includegraphics[width=4.35cm]{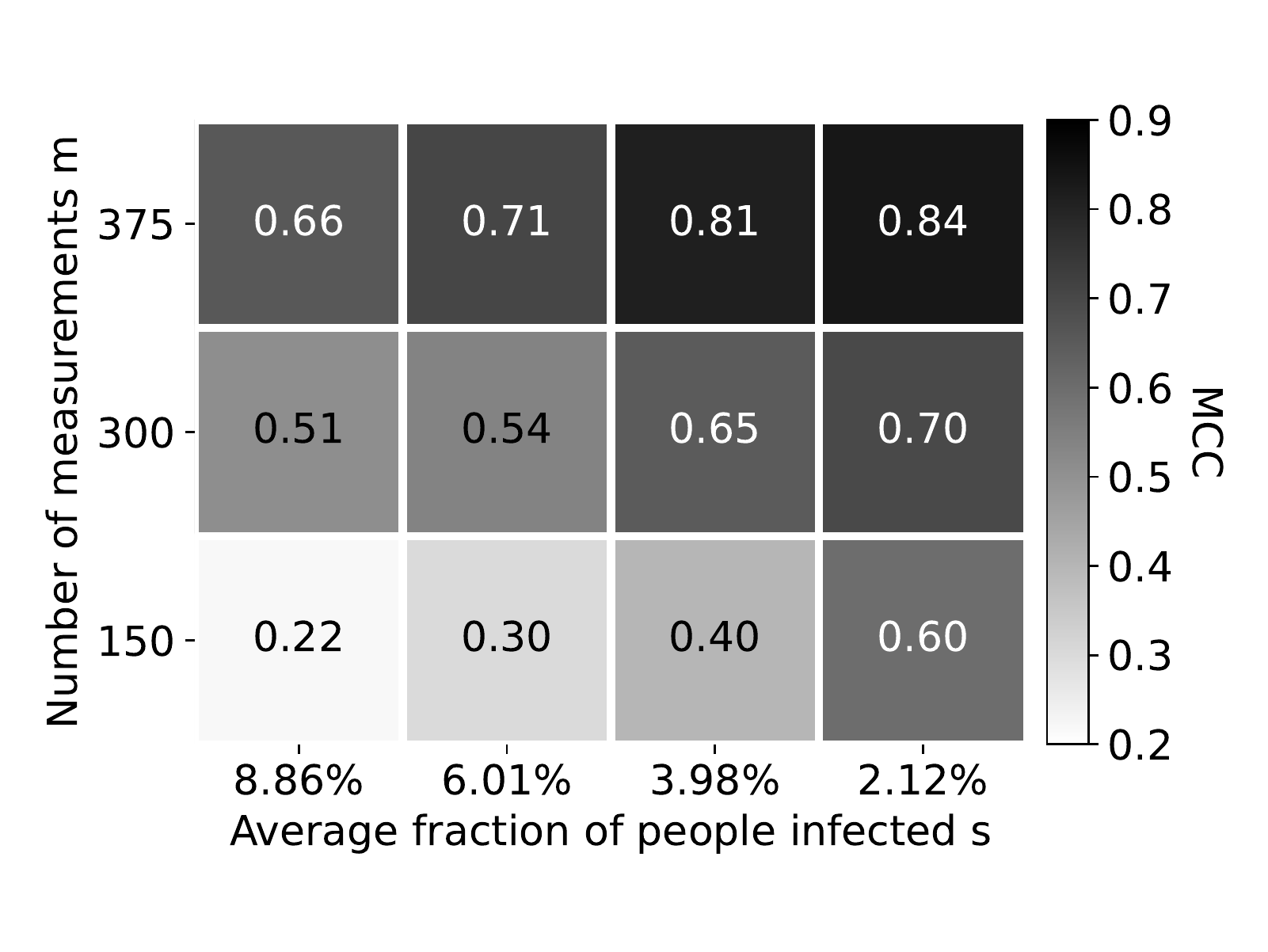}}
    \end{minipage}
    \hfill
    \begin{minipage}[b]{.24\linewidth}
      \centering
      \centerline{\includegraphics[width=4.35cm]{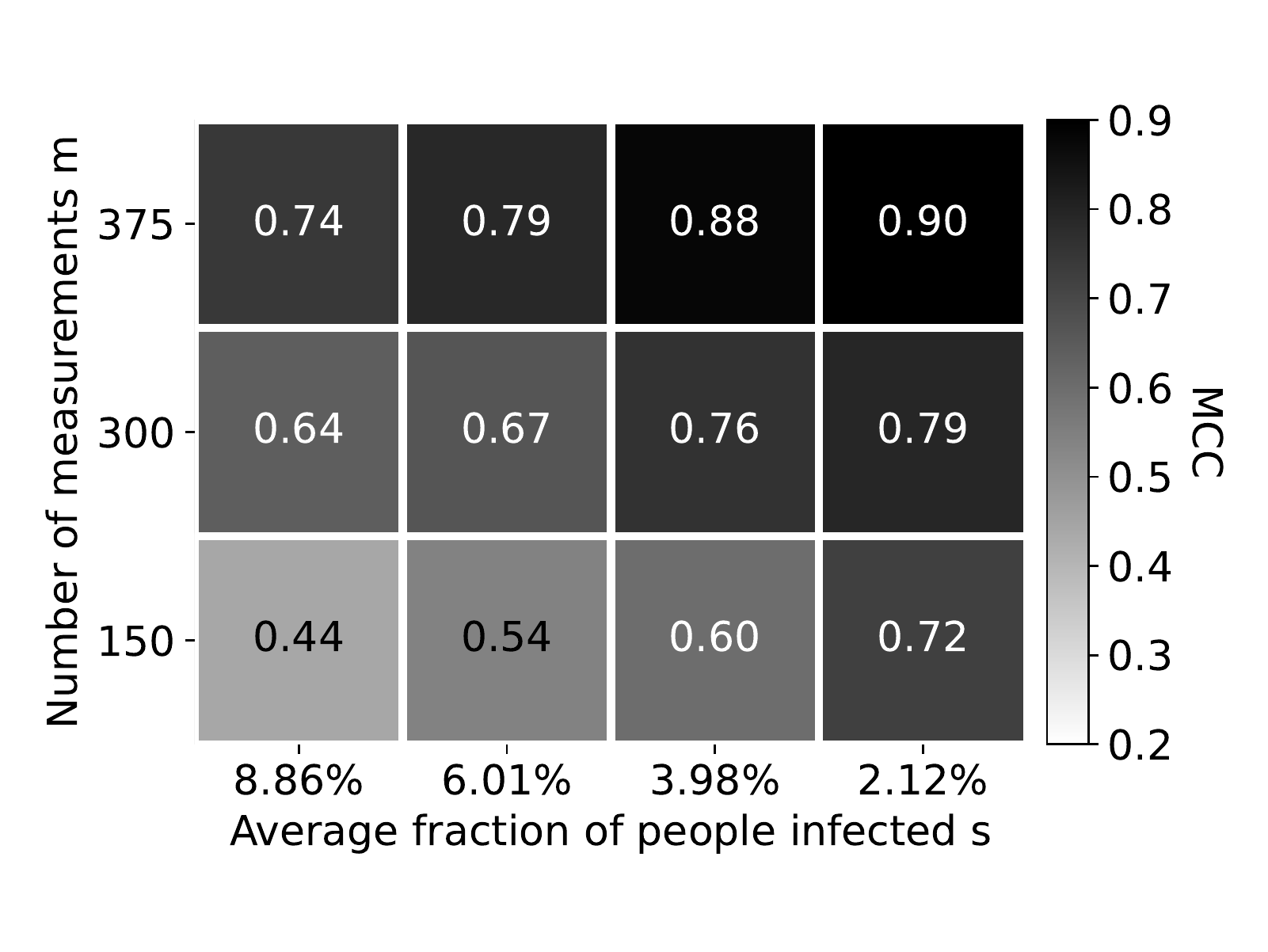}}
    \end{minipage}
    \hfill
    \begin{minipage}[b]{0.24\linewidth}
      \centering
      \centerline{\includegraphics[width=4.35cm]{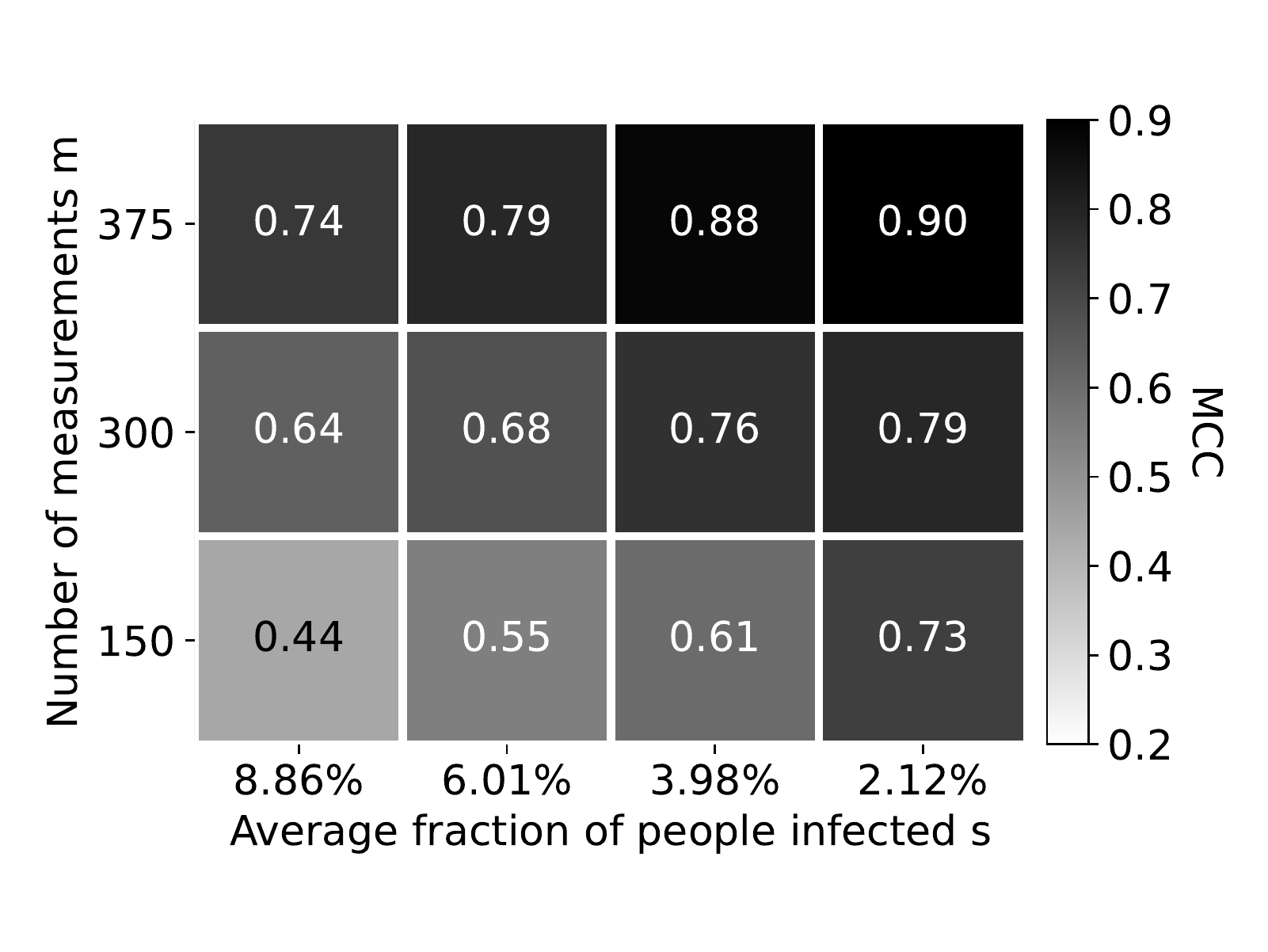}}
    \end{minipage}
    \vspace{-2mm}

    \hspace{62pt}
    \begin{minipage}[b]{.24\linewidth}
      \centering
      \centerline{\includegraphics[width=4.35cm]{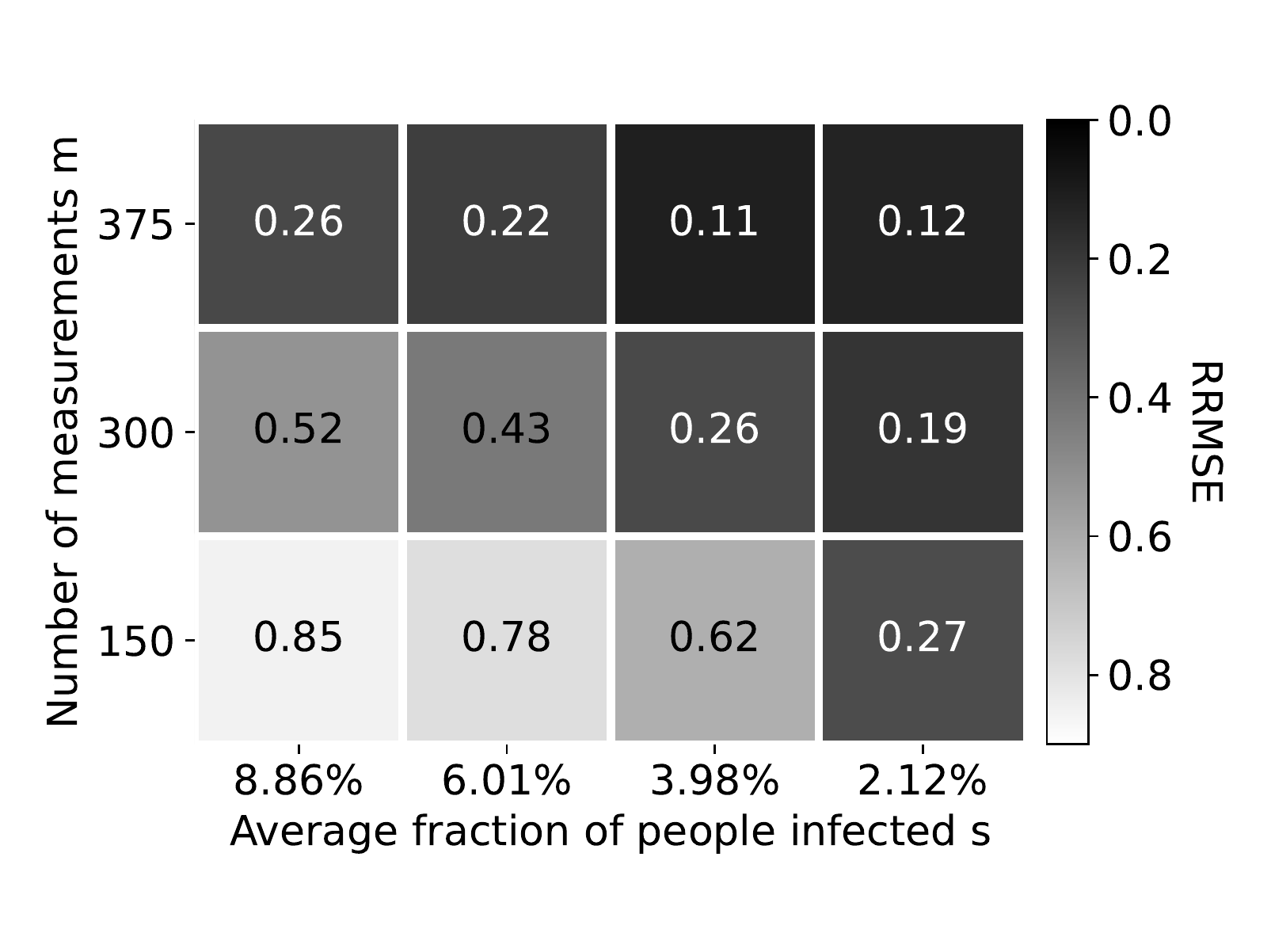}}
    \end{minipage}
    \hfill
    \begin{minipage}[b]{0.24\linewidth}
      \centering
      \centerline{\includegraphics[width=4.35cm]{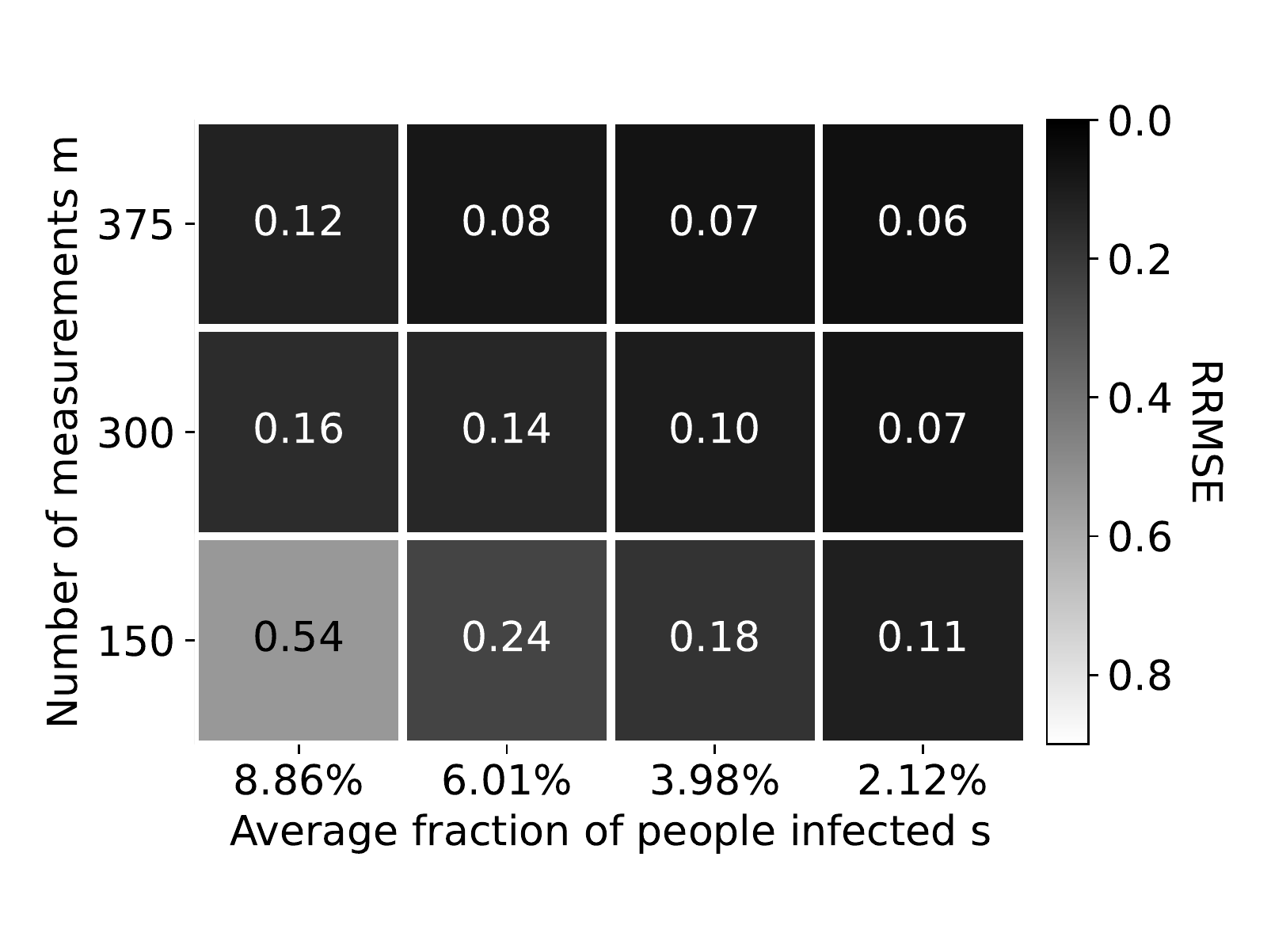}}
    \end{minipage}
    \hfill
    \begin{minipage}[b]{.24\linewidth}
      \centering
      \centerline{\includegraphics[width=4.35cm]{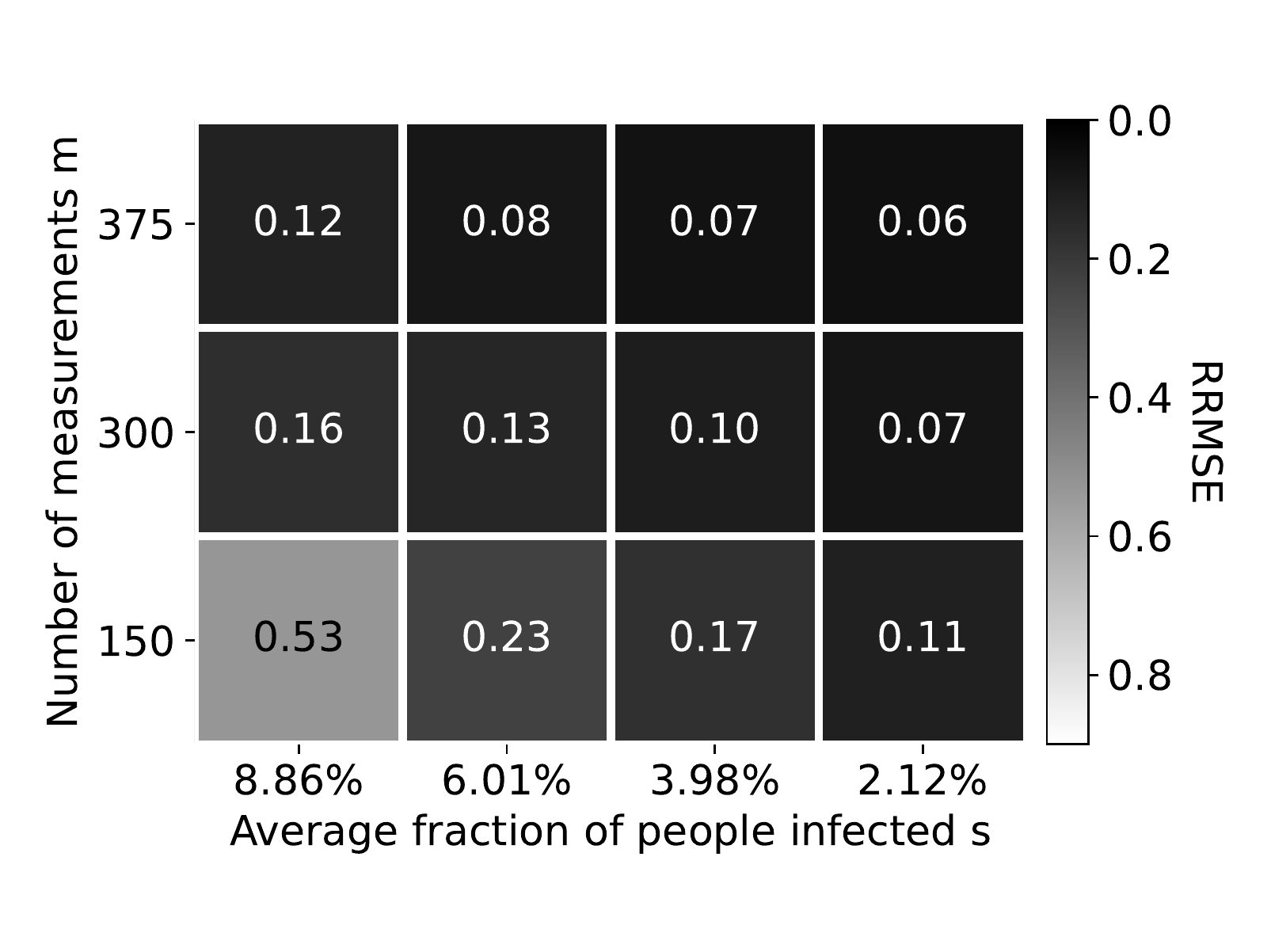}}
    \end{minipage}
    \hspace{62pt}
    \vspace{-4mm}
    \caption{Mean MCC (top row) and mean RRMSE (bottom row) obtained using \comp{}, \complasso{}, \compsqrtglasso{}, and \compsqrtoglasso{} (from left to right) for an experiment in Sec.~\ref{subsec:results_M2}. RRMSE is not applicable to \comp{} since it does not predict viral loads.}
    \label{fig:res-mcc-m2}
    \vspace{-4mm}
\end{figure*}

For model \textbf{M2}, we tested four algorithms: \comp{}, \complasso{}, \compsqrtglasso{}, and \compsqrtoglasso{}.
For \compsqrtglasso{}, we label the cliques in the block diagonal structure of the contact tracing matrix as groups. For \compsqrtoglasso{}, groups correspond to the maximal cliques in the contact tracing graph obtained by considering contacts during the seven days preceding the testing date.
The bottom row of Fig.~\ref{fig:res-m1-m2} 
compares
the performance of all four algorithms in terms of (FPR, FNR) for a variety of different values of $m$ as well as sparsity, given a population of $n = 1,000$ individuals. We also perform a comparison in terms of the Matthews correlation coefficient (MCC), 
\begin{equation}
\text{MCC} = \dfrac{\text{TP} \times \text{TN} - \text{FP} \times \text{FN}}{\sqrt{(\text{TP}\!+\!\text{FP})(\text{TP}\!+\!\text{FN})(\text{TN}\!+\!\text{FP})(\text{TN}\!+\!\text{FN})}},
\end{equation}
which is a one-number statistic to characterize the performance of binary classification algorithms \cite{Chicco2020}. Its values range from $-1$ to $+1$, where a value closer toward $+1$ is desirable. Here, $\text{TP}, \text{TN}, \text{FP}, \text{and } \text{FN}$ stand for the number of true positives, true negatives, false positives, and false negatives, respectively.

Fig.~\ref{fig:res-mcc-m2} 
compares the performance of the algorithms under consideration in terms of mean MCC and mean RRMSE values.
The results show that both \compsqrtglasso{} and \compsqrtoglasso{} outperform \complasso{} in terms of FNR and FPR, which shows 
the benefit of using CT SI.
Note that \compsqrtoglasso{} performs on par with \compsqrtglasso{}, 
even though the former infers everything on the fly from CT SI without explicit access to family SI.
The \comp{} algorithm by itself produces no false negatives (corresponding to $\text{FNR} = 0$), but many false positives. 
Further, all four algorithms yield lower (better) FNR and FPR as $m$ increases or the averaged sparsity level decreases.

\begin{figure*}[!t]
    \begin{minipage}[b]{.24\linewidth}
      \centering
      \centerline{\includegraphics[width=4.2cm]{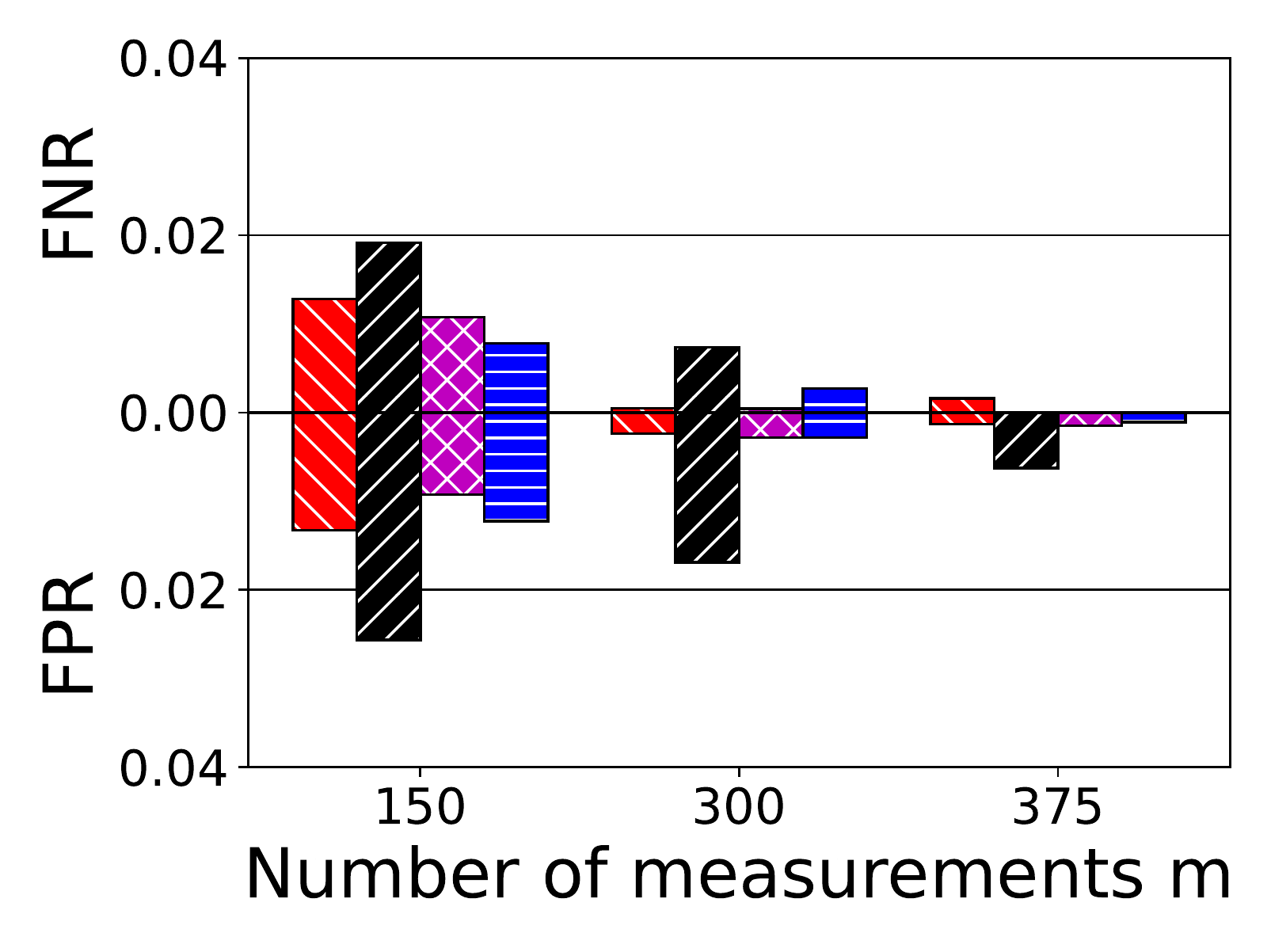}}
    \end{minipage}
    \hfill
    \begin{minipage}[b]{0.24\linewidth}
      \centering
      \centerline{\includegraphics[width=4.2cm]{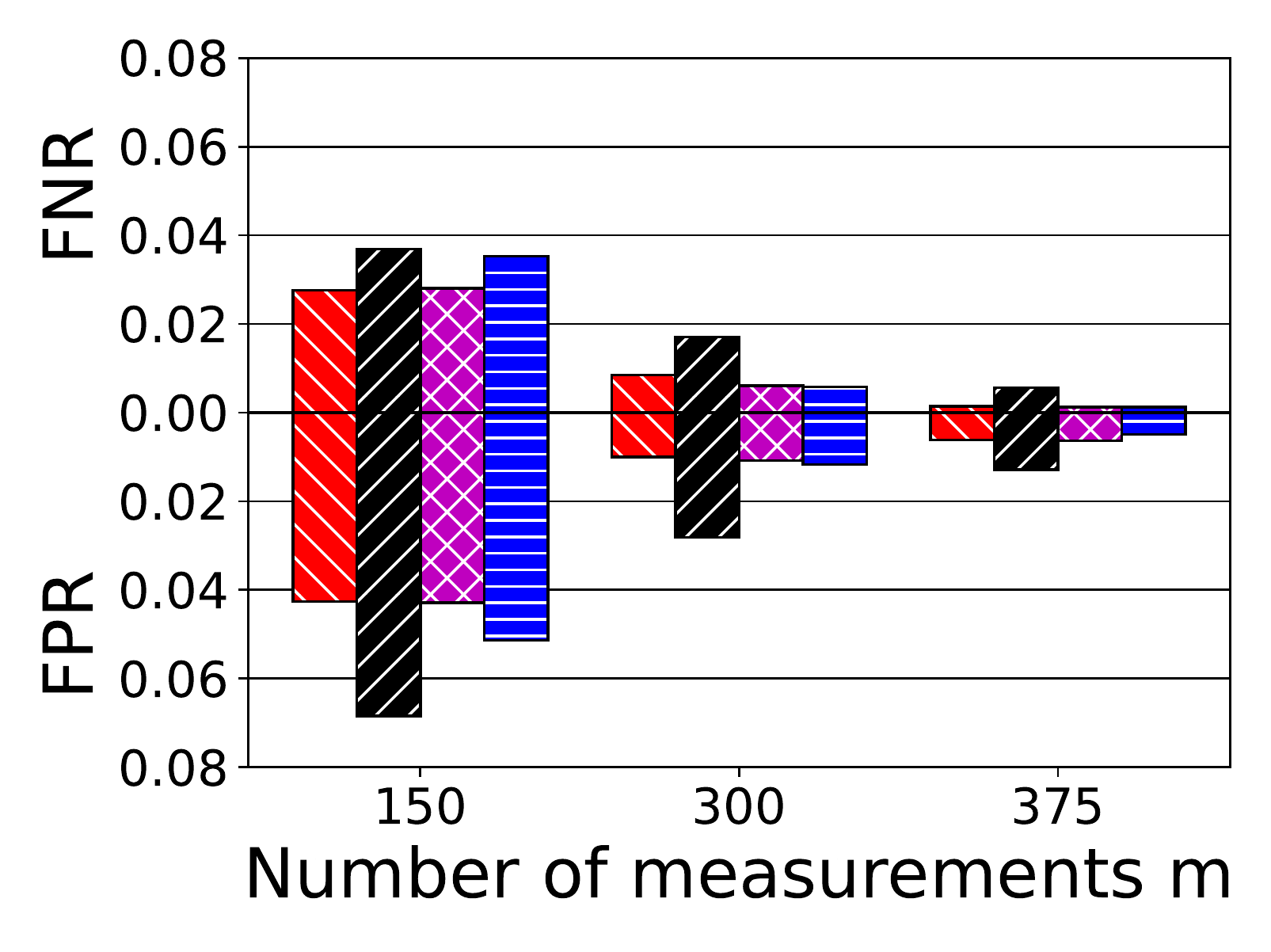}}
    \end{minipage}
    \hfill
    \begin{minipage}[b]{.24\linewidth}
      \centering
      \centerline{\includegraphics[width=4.2cm]{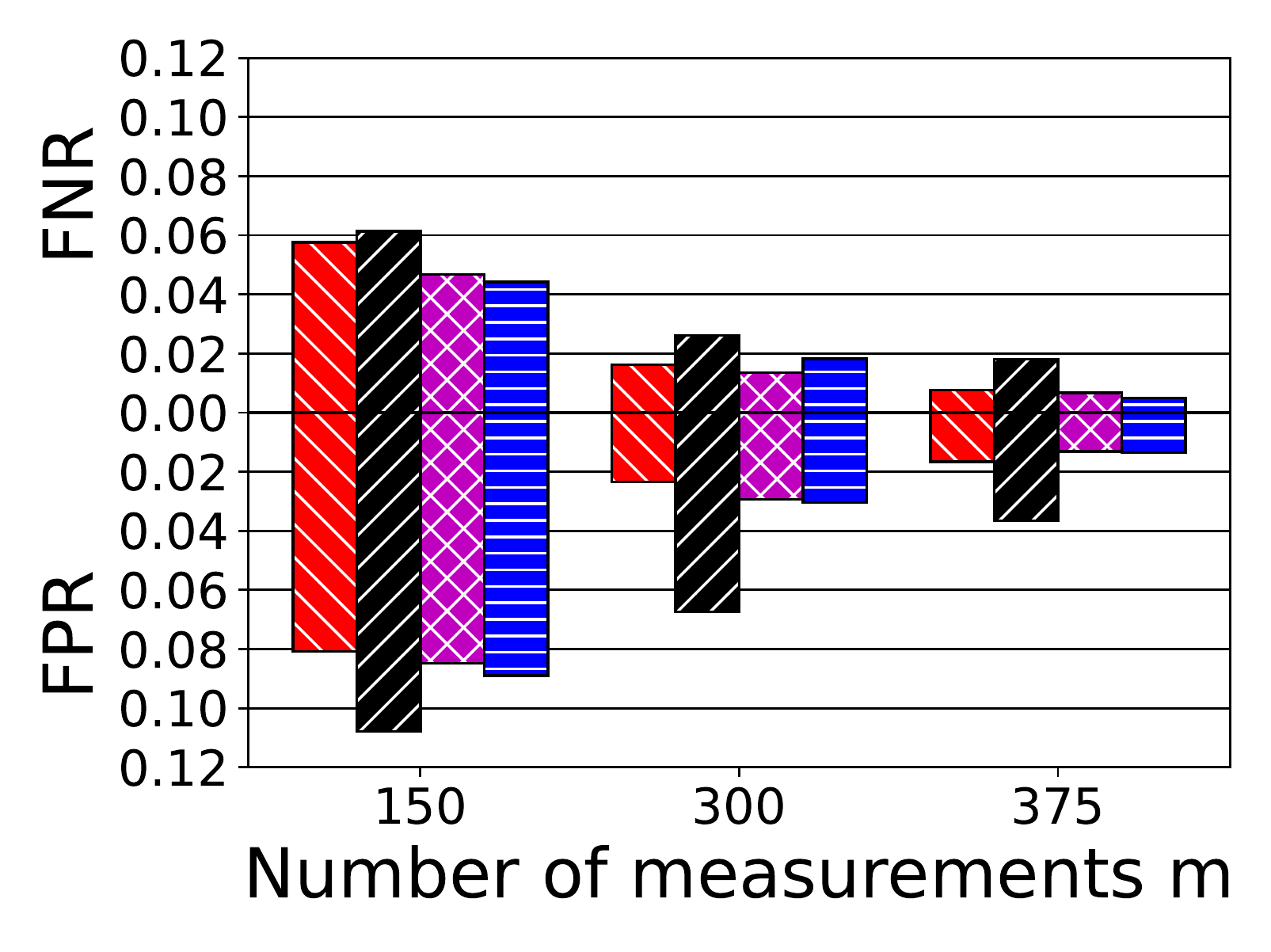}}
    \end{minipage}
    \hfill
    \begin{minipage}[b]{0.24\linewidth}
      \centering
      \centerline{\includegraphics[width=4.2cm]{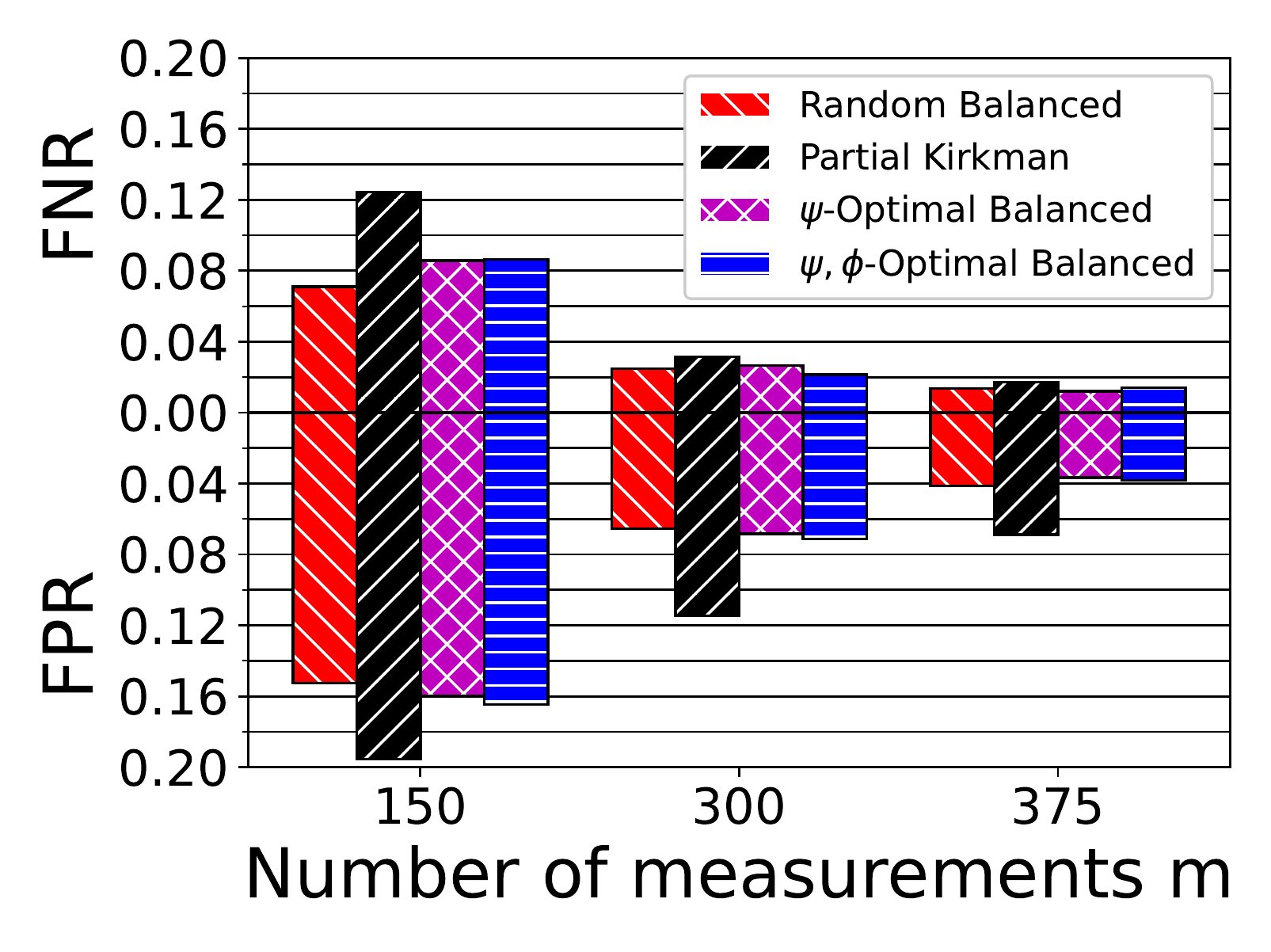}}
    \end{minipage}
    \vspace{-2mm}
    \caption{Mean FNR and FPR values obtained using \compsqrtoglasso{} with different pooling matrices, for mean sparsity levels of 2.12\%, 3.98\%, 6.01\%, and 8.86\% (from left to right) for the encoder design experiment in Sec.~\ref{subsec:results_M2}.}
    \label{fig:res-mat-m2}
    \vspace{-4mm}
\end{figure*}

\begin{figure*}[!t]
    \begin{minipage}[b]{.24\linewidth}
      \centering
      \centerline{\includegraphics[width=4.35cm]{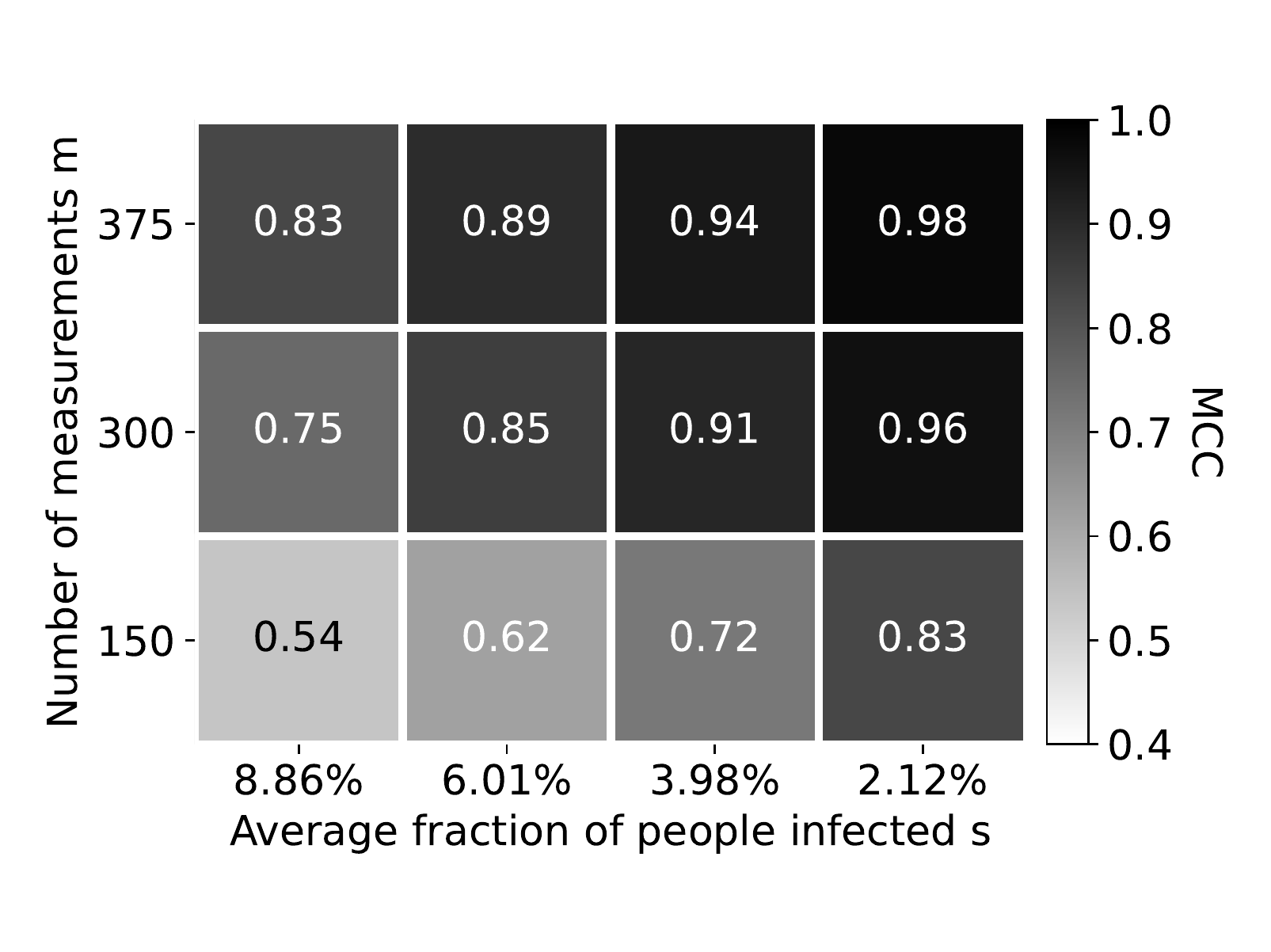}}
    \end{minipage}
    \hfill
    \begin{minipage}[b]{0.24\linewidth}
      \centering
      \centerline{\includegraphics[width=4.35cm]{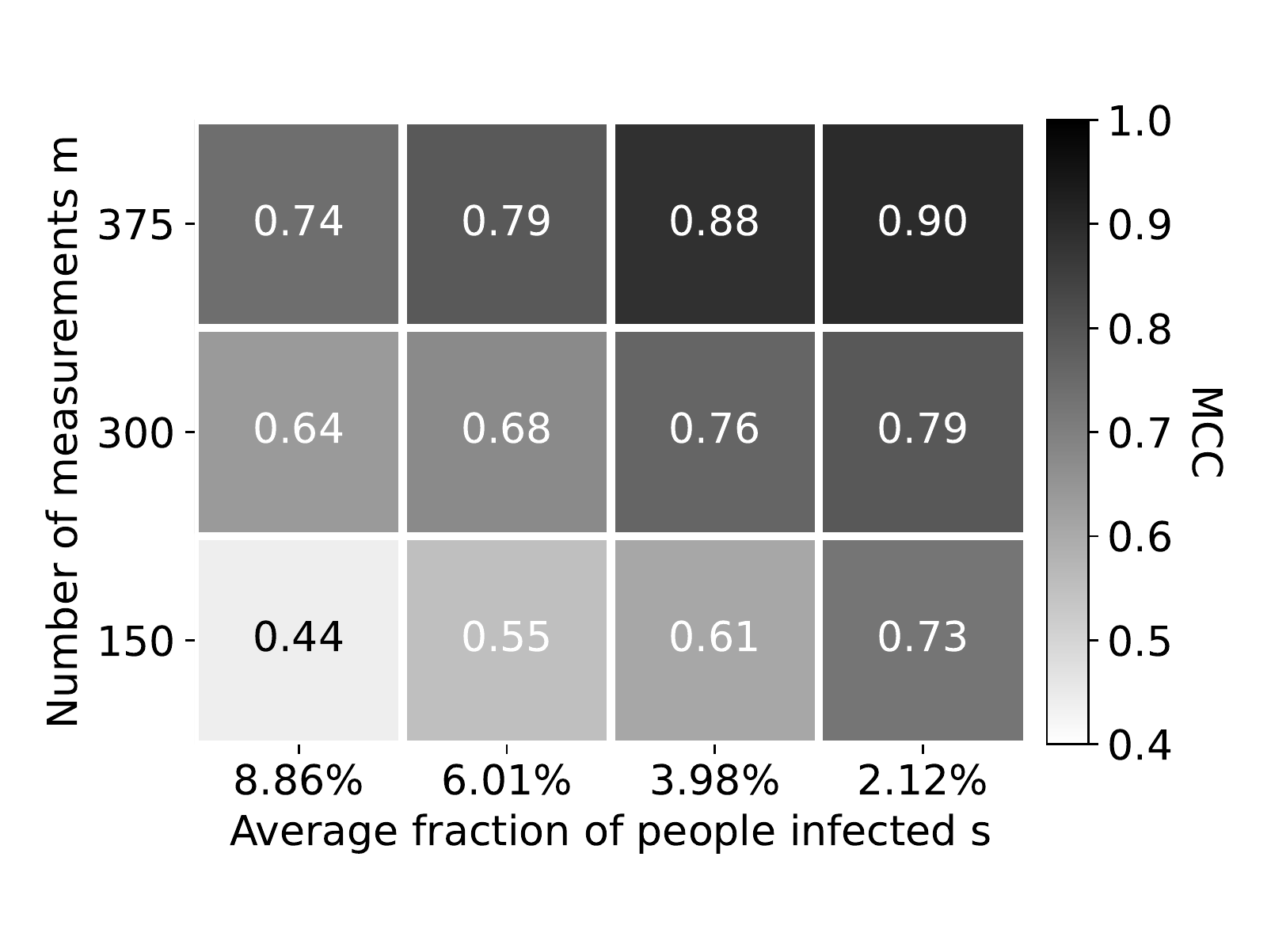}}
    \end{minipage}
    \hfill
    \begin{minipage}[b]{.24\linewidth}
      \centering
      \centerline{\includegraphics[width=4.35cm]{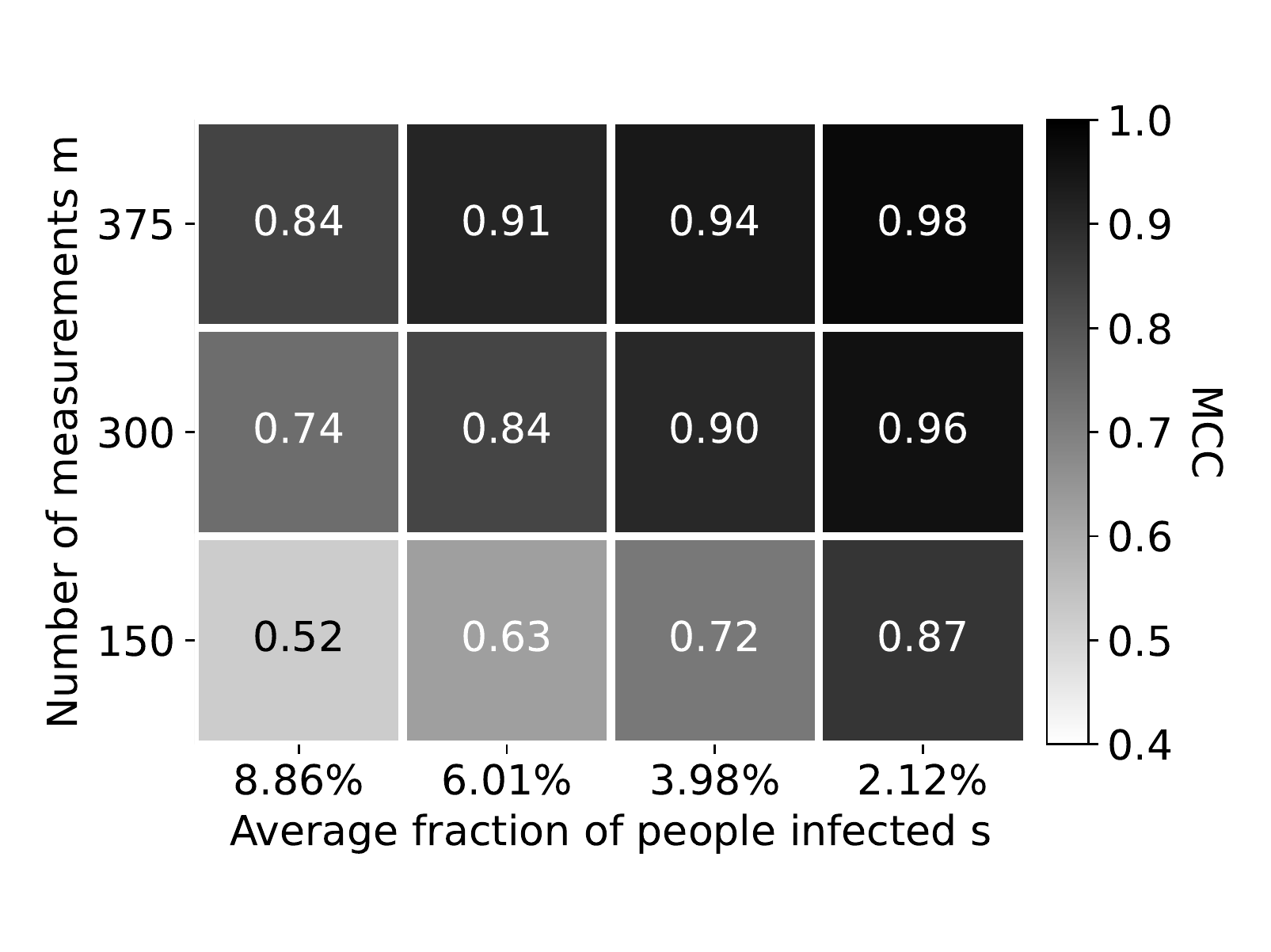}}
    \end{minipage}
    \hfill
    \begin{minipage}[b]{0.24\linewidth}
      \centering
      \centerline{\includegraphics[width=4.35cm]{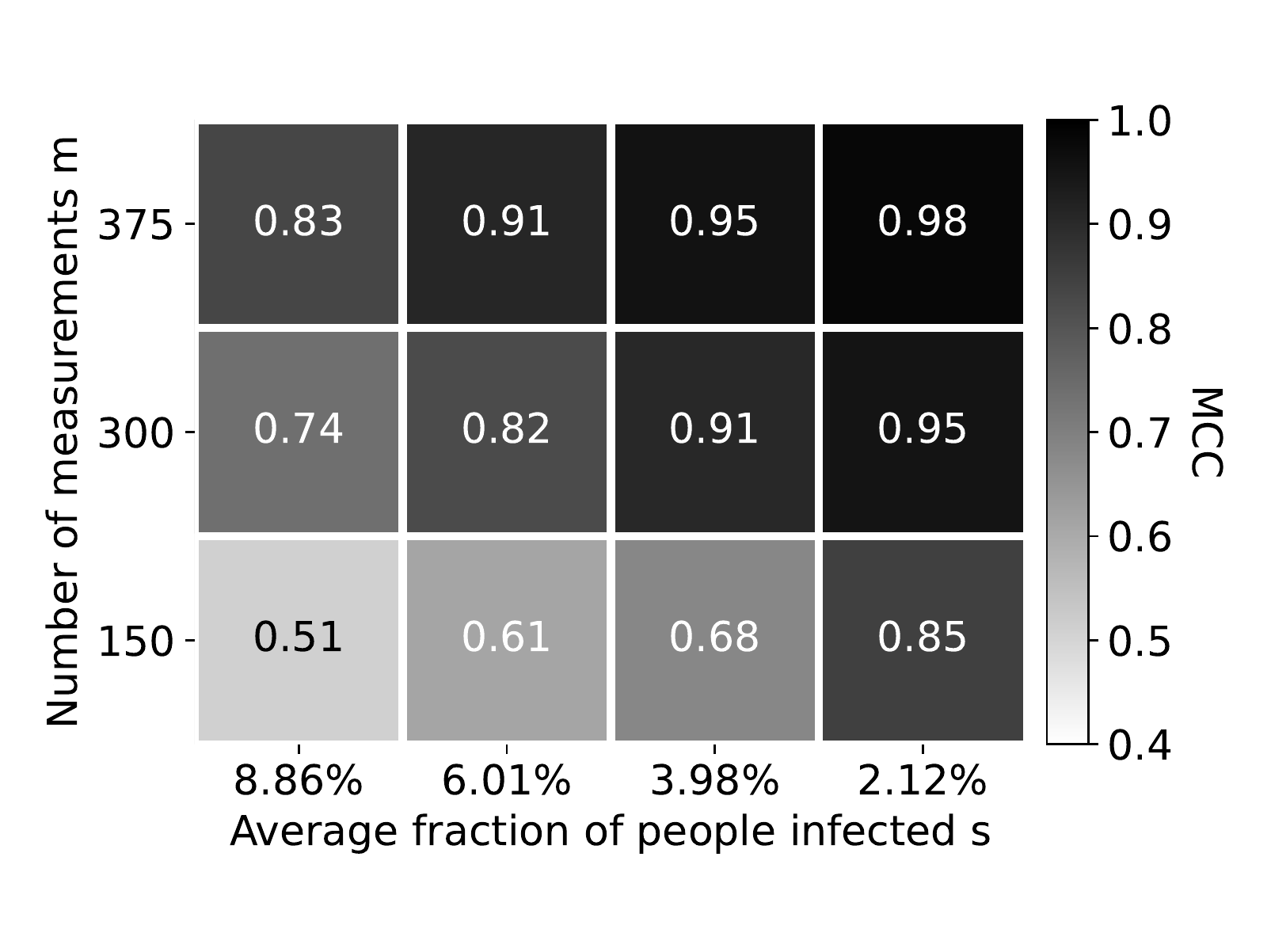}}
    \end{minipage}
    \vspace{-2mm}

    \begin{minipage}[b]{.24\linewidth}
      \centering
      \centerline{\includegraphics[width=4.35cm]{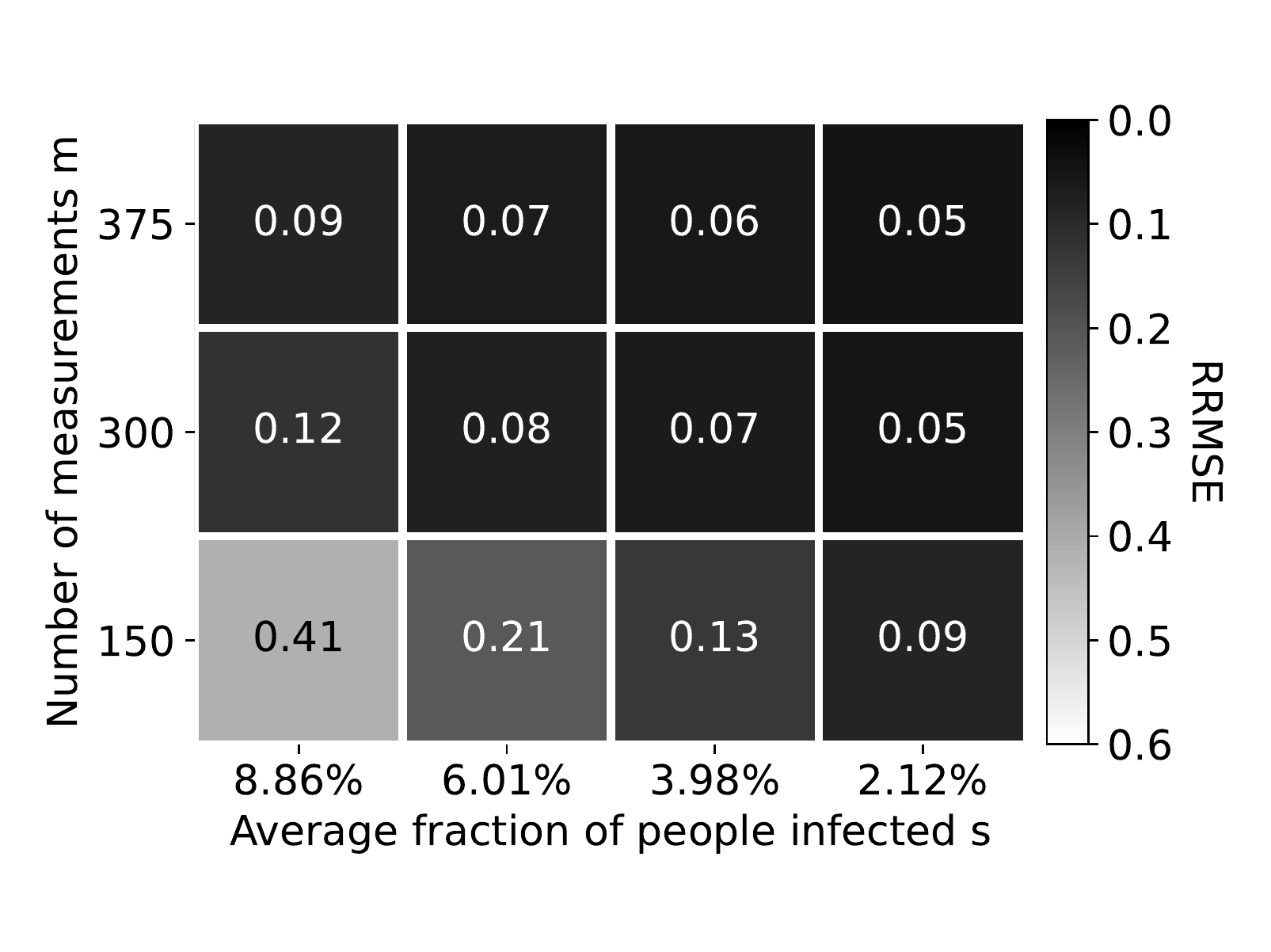}}
    \end{minipage}
    \hfill
    \begin{minipage}[b]{0.24\linewidth}
      \centering
      \centerline{\includegraphics[width=4.35cm]{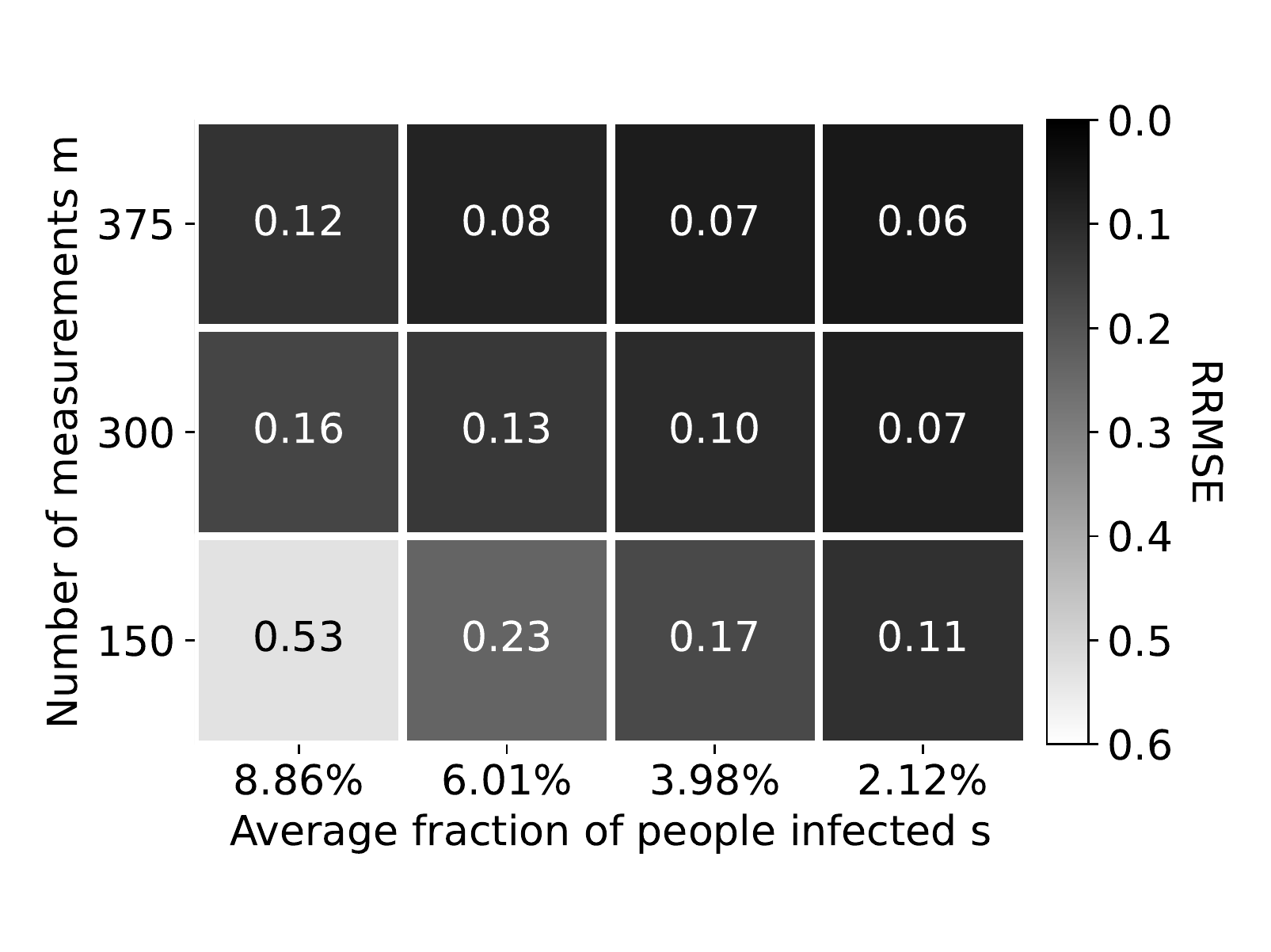}}
    \end{minipage}
    \hfill
    \begin{minipage}[b]{.24\linewidth}
      \centering
      \centerline{\includegraphics[width=4.35cm]{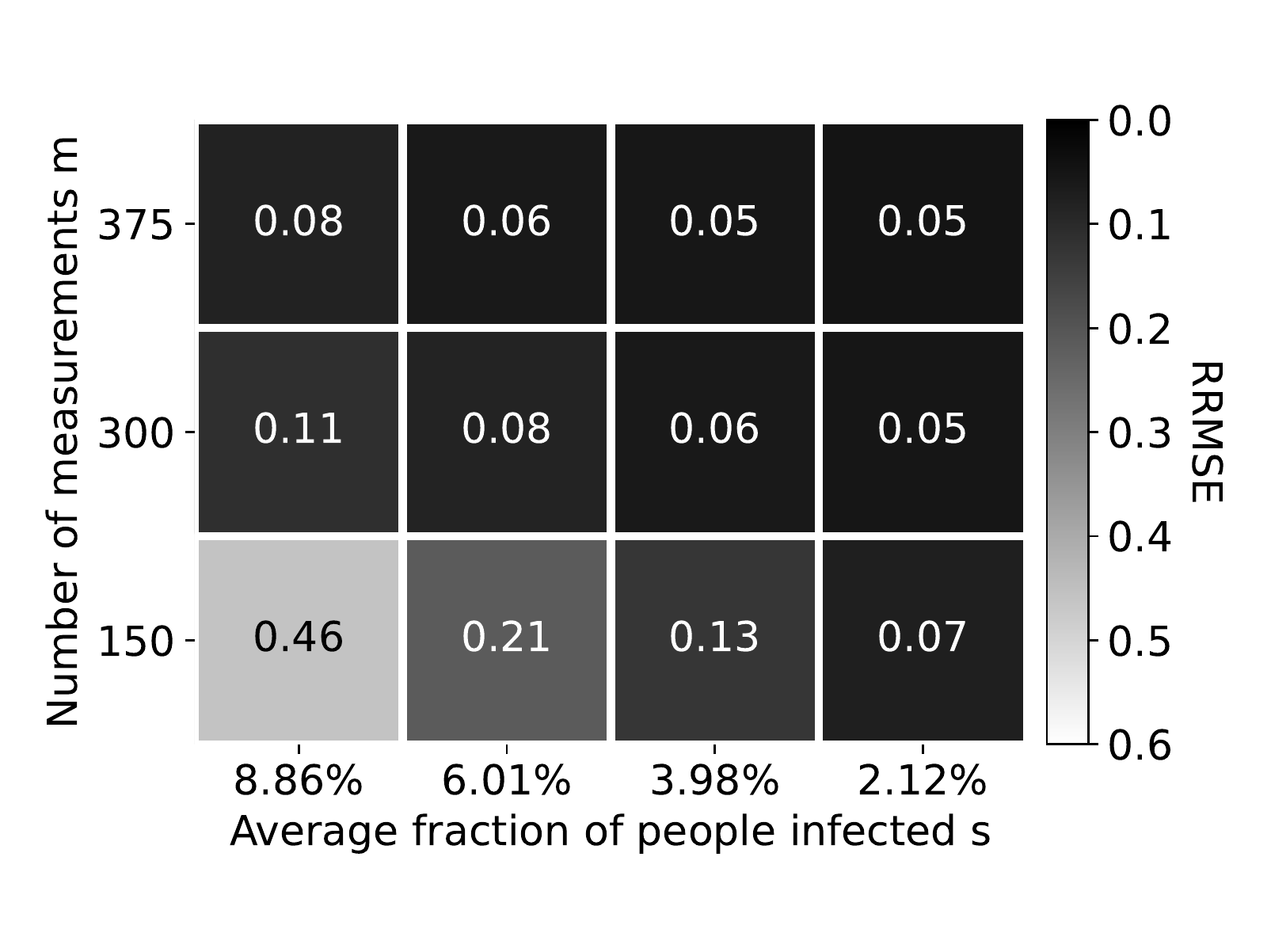}}
    \end{minipage}
    \hfill
    \begin{minipage}[b]{0.24\linewidth}
      \centering
      \centerline{\includegraphics[width=4.35cm]{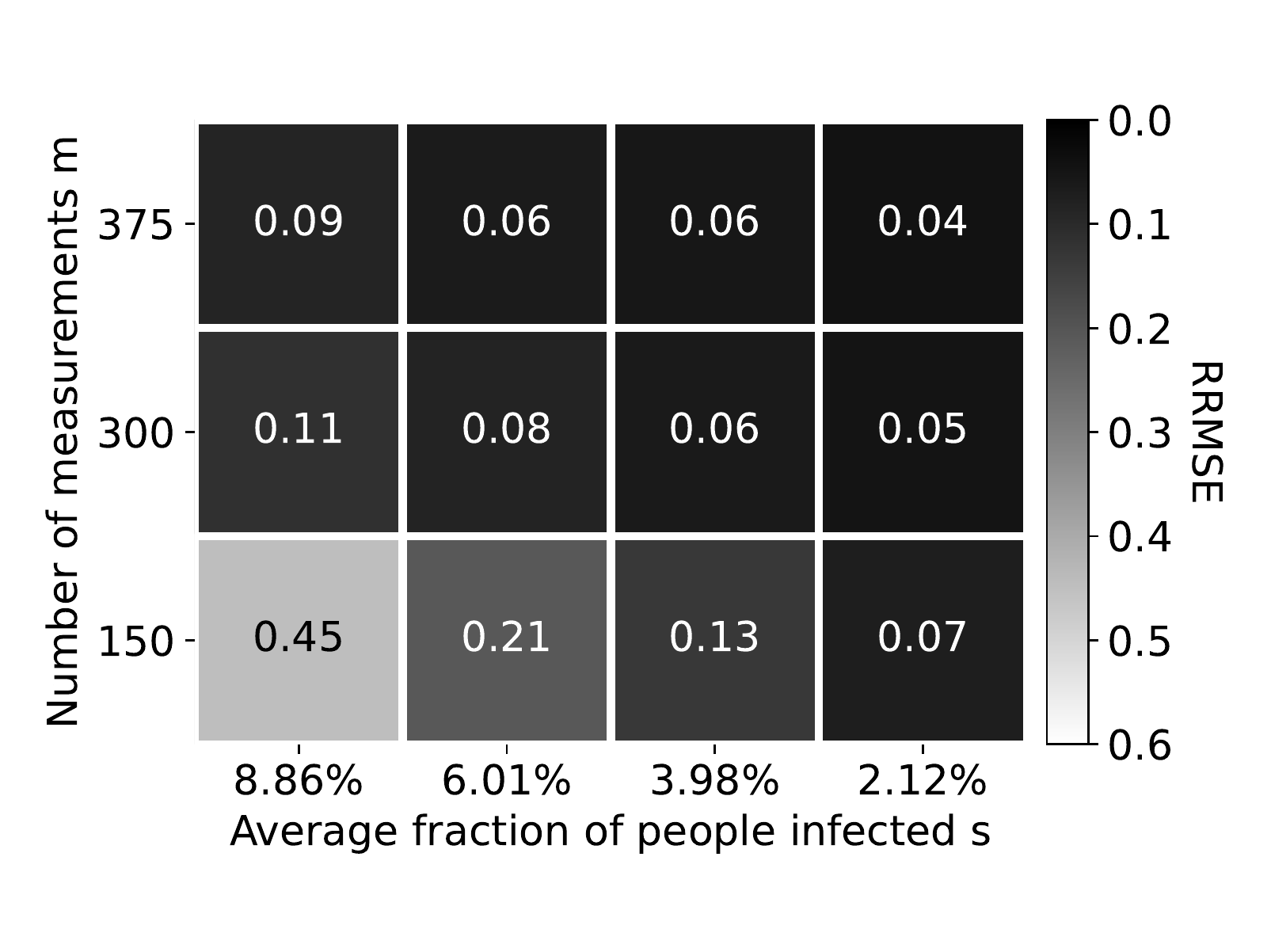}}
    \end{minipage}
    \vspace{-2mm}
    \caption{Mean MCC (top row) and mean RRMSE (bottom row) values obtained by \compsqrtoglasso{} using random balanced, partial Kirkman, $\psi$-optimal balanced, and $\psi,\phi$-optimal balanced matrices (from left to right) for the encoder design experiment in Sec.~\ref{subsec:results_M2}.}
    \label{fig:res-mat-mcc-m2}
    \vspace{+2mm}
\end{figure*}

\textbf{Results with Encoder Design Using SI:} Since \compsqrtoglasso{} is the most general and accurate decoder for \textbf{M2}, we shall use it with different pooling matrices to gain an understanding of how our designed matrices from Sec.~\ref{sec:matrix_des} perform and whether incorporating SI in encoder design \EditRevision{using our approach} yields any significant performance improvements. Fig.~\ref{fig:res-mat-m2} shows the FNR and FPR obtained using \compsqrtoglasso{} with the following matrices: ({\em i})~random balanced, ({\em ii})~partial Kirkman triple matrices, ({\em iii})~$\psi$-optimal balanced, and ({\em iv})~$\psi,\phi$-optimal balanced. Fig.~\ref{fig:res-mat-mcc-m2} shows a comparison of the performance of \compsqrtoglasso{} using the above pooling matrices in terms of mean MCC and mean RRMSE values. As before, we use matrices corresponding to three measurement levels with $m \in \{150, 300, 375\}$ and $n=1000$. 
For balanced binary matrices, ({\em i}), ({\em iii}), and ({\em iv}), we set $c = 3$. Furthermore, we design random balanced matrices by starting from the canonical matrix and performing a large number of interchange operations on it. Finally, for designing $\psi, \phi$-optimal matrices, we considered contacts during the seven days immediately preceding the testing date in order to assign the $b_{ij}$ values for computing $\phi(\A)$. While we observe no significant differences in the FNR, FPR, MCC, and RRMSE values obtained using matrices of type ({\em i}), ({\em iii}), and ({\em iv}), the performance using Kirkman matrices is slightly worse than these three in most cases. We remark here that partial Kirkman matrices are themselves $\psi$-optimal balanced matrices with $c = 3$, but the randomized nature of Alg.~\ref{algo:psi-opt}, as opposed to the deterministic algebraic design of Kirkman matrices, could explain this difference. The more interesting comparison between matrices of types~({\em iii}) and ({\em iv}) shows that \emph{we do not get significant additional gains by incorporating SI in the design of the pooling matrix}. To summarize, the results show that our $\psi$-optimal matrices perform on par with deterministic designs such as the Kirkman and that incorporating SI in encoder design yields no or minimal additional gains. \EditRevision{This may well be a limitation of the specific encoder design approaches adopted in this paper in conjunction with the decoders presented here, and there could exist approaches to encoder design that yield superior results.}

\subsection{Experiment with a Different Contact Tracing Graph for M2}
\label{subsec:results_M2_additional}

In this section, we describe an additional experiment for model \textbf{M2} which illustrates that \compsqrtoglasso{} yields better performance than \compsqrtglasso{} for general contact tracing graphs 
comprised of incomplete cliques.

\subsubsection{Data Generation}
For this experiment, we use a different and slightly more general contact tracing graph to simulate the spread of infection. Recall that the adjacency matrix of the contact graph has a block diagonal structure. However, in this case, we allow two consecutive (according to the order in which cliques appear along the diagonal of the contact matrix) nontrivial cliques (i.e., cliques with more than one node) to have an overlap of one node with probability half. This assumption is reasonable since the concept of family encompasses more general groups such as people at the same workplace, students studying in the same classroom, etc. 
Furthermore, we remove $\alpha = 5\%$ of the edges from the resulting overlapping block diagonal structure, thus converting the existing cliques into cliques with a small fraction of absent pairwise contacts. This is in line with the incomplete clique structures within many communities~\cite{Palla2005}. 
This modified block diagonal structure is kept constant over time while the cross-clique contacts are updated every day. Except for the changes in the underlying contact tracing graph, the rest of the data generation method is the same as that described in Sec.~\ref{sec:data_gen}.

\subsubsection{Inference}
We use the four algorithms (including \textsc{Comp}) for multiplicative noise described in Sec.~\ref{sec:algos}. However, instead of using maximal cliques as groups in \textsc{Comp-sqrt-oglasso}, we use the decomposition of the contact tracing graph into overlapping 3-clique communities \cite{Palla2005}, which is an intermediate notion between connected components and maximal cliques. An algorithm for detecting $k$-clique communities can be found in~\cite[Sec.~1 of Supplementary Notes]{Palla2005}. The first step of this algorithm involves finding the maximal cliques in the contact graph, for which we use the Bron-Kerbosch algorithm \cite{Bron1973}. In the next step, we detect 3-clique communities and 
label each as a group.
Further, we also label as groups the maximal cliques that are not part of any of these communities to ensure that every contact is taken into account. The advantage of using 3-clique communities over just maximal cliques is that the former can capture incomplete cliques as groups.

\subsubsection{Numerical Results} 

\begin{figure*}[!t]
    \begin{minipage}[b]{.24\linewidth}
      \centering
      \centerline{\includegraphics[width=4.2cm]{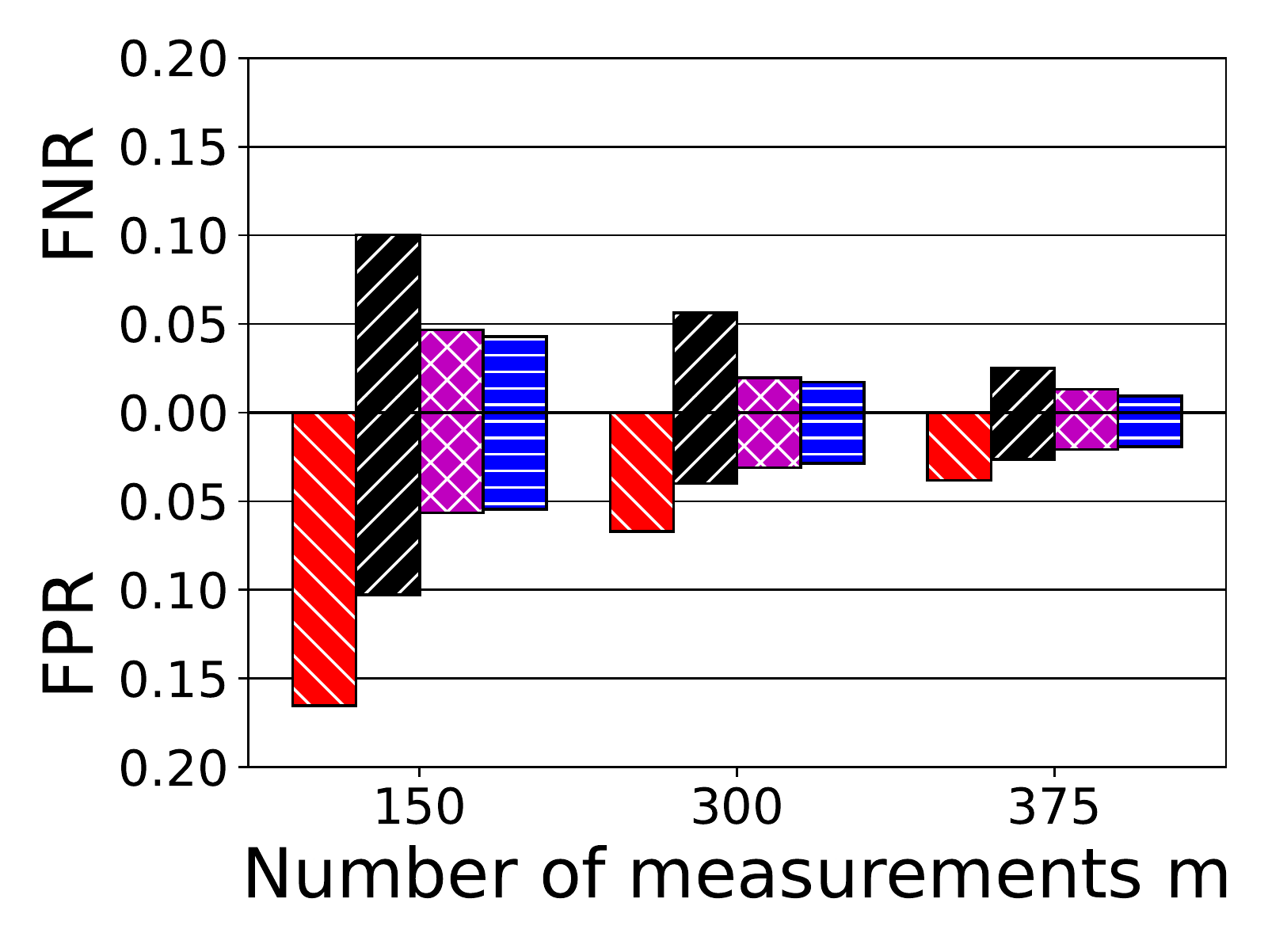}}
    \end{minipage}
    \hfill
    \begin{minipage}[b]{0.24\linewidth}
      \centering
      \centerline{\includegraphics[width=4.2cm]{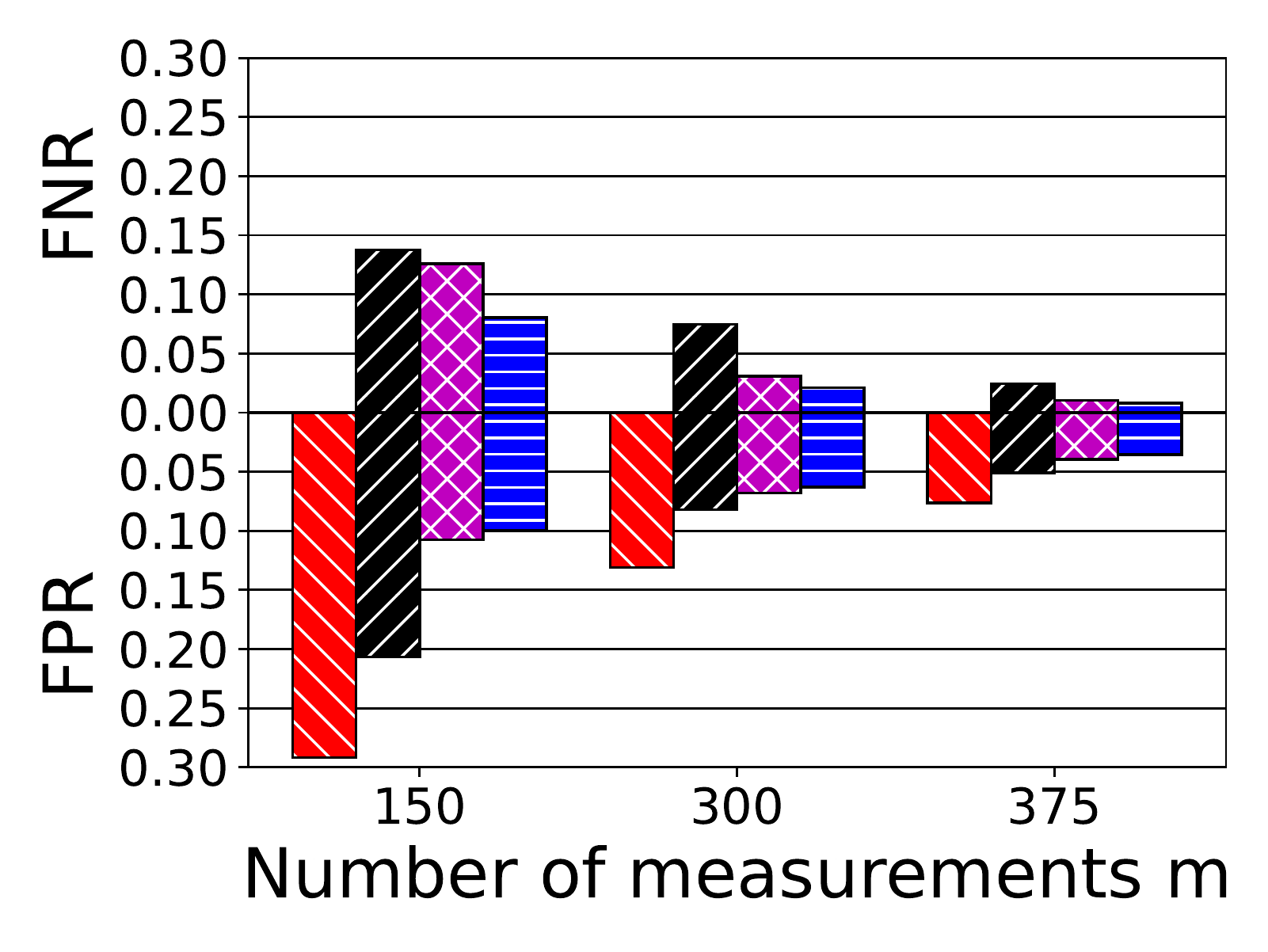}}
    \end{minipage}
    \hfill
    \begin{minipage}[b]{.24\linewidth}
      \centering
      \centerline{\includegraphics[width=4.2cm]{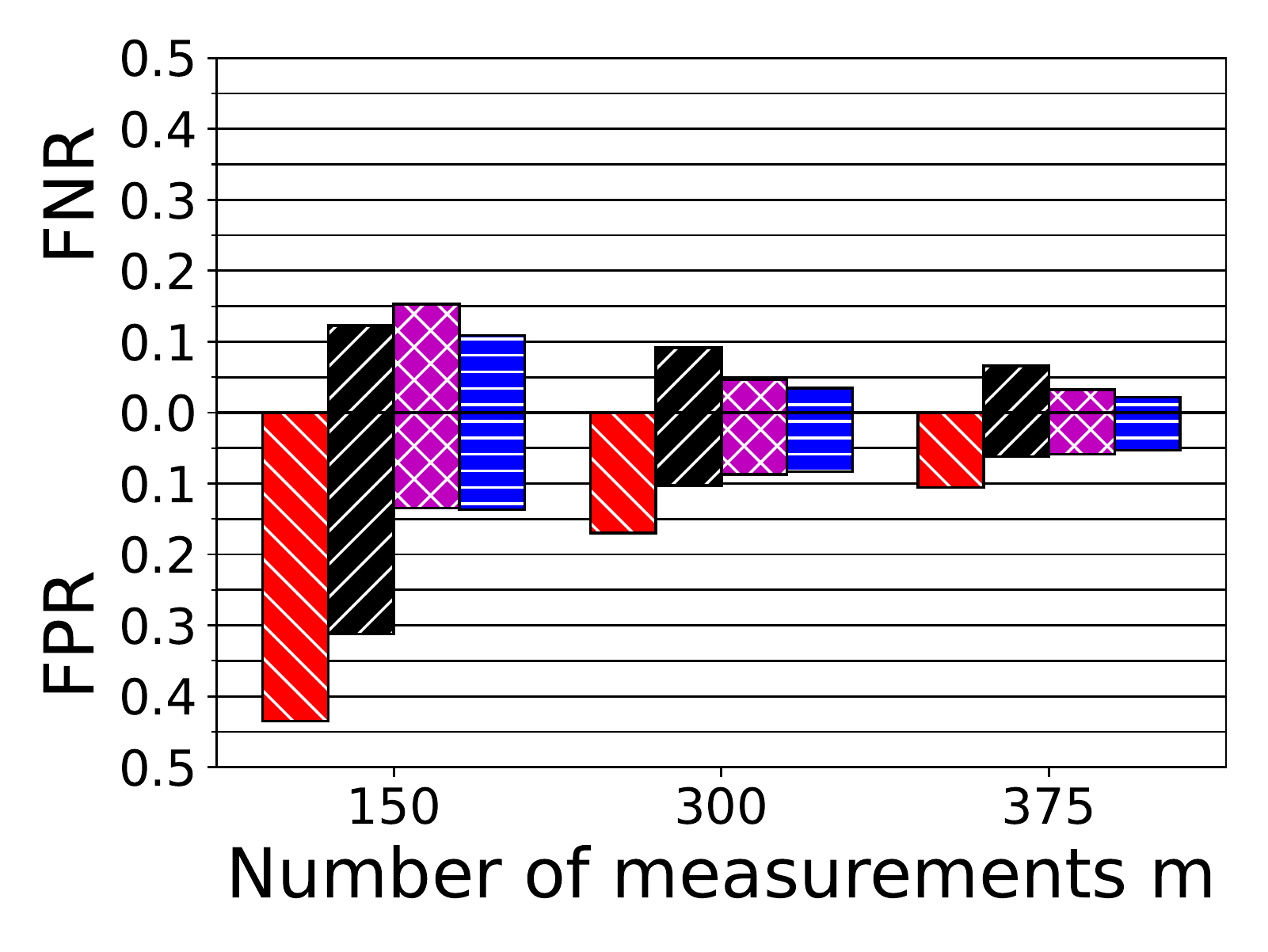}}
    \end{minipage}
    \hfill
    \begin{minipage}[b]{0.24\linewidth}
      \centering
      \centerline{\includegraphics[width=4.2cm]{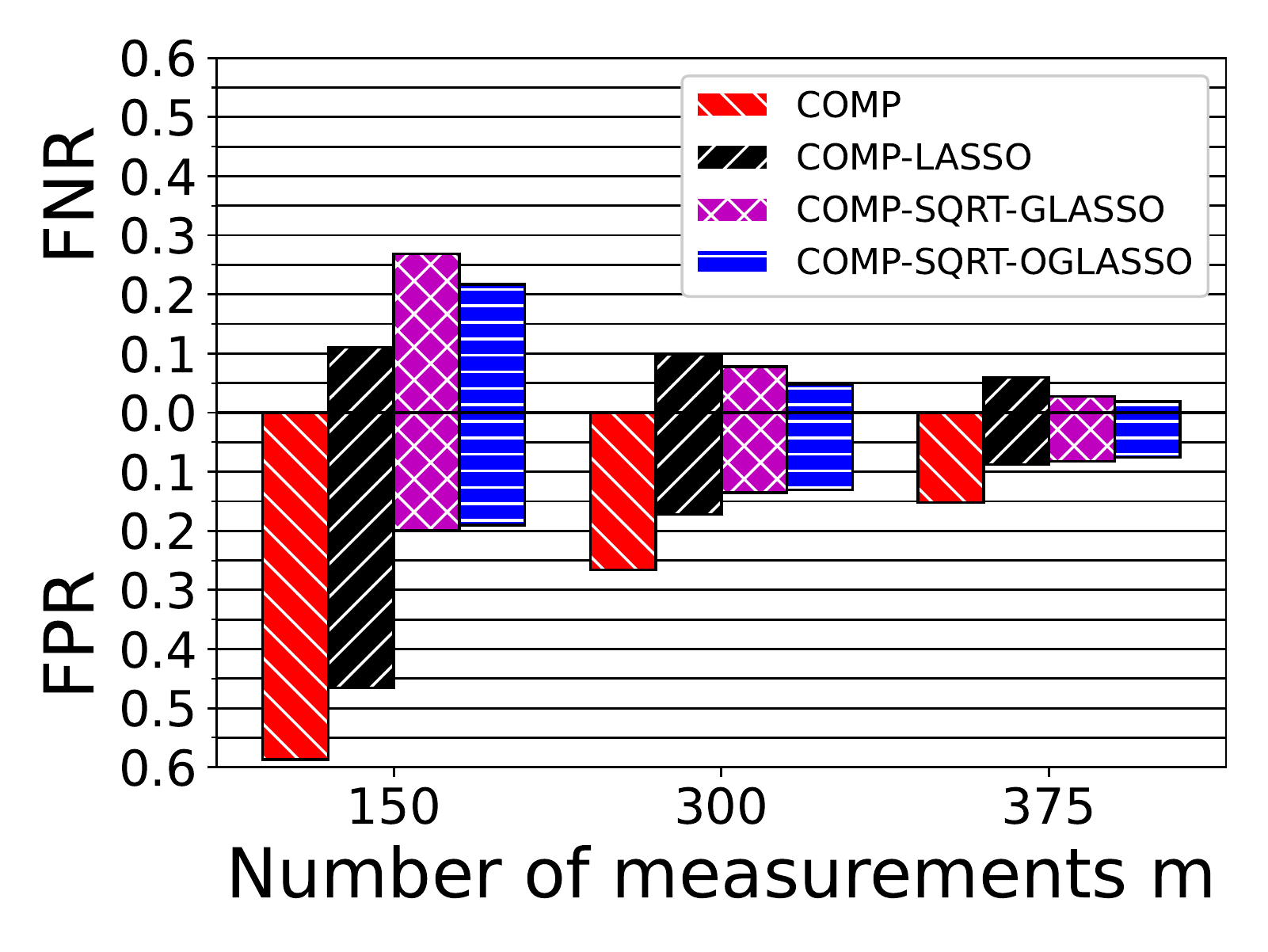}}
    \end{minipage}
    \vspace{-2mm}
    \caption{Mean FNR and FPR values for the experiments in Sec.~\ref{subsec:results_M2_additional}, for mean sparsity levels of 3.20\%, 4.84\%, 6.25\%, and 8.66\% (from left to right).}
    \label{fig:res-gen-m2}
    \vspace{-4mm}
\end{figure*}

\begin{figure*}[!t]
    \begin{minipage}[b]{.24\linewidth}
      \centering
      \centerline{\includegraphics[width=4.35cm]{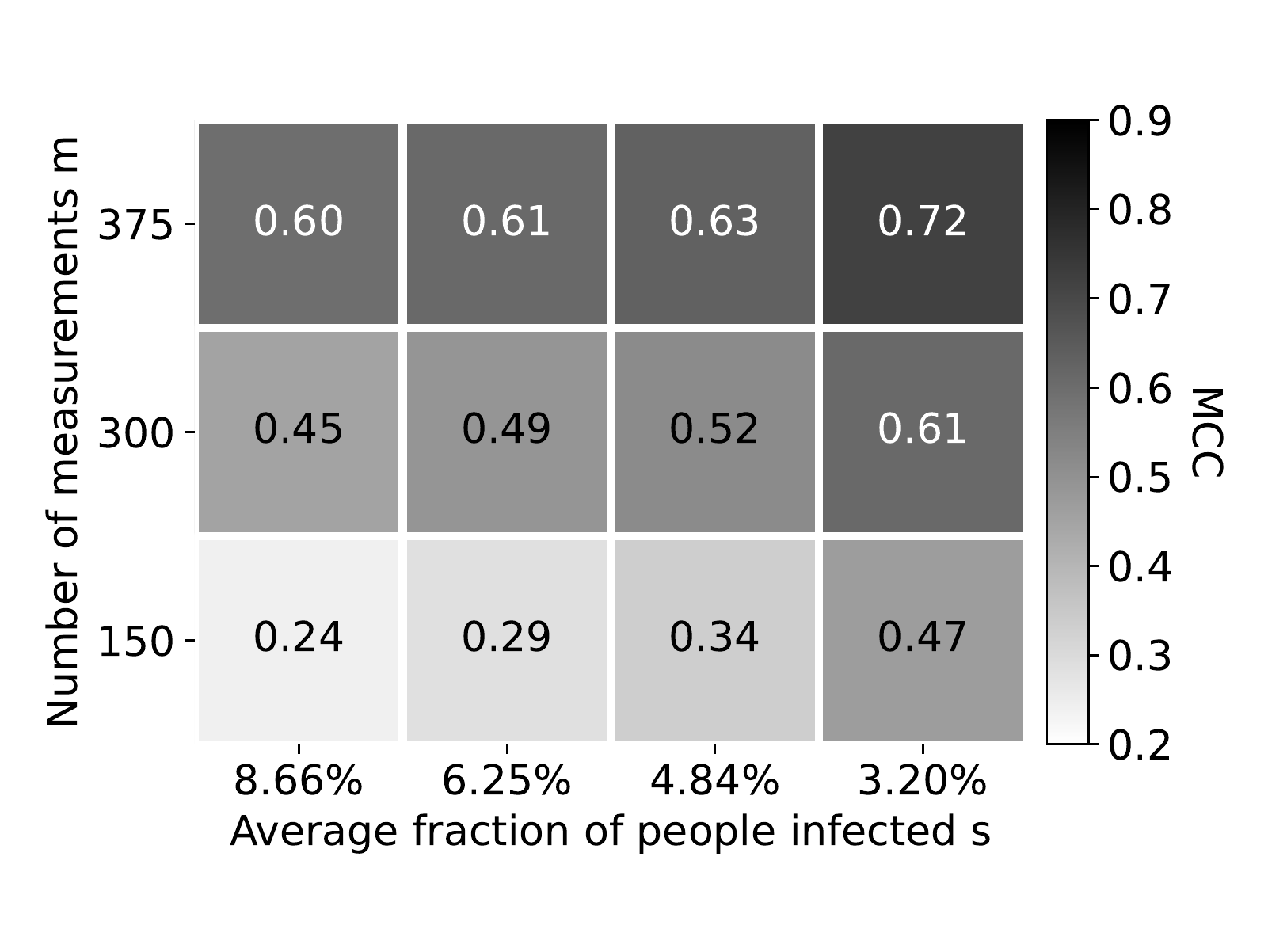}}
    \end{minipage}
    \hfill
    \begin{minipage}[b]{0.24\linewidth}
      \centering
      \centerline{\includegraphics[width=4.35cm]{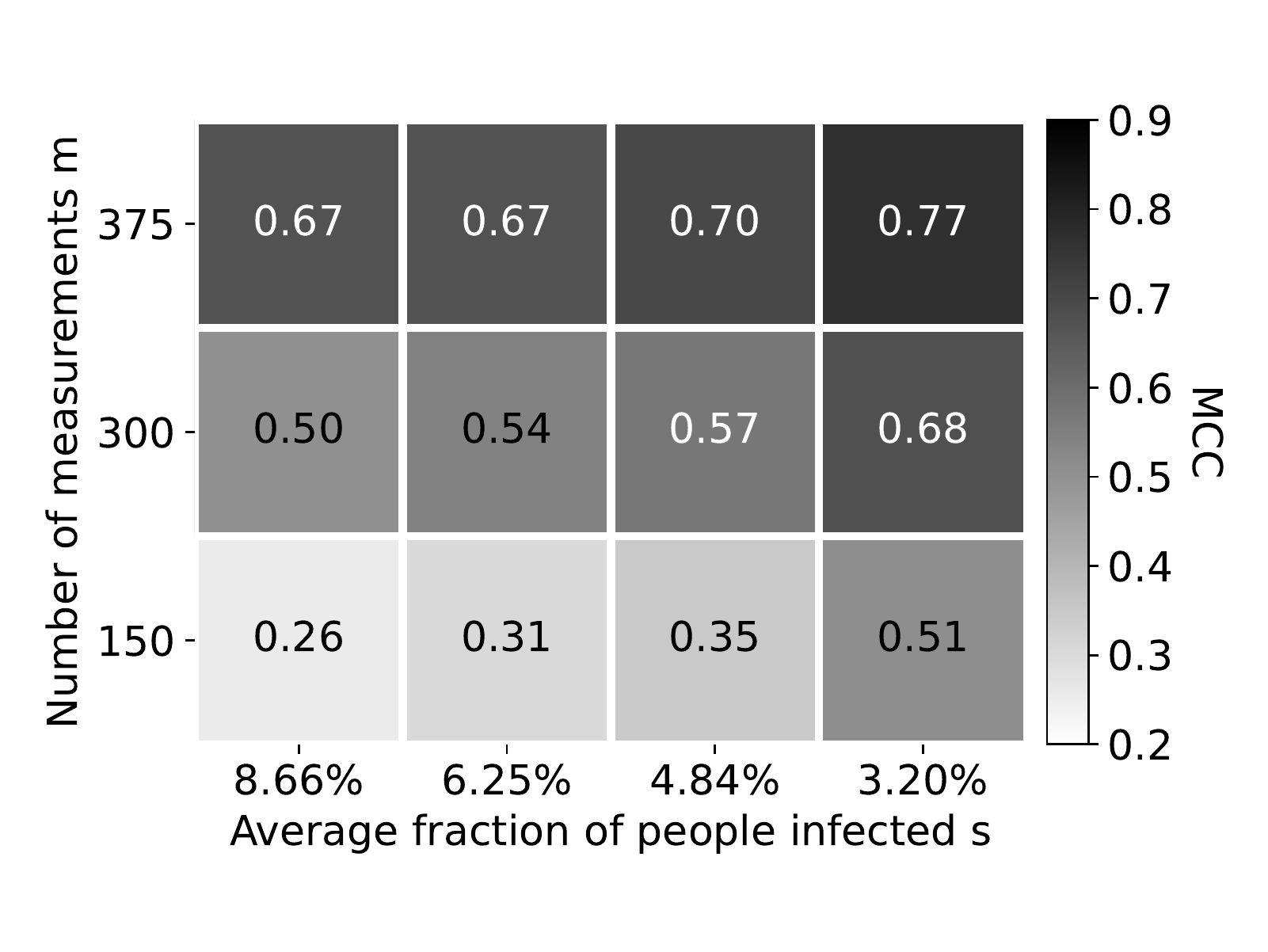}}
    \end{minipage}
    \hfill
    \begin{minipage}[b]{.24\linewidth}
      \centering
      \centerline{\includegraphics[width=4.35cm]{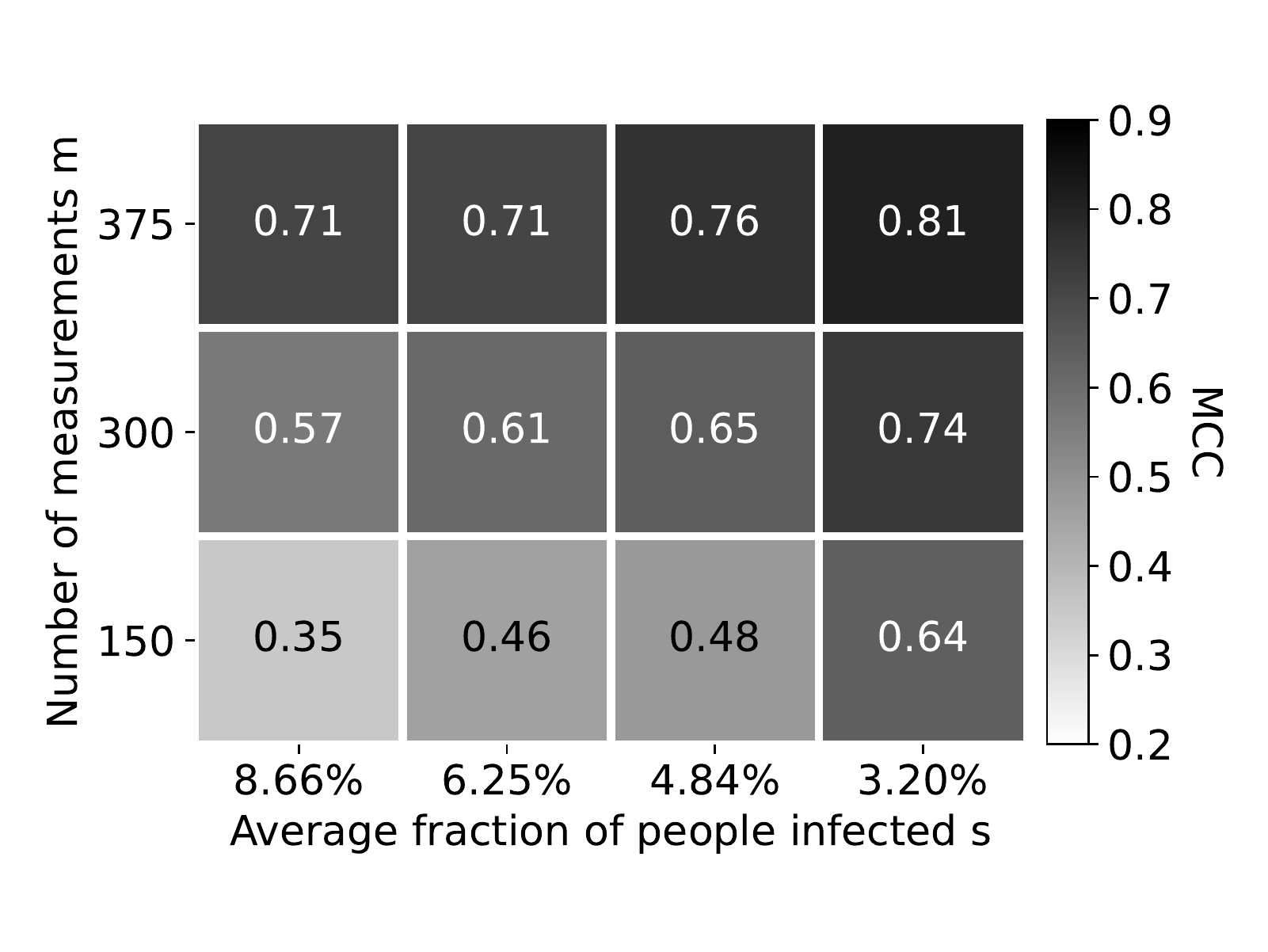}}
    \end{minipage}
    \hfill
    \begin{minipage}[b]{0.24\linewidth}
      \centering
      \centerline{\includegraphics[width=4.35cm]{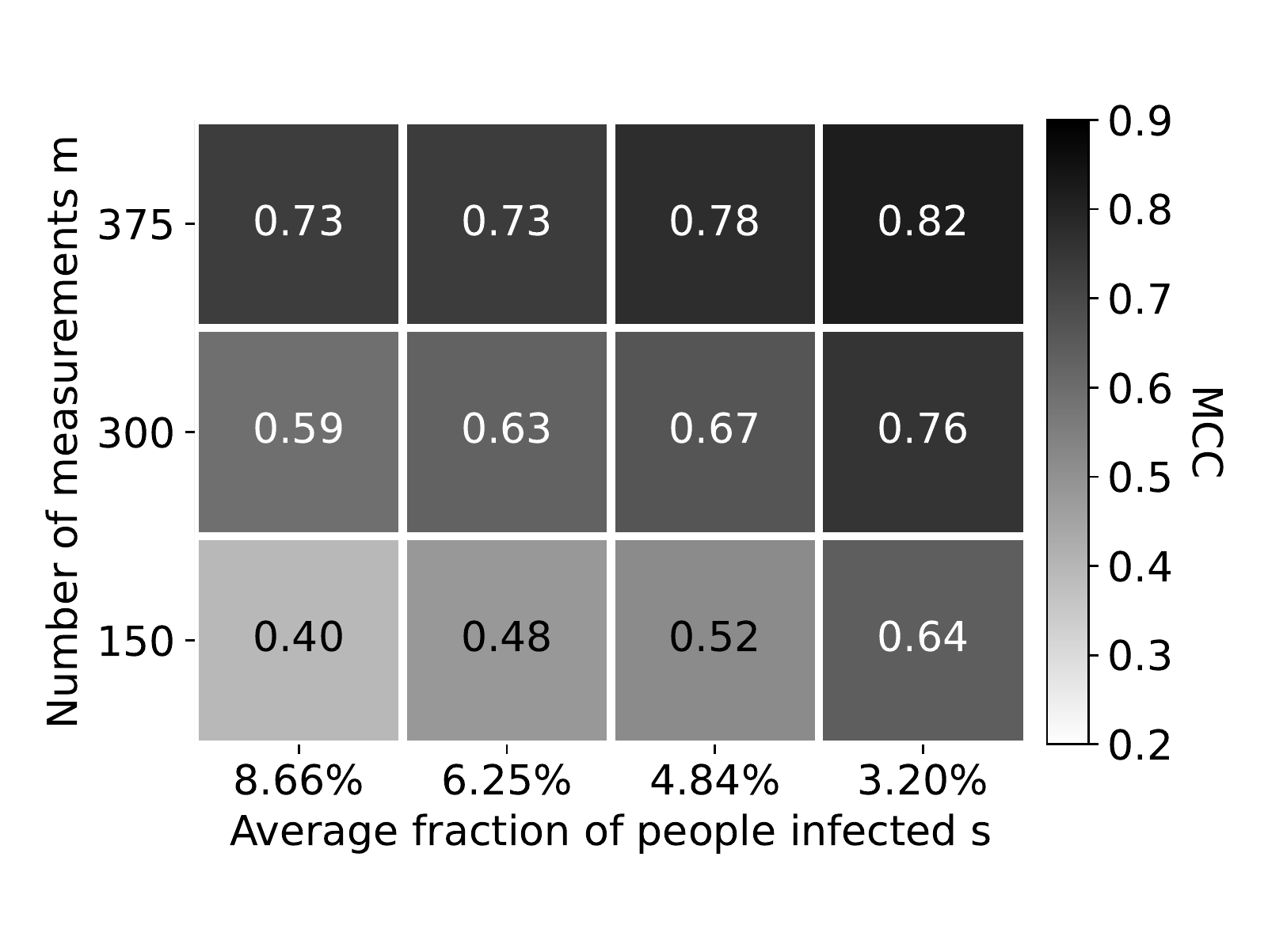}}
    \end{minipage}
    \vspace{-2mm}
    
    \hspace{62pt}
    \begin{minipage}[b]{.24\linewidth}
      \flushright
      \centerline{\includegraphics[width=4.35cm]{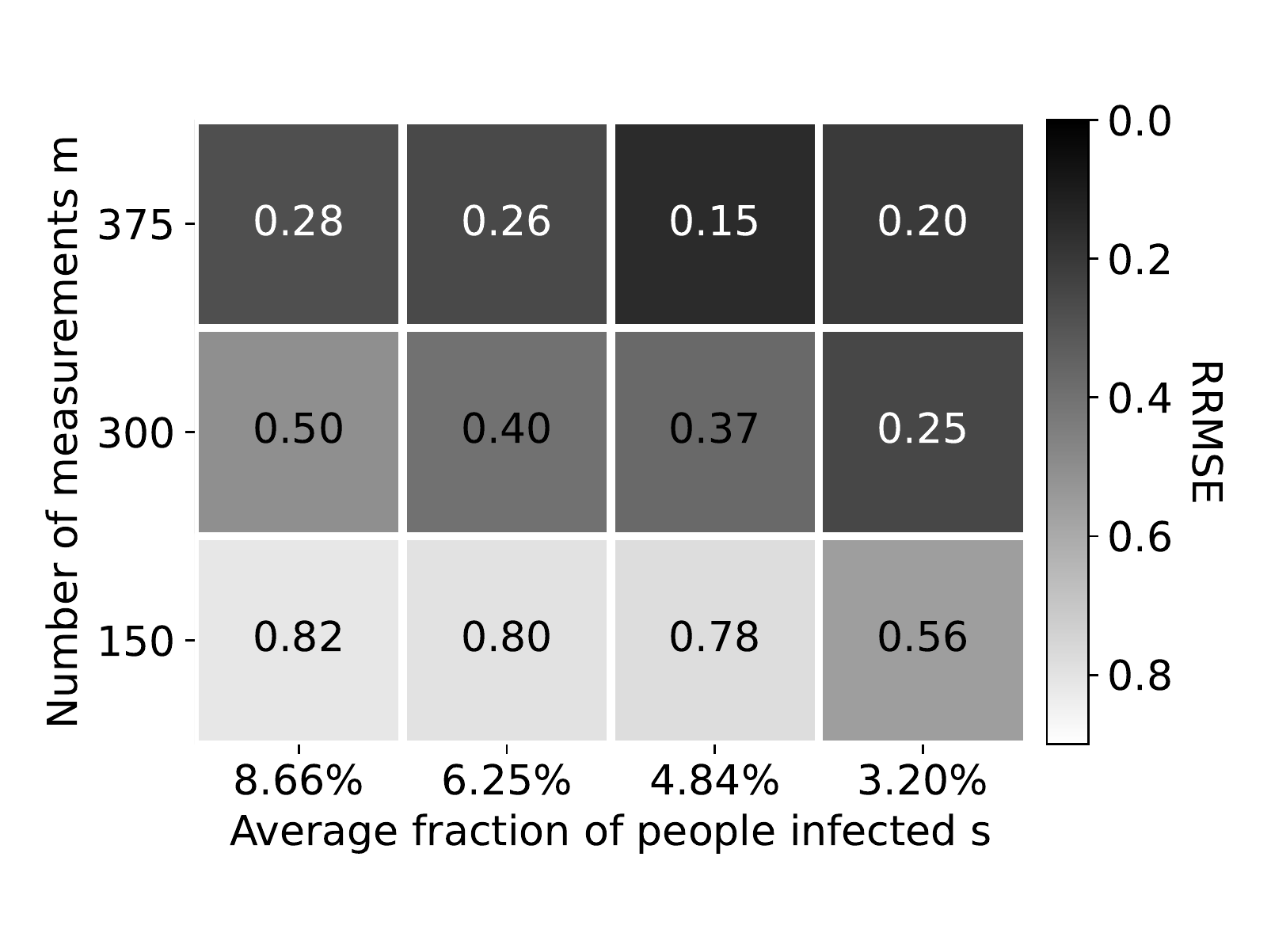}}
    \end{minipage}
    \hfill
    \begin{minipage}[b]{0.24\linewidth}
      \centering
      \centerline{\includegraphics[width=4.35cm]{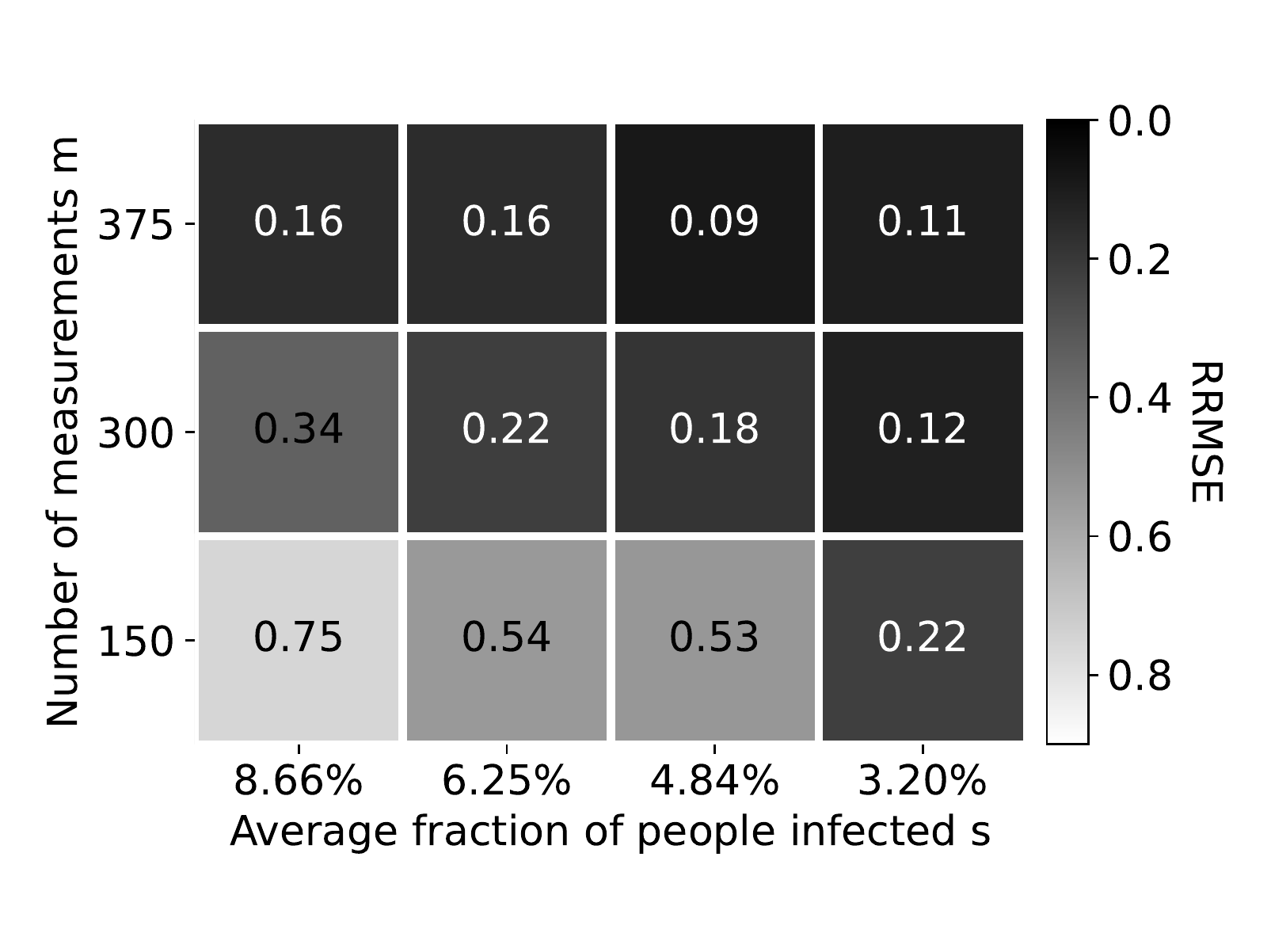}}
    \end{minipage}
    \hfill
    \begin{minipage}[b]{.24\linewidth}
      \flushleft
      \centerline{\includegraphics[width=4.35cm]{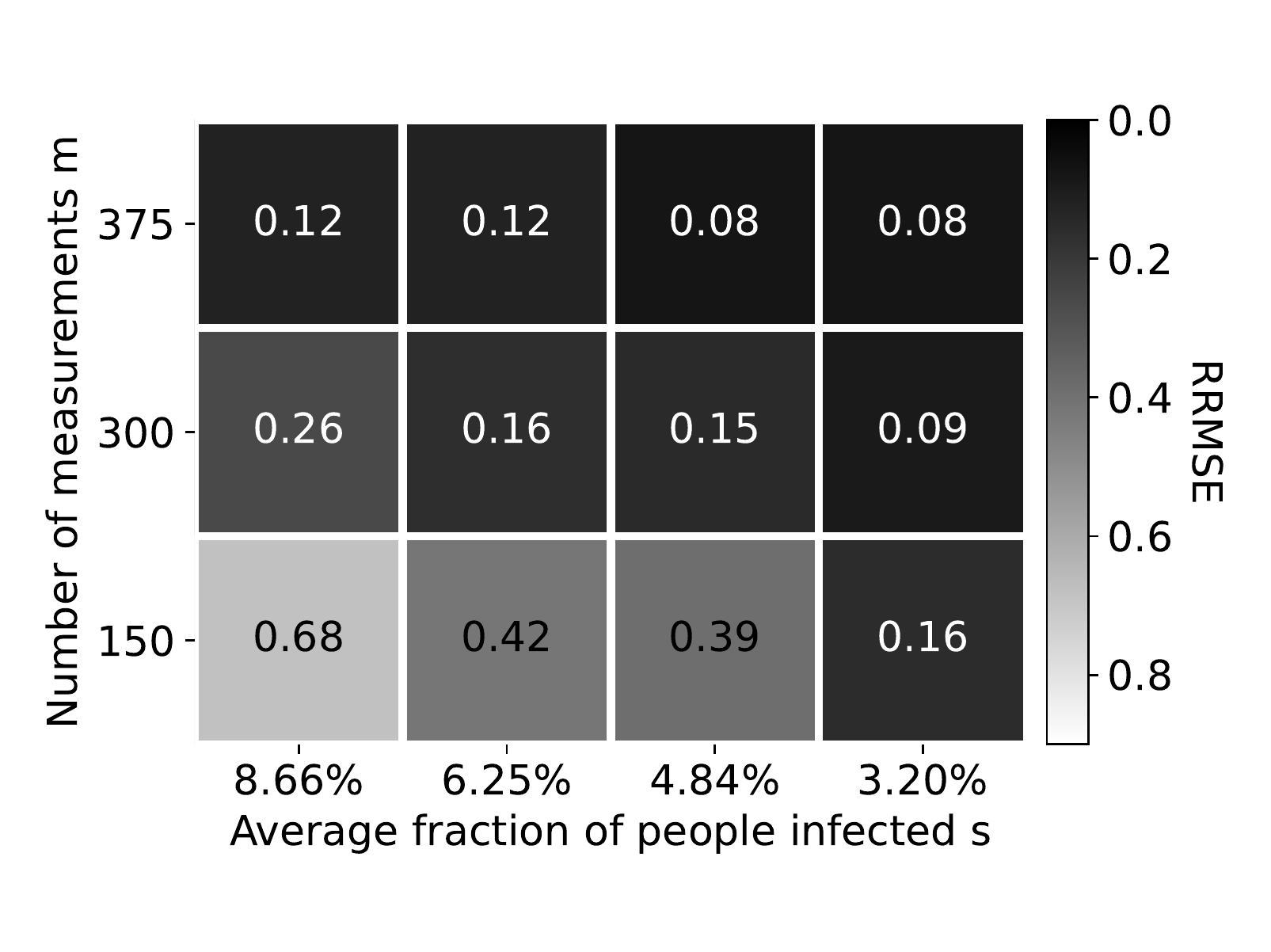}}
    \end{minipage}
    \hspace{62pt}
    \vspace{-2mm}
    \caption{Mean MCC (top row) and mean RRMSE (bottom row) values for the experiment in Sec.~\ref{subsec:results_M2_additional} obtained using \comp{} (no RRMSE), \complasso{}, \compsqrtglasso{}, and \compsqrtoglasso{} (from left to right) for the experiment from Sec.~\ref{subsec:results_M2_additional}. 
    }
    \label{fig:res-gen-mcc-m2}
    \vspace{-0mm}
\end{figure*}

Fig.~\ref{fig:res-gen-m2} shows the mean values (across 50 signals) of FNR and FPR obtained for four different sparsity levels. The sparsity levels were obtained by varying the amount of cross-clique contacts.
Fig.~\ref{fig:res-gen-mcc-m2} shows a comparison of the performance of the algorithms under consideration in terms of mean MCC and mean RRMSE.
It is evident from Fig.~\ref{fig:res-gen-m2} that the FNR and FPR values obtained using \compsqrtoglasso{} are significantly better than \compsqrtglasso{} in almost every case. More prominently, the values in Fig.~\ref{fig:res-gen-mcc-m2} show that the mean MCC and RRMSE values obtained by \compsqrtoglasso{} are always better than \compsqrtglasso{}.

\subsection{\EditRevision{Experimental Results with Erroneously Specified CT SI}}
\label{subsec:noisy-CT-info}
\EditRevision{
In certain situations, there may be errors in specifying the contact-tracing information accurately, leading to errors in the specifications of the graph (also see Sec.~\ref{sec:discussion}). For algorithms for \textbf{M2} such as \compsqrtglasso{}, we have observed that errors in the CT graph do not excessively influence the performance in terms of RMSE, sensitivity, as well as specificity. In Table~\ref{tab:errors_ctsi}, we provide numerical results from an experiment to illustrate this for an average signal sparsity of $6.01\%$ and a $\psi$-optimal balanced binary matrix of size $300 \times 1000$. The results are averaged over $50$ simulated signals. Errors were introduced in the CT graph, either by deliberately adding a certain number of fake contact edges to the CT graph between randomly chosen pairs of nodes, or by deliberately removing a certain number of edges at random from the CT graph, or both. These are respectively referred to as ``Add $(100 \cdot f_e)\%$'' and ``Remove $(100 \cdot f_e)\%$'' in TABLE~\ref{tab:errors_ctsi} where $f_e$ refers to a fraction of the total number of edges present in the ground-truth CT graph. TABLE~\ref{tab:errors_ctsi} also reports results on a combination of both types of errors in the CT graph, i.e., addition as well as removal of edges. Referring to TABLE~\ref{tab:errors_ctsi}, we notice that the degradation of the sensitivity and specificity is roughly linear. For example, the sensitivity of \complasso{} (i.e., $100\%$ of edges from the CT graph removed) and \compsqrtoglasso{} with an accurate CT graph is $0.964$ and $0.986$, respectively. When $12\%$ of the edges in the contact graph are removed, the sensitivity drops to $0.983$. We see that the ratio $(0.986-0.983)/(0.986-0.964) = 13.6\%$ is slightly more than the fraction of edges deleted (i.e., $12\%$). Likewise, the specificity of \compsqrtoglasso{} also shows similar roughly linear decrease.} 

\begin{table*}
\centering
\caption{\EditRevision{The effect of errors in CT SI on the performance of \compsqrtoglasso{} for average sparsity level $6.01\%$ and matrix size $300 \times 1000$. Performance is measured in terms of average RMSE, FN, FP, sensitivity, and specificity along with their standard deviation, computed across $50$ signals.}
}
\begin{tabular}{|c|c|c|c|c|c|}
\hline
Experiment & RMSE & FN & FP & Sensitivity & Specificity \\\hline
No Noise & $0.083 \pm 0.024$ & $1.02 \pm 1.24$ & $27.16 \pm 24.06$ & $0.986 \pm 0.016$ & $0.971 \pm 0.027$ \\ \hline

Add $1.5\%$ & $0.079 \pm 0.032$ & $1.02 \pm 1.94$ & $26.10 \pm 22.79$ & $0.987 \pm 0.022$ & $0.972 \pm 0.025$ \\ \hline
Remove $1.5\%$ & $0.081 \pm 0.023$ & $1.20 \pm 1.43$ & $25.96 \pm 23.09$ & $0.984 \pm 0.018$ & $0.972 \pm 0.025$ \\ \hline
Add $1.5\%$, remove $1.5\%$ & $0.081 \pm 0.023$ & $1.04 \pm 1.26$ & $27.04 \pm 21.92$ & $0.986 \pm 0.017$ & $0.971 \pm 0.024$ \\ \hline

Add $7.5\%$ & $0.082 \pm 0.027$ & $0.98 \pm 1.24$ & $28.24 \pm 26.02$ & $0.987 \pm 0.016$ & $0.969 \pm 0.029$ \\ \hline
Remove $7.5\%$ & $0.083 \pm 0.023$ & $1.14 \pm 1.33$ & $28.34 \pm 21.74$ & $0.984 \pm 0.018$ & $0.969 \pm 0.024$ \\ \hline
Add $7.5\%$, remove $7.5\%$ & $0.081 \pm 0.019$ & $1.12 \pm 1.32$ & $27.94 \pm 22.12$ & $0.985 \pm 0.017$ & $0.970 \pm 0.024$ \\ \hline

Add $10\%$ & $0.081 \pm 0.027$ & $1.16 \pm 1.65$ & $27.42 \pm 23.89$ & $0.985 \pm 0.020$ & $0.970 \pm 0.026$ \\ \hline
Remove $10\%$ & $0.083 \pm 0.023$ & $1.14 \pm 1.46$ & $29.66 \pm 20.97$ & $0.984 \pm 0.019$ & $0.968 \pm 0.023$ \\ \hline
Add $10\%$, remove $10\%$ & $0.085 \pm 0.038$ & $1.38 \pm 2.03$ & $27.82 \pm 25.45$ & $0.982 \pm 0.024$ & $0.970 \pm 0.028$ \\ \hline

Add $12\%$ & $0.081 \pm 0.024$ & $1.24 \pm 1.52$ & $28.80 \pm 24.58$ & $0.983 \pm 0.019$ & $0.969 \pm 0.027$ \\ \hline
Remove $12\%$ & $0.082 \pm 0.022$ & $1.26 \pm 1.38$ & $28.28 \pm 22.86$ & $0.983 \pm 0.017$ & $0.969 \pm 0.025$ \\ \hline
Add $12\%$, remove $12\%$ & $0.086 \pm 0.027$ & $1.32 \pm 1.56$ & $29.08 \pm 22.79$ & $0.982 \pm 0.019$ & $0.969 \pm 0.025$ \\ \hline
\end{tabular}
\label{tab:errors_ctsi}
\end{table*}

\section{Discussion} 
\label{sec:discussion}

\subsection{Logarithmic Transformation for Multiplicative Noise} 
Given the multiplicative noise model \mTwo{}, $y_i = \a_i^T \x \, (1+q_a)^{\eta_i}$,
it is natural to separate the noise term by applying a logarithmic transformation.
A logarithmic transformation coincides with the practice in  
applied statistics of applying the variance-stabilizing transformation to the response variable $y_i$ for decoupling the signal and noise~\cite{everitt2002cambridge}.
Such a decoupling step can enable the application of computationally efficient least-squares style estimators~\cite{morgenthaler2012advantages}.
A typical example is to apply the logarithmic transformation to gamma regression~\cite{Czado-gamma}. 

However, applying the logarithmic transformation to \mTwo{} does not allow us to use least-squares style estimators such as \lasso{}. 
More specifically, $\E \left[ \log y_i \right] = \log \left( \a_i^T \x \right)$ indicates that the mean of $\log y_i$ is no longer linear in $\x$ after the transformation. 
This could lead to inaccurate estimates of $\x$ in conjunction with numerical optimization, 
should any mismatch exist between the data and the assumed model.

The other option, adopted in Sec.~\ref{subsec:alg_mult}, is to linearize the multiplicative noise into additive noise at the cost of introducing unequal variances among $m$ different pooled measurements. It can be shown that $y_i \approx \a_i^T \x  + \left[ \a_i^T \x \ln(1+q_a) \right] \eta_i,$
where the approximation sign is precise due to $\var(\eta_i) = \sigma^2 \ll 1$.
We note that $\var(y_i) = \left[ \a_i^T \x \log(1+q_a) \right]^2 \sigma^2$ is dependent on the true data $\a_i^T \x$, which will make least-squares style \lasso{} algorithms suboptimal due to the bias introduced by the unequal variances. 

We had experimented with the logarithmic transformation approach using an estimator of the form $\widehat{\x} = \operatorname{argmin}_{\x} \|\log \y - \log(\A\x)\|^2 + \rho \|\x\|_1$. This estimator was invoked after the \textsc{Comp} step that eliminated zero-valued measurements in $\y$ or $\A\x$.
Experimental results show that this log-transformed approach does not outperform \textsc{Comp-Lasso} in accuracy and speed. 
Another approach to multiplicative noise involves deriving an
appropriate output channel denoiser in GAMP~\cite{Zhu2020,rangan2011generalized}; 
we leave this direction for future work.

\subsection{\EditRevision{Viral Load and Infection Dynamics}}
\EditRevision{In Sec.~\ref{sec:data_gen}, 
we have made a simplified assumption that the viral load of an infected/infectious node remains constant throughout the combined 14-day period of infection.
We discuss why it is reasonable to make such simplification while we explicit model the infection dynamics.}

\EditRevision{The viral load of an individual is a function of time. 
This time-varying characteristic also naturally manifests itself at the population level in terms of the infection dynamics, namely, an individual in a population may go through the following evolution in states: susceptible (zero viral load) $\rightarrow$ infected (increasing but low viral load) $\rightarrow$ infectious (high viral load)~\cite{marc2021quantifying} $\rightarrow$ infected (decaying viral load) $\rightarrow$ recovered (zero viral load). Considering that our work is focusing on exploiting infection dynamics for group testing rather than tracking the change of viral load, we choose to directly model the aforementioned infection dynamics. 
Our approach is less complicated than the biomedical approach that uses the time-varying viral load to infer the probability of infection from bottom up~\cite{marc2021quantifying}.}

\EditRevision{Using the time-varying or fixed viral load is not likely to affect the performance of the M1 algorithms, because the notion of the viral load is not defined in model M1. Although model M2 involves viral load, the group-lasso-based algorithms are not designed to exploit prior distribution of the viral load hence its impact on M2 algorithms should also be minimal.}

\subsection{Use Cases for M1 and M2}
The different ways in which the proposed algorithms for models \textbf{M1} and \textbf{M2} utilize SI make them applicable in different scenarios. The algorithms for model \textbf{M2} do not make use of previous inference results, thereby making it useful when a population has to be tested once without access to previous results. On the other hand, the CT denoiser for model \textbf{M1} has the ability to incorporate previous testing results into the prior, thereby making it suitable for scenarios where the same population needs to be tested regularly, e.g., warehouse employees.
Furthermore, while model \textbf{M1} performs well in the presence of erroneous binary tests, it does not yield viral load estimates unlike the algorithms for model \textbf{M2}. Viral load estimates could prove to be useful since there is a positive correlation between mortality and viral loads~\cite{Westblade2020,Pujadas2020}. 

\subsection{\EditRevision{Comparison of Balanced and Unbalanced Binary Matrices}}
\label{subsec:unbalanced_mat_result}
\EditRevision{
We compare the performance of balanced matrices to ({\em i})~randomly generated Bernoulli matrices with different proportions of ones, and ({\em ii})~random column-balanced binary matrices. 
\EditRevision{A column-balanced binary matrix has an equal number of ones at randomly selected locations within a column, without ensuring an equal number of ones in each row.}
We observed that in terms of sensitivity 
and specificity, the balanced binary matrices \EditRevision{outperform the Bernoulli matrices while performing 
on par with random column-balanced matrices, across a range of sparsity levels. These results are summarized in 
TABLES~\ref{tab:balanced_binary},~\ref{tab:random_bernoulli0.1},~\ref{tab:random_bernoulli0.5}, and ~\ref{tab:column_balanced}
of the appendix. 
}} \EditRevision{There could exist other unbalanced designs, apart from the ones we considered, that could outperform our balanced designs, even more so considering non-i.i.d. priors. 
A full investigation of this is nontrivial and left to future work. However, there are hard physical constraints on the design of the pooling matrix (articulated earlier in 
Sec.~\ref{subsec:justify-balanced-mat}), due to which unbalanced designs may not be operationally feasible even if it turns out that they give better performance.}

\subsection{\EditRevision{Errors in Contact-Tracing Information}}
\EditRevision{Recent studies have shown that contact-tracing information via Bluetooth can contain many errors besides the privacy concern~\cite{Kleinman2020}. However, we note that there exist 
methods for obtaining CT information, which have been shown to be quite effective, including inquiries by social workers, cellphone based localization, and the analyses of closed-circuit television footages and financial transactions~\cite{Hohman2021,Ross2020,CDC_CT}.
A fusion of these modalities can lead to more accurate contact-tracing graphs~\cite{Kleinman2020}. Contact tracing has been shown to be useful in previous epidemics~\cite{Kwok2021}. 
As shown in Sec.~\ref{subsec:noisy-CT-info},
our algorithms are robust to errors in the CT SI. 
However, we leave a full-fledged investigation of SI errors and their impact on our algorithms to future work. We would like to point out that at the peak of a pandemic, testing resources (including time, skilled manpower, testing kits, and reagents) are known to be scarce.
In such a scenario, it is important to exploit as much information as is reasonably available in order to improve the performance of group testing 
algorithms, with the aim of saving critical testing resources.} 

\section{Conclusion} \label{sec:conclusion}
In this paper, we have presented numerical evidence that SI from family structures and contact-tracing data can significantly improve the efficiency of group testing.
Consistent gains have been observed from the algorithms proposed for both the binary noise model \mOne{} and the multiplicative noise model \mTwo{}.
We have also presented an approach to design group testing matrices of flexible sizes with low mutual coherence, and have demonstrated how to incorporate SI in the matrix design by optimizing on a modified form of the mutual coherence quality measure.
An interesting observation is that incorporating SI into the design of the pooling matrix \EditRevision{using our approach} yielded only modest gains, once SI has already been incorporated into the decoder. \EditRevision{There could exist other approaches that produce better reconstruction performance upon incorporating SI at the encoder end, even in conjunction with decoders that use SI. However, we leave a full investigation of this aspect to future work.}

Finally, due to the exploratory nature of our work and our deliberate efforts to bring algorithmic advances closer to practice, finding publicly available datasets with associated SI proved to be challenging.
In light of this, we crafted a generative model that closely reflects the key characteristics of COVID-19 transmission for generating the data used in our investigation.
Our numerical results present strong empirical evidence that contact-tracing supported group testing is a viable option to optimize the use of resources for a widespread testing program during a pandemic. 

\section*{Acknowledgment}
The authors thank Dr. Junan Zhu for allowing them to use his implementation of GAMP with SI in their implementation of the family and CT denoisers.

\begin{table*}[!t]
\centering
\caption{Performance of \compsqrtoglasso{} using $300 \times 1000$ $\psi$-optimal balanced matrix. For each criterion and each sparsity ($s$) value, mean, and standard deviation values are reported across $50$ signals.}
\label{tab:balanced_binary}
\begin{tabular}{|c|c|c|c|c|c|}
\hline
\textbf{s}    & \textbf{RMSE}       & \textbf{\#FN}   & \textbf{\#FP}     & \textbf{Sensitivity}        & \textbf{Specificity}        \\ \hline
$2.12\%$ & $0.05294 \pm 0.02025$ & $0.02 \pm 0.14$ & $2.72 \pm 3.36$ & $0.99956 \pm 0.00314$ & $0.99719 \pm 0.00350$ \\ \hline
$3.98\%$ & $0.06005 \pm 0.01646$ & $0.36 \pm 0.66$ & $10.24 \pm 11.84$ & $0.99394 \pm 0.01111$ & $0.98916 \pm 0.01274$ \\ \hline
$6.01\%$ & $0.08309 \pm 0.02442$ & $1.02 \pm 1.24$ & $27.16 \pm 24.06$ & $0.98642 \pm 0.01580$ & $0.97066 \pm 0.02651$ \\ \hline
$8.86\%$ & $0.11189 \pm 0.03171$ & $2.40 \pm 1.55$ & $61.46 \pm 32.92$ & $0.97348 \pm 0.01559$ & $0.93170 \pm 0.03760$ \\ \hline
\end{tabular}
\end{table*}

\begin{table*}[!t]
\centering
\caption{Performance of \compsqrtoglasso{} using $300 \times 1000$ Bernoulli($0.1$) matrix. For each criterion and each sparsity ($s$) value, mean, and standard deviation values are reported across $50$ signals.}
\label{tab:random_bernoulli0.1}
\begin{tabular}{|c|c|c|c|c|c|}
\hline
\textbf{s}    & \textbf{RMSE}       & \textbf{\#FN}   & \textbf{\#FP}     & \textbf{Sensitivity}        & \textbf{Specificity}        \\ \hline
$2.12\%$ & $0.04745 \pm 0.02610$ & $0.14 \pm 0.35$ & $19.76 \pm 30.60$ & $0.99592 \pm 0.01039$ & $0.97954 \pm 0.03184$ \\ \hline
$3.98\%$ & $0.09122 \pm 0.04046$ & $1.06 \pm 1.24$ & $80.28 \pm 55.91$ & $0.97629 \pm 0.02488$ & $0.91546 \pm 0.05990$ \\ \hline
$6.01\%$ & $0.14744 \pm 0.04986$ & $2.90 \pm 2.32$ & $140.48 \pm 48.07$ & $0.95504 \pm 0.02863$ & $0.84965 \pm 0.05374$ \\ \hline
$8.86\%$ & $0.23501 \pm 0.06415$ & $4.38 \pm 2.18$ & $193.76 \pm 38.97$ & $0.95108 \pm 0.02059$ & $0.78655 \pm 0.04615$ \\ \hline
\end{tabular}
\end{table*}

\begin{table*}[!t]
\centering
\caption{Performance of \compsqrtoglasso{} using $300 \times 1000$ Bernoulli($0.5$) matrix. For each criterion and each sparsity ($s$) value, mean, and standard deviation values are reported across $50$ signals.}
\label{tab:random_bernoulli0.5}
\begin{tabular}{|c|c|c|c|c|c|}
\hline
\textbf{s}    & \textbf{RMSE}       & \textbf{\#FN}   & \textbf{\#FP}     & \textbf{Sensitivity}        & \textbf{Specificity}        \\ \hline
$2.12\%$ & $0.14077 \pm 0.07266$ & $0.78 \pm 1.00$ & $83.00 \pm 49.06$ & $0.96968 \pm 0.03883$ & $0.91475 \pm 0.05078$ \\ \hline
$3.98\%$ & $0.23440 \pm 0.07964$ & $3.08 \pm 3.28$ & $125.60 \pm 33.22$ & $0.93383 \pm 0.04910$ & $0.86881 \pm 0.03608$ \\ \hline
$6.01\%$ & $0.34310 \pm 0.08414$ & $5.52 \pm 3.49$ & $161.70 \pm 29.93$ & $0.91202 \pm 0.05027$ & $0.82755 \pm 0.03391$ \\ \hline
$8.86\%$ & $0.50454 \pm 0.11203$ & $10.04 \pm 6.88$ & $194.80 \pm 32.48$ & $0.89243 \pm 0.05364$ & $0.78566 \pm 0.03837$ \\ \hline
\end{tabular}
\end{table*}

\begin{table*}[!t]
\centering
\caption{Performance of \compsqrtoglasso{} using $300 \times 1000$ random column-balanced matrix. For each criterion and each sparsity ($s$) value, mean, and standard deviation values are reported across $50$ signals.}
\label{tab:column_balanced}
\begin{tabular}{|c|c|c|c|c|c|}
\hline
\textbf{s}    & \textbf{RMSE}       & \textbf{\#FN}   & \textbf{\#FP}     & \textbf{Sensitivity}        & \textbf{Specificity}        \\ \hline
$2.12\%$ & $0.05251 \pm 0.01741$ & $0.02 \pm 0.14$ & $2.94 \pm 3.23$ & $0.99956 \pm 0.00314$ & $0.99697 \pm 0.00334$ \\ \hline
$3.98\%$ & $0.06646 \pm 0.01815$ & $0.26 \pm 0.56$ & $10.30 \pm 11.52$ & $0.99589 \pm 0.00874$ & $0.98910 \pm 0.01235$ \\ \hline
$6.01\%$ & $0.09021 \pm 0.03150$ & $1.42 \pm 1.50$ & $29.86 \pm 23.26$ & $0.97906 \pm 0.01913$ & $0.96781 \pm 0.02561$ \\ \hline
$8.86\%$ & $0.12261 \pm 0.03385$ & $2.30 \pm 1.68$ & $58.88 \pm 28.75$ & $0.97520 \pm 0.01614$ & $0.93468 \pm 0.03301$ \\ \hline
\end{tabular}
\end{table*}

\appendix[Comparisons for Different Types of Binary Matrices]
In this section, we present an experimental comparison between different types of binary matrices: ({\em i})~balanced binary matrices (as proposed in the main paper), ({\em ii})~random Bernoulli matrices containing equal number of ones and zeros, ({\em iii})~random Bernoulli matrices for which the number of ones is only $10\%$ of the total number of entries, and ({\em iv})~column-balanced binary matrices. As defined in the main paper, a column-balanced binary matrix has an equal number of ones at randomly selected locations within a column, without ensuring an equal number of ones in each row. We report the performance for the \compsqrtoglasso{} estimator in terms of RMSE, sensitivity, specificity, the number of false negative, and the number of false positives. For each average sparsity level, we report the mean and the standard deviation of each metric across 50 signals (same set of signals that are used for the experiments in Sec. VI-B and Sec. VI-D of the main paper). From these results shown in TABLES~\ref{tab:balanced_binary},~\ref{tab:random_bernoulli0.1},~\ref{tab:random_bernoulli0.5}, and ~\ref{tab:column_balanced}, it is clear that the performance of balanced binary matrices is the best, closely followed by column-balanced binary matrices.

\bibliographystyle{IEEEtran}
\bibliography{IEEEabrv,refs}

\begin{thebibliography}{10}
\providecommand{\url}[1]{#1}
\csname url@samestyle\endcsname
\providecommand{\newblock}{\relax}
\providecommand{\bibinfo}[2]{#2}
\providecommand{\BIBentrySTDinterwordspacing}{\spaceskip=0pt\relax}
\providecommand{\BIBentryALTinterwordstretchfactor}{4}
\providecommand{\BIBentryALTinterwordspacing}{\spaceskip=\fontdimen2\font plus
\BIBentryALTinterwordstretchfactor\fontdimen3\font minus
  \fontdimen4\font\relax}
\providecommand{\BIBforeignlanguage}[2]{{%
\expandafter\ifx\csname l@#1\endcsname\relax
\typeout{** WARNING: IEEEtran.bst: No hyphenation pattern has been}%
\typeout{** loaded for the language `#1'. Using the pattern for}%
\typeout{** the default language instead.}%
\else
\language=\csname l@#1\endcsname
\fi
#2}}
\providecommand{\BIBdecl}{\relax}
\BIBdecl

\bibitem{Goenka2021}
R.~Goenka, S.-J. Cao, C.-W. Wong, A.~Rajwade, and D.~Baron, ``Contact tracing
  enhances the efficiency of {COVID-19} group testing,'' in \emph{{IEEE} Int.
  Conf. Acoust., Speech, Signal Process. (ICASSP)}, 2021, pp. 8168--8172.

\bibitem{Dorfman1943}
R.~Dorfman, ``The detection of defective members of large populations,''
  \emph{Ann. Math. Stat.}, vol.~14, no.~4, p. 436–440, 1943.

\bibitem{aldridge2019group}
M.~Aldridge, O.~Johnson, and J.~Scarlett, \emph{Group Testing: {An} Information
  Theory Perspective}, 2019.

\bibitem{Hogan2020}
C.~Hogan, M.~Sahoo, and B.~Pinsky, ``Sample pooling as a strategy to detect
  community transmission of {SARS-CoV-2},'' \emph{J. Am. Med. Assoc.}, vol.
  323, no.~19, pp. 1967--1969, Apr. 2020.

\bibitem{Abdelhamid2020}
B.~Abdalhamid \emph{et~al.}, ``Assessment of specimen pooling to conserve {SARS
  CoV-2} testing resources,'' \emph{Am. J. Clin. Pathol.}, vol. 153, no.~6, pp.
  715--718, May 2020.

\bibitem{Zhu2020}
J.~Zhu, K.~Rivera, and D.~Baron, ``Noisy pooled {PCR} for virus testing,''
  \url{https://arxiv.org/abs/2004.02689}, Apr. 2020.

\bibitem{Yi_arxiv}
J.~Yi, R.~Mudumbai, and W.~Xu, ``Low-cost and high-throughput testing of
  {COVID-19} viruses and antibodies via compressed sensing: System concepts and
  computational experiments,'' \url{https://arxiv.org/abs/2004.05759}, Apr.
  2020.

\bibitem{Ghosh2021}
S.~Ghosh \emph{et~al.}, ``A compressed sensing approach to pooled {RT-PCR}
  testing for {COVID-19} detection,'' \emph{IEEE Open J. Signal Process.},
  2021.

\bibitem{zhu2020paris}
\BIBentryALTinterwordspacing
J.~Zhu, K.~Rivera, C.~Rush, and D.~Baron, ``Noisy pooled {PCR} for {COVID-19}
  testing,'' \emph{Paris Machine Learning Meetup}, May 2020. [Online].
  Available: \url{https://youtu.be/gYJqnXbi1Bg}
\BIBentrySTDinterwordspacing

\bibitem{Heiderzadeh2020}
A.~Heidarzadeh and K.~Narayanan, ``Two-stage adaptive pooling with {RT-qPCR}
  for {COVID-19} screening,'' \url{https://arxiv.org/abs/2007.02695}, Jul.
  2020.

\bibitem{Nikolopoulos2020}
P.~Nikolopoulos, T.~Guo, C.~Fragouli, and S.~Diggavi, ``Community aware group
  testing,'' \url{https://arxiv.org/abs/2007.08111}, Jul. 2020.

\bibitem{Shental2020}
N.~Shental \emph{et~al.}, ``Efficient high throughput {SARS-CoV-2} testing to
  detect asymptomatic carriers,'' \emph{Sci. Adv.}, vol.~6, no.~37, Sep. 2020.

\bibitem{lin2020comparisons}
Y.-J. Lin, C.-H. Yu, T.-H. Liu, C.-S. Chang, and W.-T. Chen, ``Comparisons of
  pooling matrices for pooled testing of {COVID-19},''
  \url{https://arxiv.org/abs/2010.00060}, Sep. 2020.

\bibitem{cohen2020multilevel}
A.~Cohen, N.~Shlezinger, A.~Solomon, Y.~C. Eldar, and M.~Médard, ``Multi-level
  group testing with application to one-shot pooled {COVID}-19 tests,''
  \url{https://arxiv.org/abs/2010.06072}, Oct. 2020.

\bibitem{lin2020positively}
Y.-J. Lin, C.-H. Yu, T.-H. Liu, C.-S. Chang, and W.-T. Chen, ``Positively
  correlated samples save pooled testing costs,''
  \url{https://arxiv.org/abs/2011.09794}, Nov. 2020.

\bibitem{Nikolopoulos2021b}
P.~Nikolopoulos, S.~R. Srinivasavaradhan, T.~Guo, C.~Fragouli, and S.~Diggavi,
  ``Group testing for overlapping communities,'' in \emph{IEEE Int. Conf.
  Commun.}, 2021.

\bibitem{ahn2021adaptive}
S.~Ahn, W.-N. Chen, and A.~Ozgur, ``Adaptive group testing on networks with
  community structure,'' \url{https://arxiv.org/abs/2101.02405}, Jan. 2021.

\bibitem{arasli2021group}
B.~Arasli and S.~Ulukus, ``Group testing with a graph infection spread model,''
  \url{https://arxiv.org/abs/2101.05792}, Jan. 2021.

\bibitem{Benatia2020}
D.~Benatia, R.~Godefroy, and J.~Lewis, ``Estimating {COVID-19} prevalence in
  the {United States}: A sample selection model approach,''
  \url{https://doi.org/10.1101/2020.04.20.20072942}.

\bibitem{cdc_contact_tracing}
{Center for Disease Control and Prevention}, ``Contact tracing for
  {COVID-19},''
  \url{https://www.cdc.gov/coronavirus/2019-ncov/php/contact-tracing/contact-tracing-plan/contact-tracing.html}.

\bibitem{Hekmati2020}
A.~Hekmati, G.~Ramachandran, and B.~Krishnamachari, ``{CONTAIN:
  P}rivacy-oriented contact tracing protocols for epidemics,''
  \url{https://arxiv.org/abs/2004.05251}.

\bibitem{Kleinman2020}
R.~Kleinman and C.~Merkel, ``Digital contact tracing for {COVID}-19,''
  \emph{Canadian Medical Association Journal}, 2020.

\bibitem{Hohman2021}
M.~Hohman, F.~McMaster, and S.~I. Woodruff, ``Contact tracing for covid-19: The
  use of motivational interviewing and the role of social work,''
  \emph{Clinical Social Work Journal}, 2021.

\bibitem{Ross2020}
A.~M. Ross, L.~Zerden, B.~Ruth, J.~Zelnick, and J.~Cederbaum, ``Contact
  tracing: An opportunity for social work to lead,'' \emph{Social Work in
  Public Health}, vol.~35, no.~7, pp. 533--545, 2020.

\bibitem{CDC_CT}
``Case investigation and contact tracing: Part of a multipronged approach to
  fight the {COVID}-19 pandemic,''
  \url{https://www.cdc.gov/coronavirus/2019-ncov/php/principles-contact-tracing.html},
  2021.

\bibitem{rangan2011generalized}
S.~Rangan, ``Generalized approximate message passing for estimation with random
  linear mixing,'' in \emph{IEEE Int. Symp. Inf. Theory}, 2011, pp. 2168--2172.

\bibitem{Yuan2006}
M.~Yuan and Y.~Lin, ``Model selection and estimation in regression with grouped
  variables,'' \emph{J. Royal Stat. Soc. Series B}, vol.~68, no.~1, 2006.

\bibitem{Jacob2009}
L.~Jacob, G.~Obozinski, and J.-P. Vert, ``Group {LASSO} with overlap and graph
  {LASSO},'' in \emph{Int. Conf. Mach. Learning}, 2009.

\bibitem{Deckert2020}
A.~Deckert, T.~Barnighausen, and N.~N. Kyei, ``Simulation of pooled-sample
  analysis strategies for {COVID}-19 mass testing,'' \emph{Bulletin of the
  World Health Organization}, vol.~98, no.~9, 2020.

\bibitem{McMahan2012}
C.~S. McMahan, J.~M. Tebbs, and C.~R. Bilder, ``Informative {Dorfman}
  screening,'' \emph{Biometrics}, vol.~68, no.~1, pp. 287--296, 2012.

\bibitem{Bilder2010}
C.~R. Bilder, J.~M. Tebbs, and P.~Chen, ``Informative retesting,'' \emph{J. Am.
  Stat. Assoc.}, vol. 105, no. 491, pp. 942--955, 2010.

\bibitem{Nikolopoulos2021a}
P.~Nikolopoulos, S.~Rajan~Srinivasavaradhan, T.~Guo, C.~Fragouli, and
  S.~Diggavi, ``Group testing for connected communities,'' in \emph{24th Int.
  Conf. Artificial Intell. Stat.}, ser. Proc. Mach. Learning Res., vol.
  130.\hskip 1em plus 0.5em minus 0.4em\relax PMLR, 2021, pp. 2341--2349.

\bibitem{Baron2010}
D.~Baron, S.~Sarvotham, and R.~G. Baraniuk, ``Bayesian compressive sensing via
  belief propagation,'' \emph{{IEEE} Trans. Signal Process.}, vol.~58, no.~1,
  pp. 269--280, 2010.

\bibitem{Lendle2012}
S.~D. Lendle, M.~G. Hudgens, and B.~F. Qaqish, ``Group testing for case
  identification with correlated responses,'' \emph{Biometrics}, vol.~68, p.
  532–540, 2012.

\bibitem{Gilbert2008}
A.~C. {Gilbert}, M.~A. {Iwen}, and M.~J. {Strauss}, ``Group testing and sparse
  signal recovery,'' in \emph{42nd Asilomar Conf. Signals, Syst. and Comput.},
  2008, pp. 1059--1063.

\bibitem{Davenport2012}
M.~A. Davenport, M.~F. Duarte, Y.~C. Eldar, and G.~Kutyniok, \emph{Introduction
  to compressed sensing}.\hskip 1em plus 0.5em minus 0.4em\relax Cambridge
  University Press, 2012, p. 1–64.

\bibitem{Abolghasemi2010}
V.~{Abolghasemi}, S.~{Ferdowsi}, B.~{Makkiabadi}, and S.~{Sanei}, ``On
  optimization of the measurement matrix for compressive sensing,'' in
  \emph{18th European Signal Process. Conf.}, 2010, pp. 427--431.

\bibitem{Carin2011}
L.~{Carin}, D.~{Liu}, and B.~{Guo}, ``Coherence, compressive sensing, and
  random sensor arrays,'' \emph{{IEEE} Trans. Antennas Propag.}, vol.~53,
  no.~4, pp. 28--39, 2011.

\bibitem{Arguello2014}
H.~{Arguello} and G.~R. {Arce}, ``Colored coded aperture design by
  concentration of measure in compressive spectral imaging,'' \emph{{IEEE}
  Trans. Image Process.}, vol.~23, no.~4, pp. 1896--1908, 2014.

\bibitem{Devore2007}
R.~A. DeVore, ``Deterministic constructions of compressed sensing matrices,''
  \emph{J. Complexity}, vol.~23, no.~4, pp. 918--925, 2007.

\bibitem{Naidu2016}
R.~R. {Naidu}, P.~{Jampana}, and C.~S. {Sastry}, ``Deterministic compressed
  sensing matrices: Construction via {Euler} squares and applications,''
  \emph{{IEEE} Trans. Signal Process.}, vol.~64, no.~14, pp. 3566--3575, 2016.

\bibitem{Naidu2017}
R.~R. {Naidu} and C.~R. {Murthy}, ``Construction of binary sensing matrices
  using extremal set theory,'' \emph{{IEEE} Signal Process. Lett.}, vol.~24,
  no.~2, pp. 211--215, 2017.

\bibitem{Li2014}
S.~{Li} and G.~{Ge}, ``Deterministic construction of sparse sensing matrices
  via finite geometry,'' \emph{{IEEE} Trans. Signal Process.}, vol.~62, no.~11,
  pp. 2850--2859, 2014.

\bibitem{Haseltine2020}
W.~Haseltine, ``What {COVID}-19 reinfection means for vaccines,''
  \url{https://www.scientificamerican.com/article/what-covid-19-reinfection-means-for-vaccines/}.

\bibitem{Carcione2020}
J.~M. Carcione, J.~E. Santos, C.~Bagaini, and J.~Ba, ``A simulation of a
  {COVID-19} epidemic based on a deterministic {SEIR} model,'' \emph{Frontiers
  in Public Health}, vol.~8, p. 230, 2020.

\bibitem{WHOreport}
{World Health Organization}, ``Coronavirus disease 2019 {(COVID-19)} situation
  report--73,'' \url{https://tinyurl.com/ybnbky8m}, Apr. 2020.

\bibitem{Buchan2020}
B.~Buchan \emph{et~al.}, ``Distribution of {SARS-CoV-2 PCR} cycle threshold
  values provide practical insight into overall and target-specific sensitivity
  among symptomatic patients,'' \emph{Am. J. Clin. Pathol.}, 2020.

\bibitem{UNDoc}
{United Nations}, ``Household size and composition around the world 2017,''
  \url{https://tinyurl.com/vo7hrlv}.

\bibitem{LaGatta2021}
V.~L. Gatta, V.~Moscato, M.~Postiglione, and G.~Sperli, ``An epidemiological
  neural network exploiting dynamic graph structured data applied to the
  {COVID}-19 outbreak,'' \emph{IEEE Trans. Big Data}, vol.~7, no.~1, 2021.

\bibitem{dror_plugin}
Y.~Ma, J.~Tan, N.~Krishnan, and D.~Baron, ``Empirical {B}ayes and full {B}ayes
  for signal estimation,'' in \emph{Inf. Theory App. Workshop}, San Diego, CA,
  Feb. 2014, pp. 994--1001.

\bibitem{THW2015}
T.~Hastie, R.~Tibshirani, and M.~Wainwright, \emph{Statistical Learning with
  Sparsity: The {LASSO} and Generalizations}.\hskip 1em plus 0.5em minus
  0.4em\relax {CRC} Press, 2015.

\bibitem{Belloni2011}
A.~Belloni, V.~Chernuzhukov, and L.~Wang, ``Square-root {LASSO}: {P}ivotal
  recovery of sparse signals via conic programming,'' \emph{Biometrika},
  vol.~98, no.~4, pp. 791--806, 2011.

\bibitem{Bron1973}
C.~Bron and J.~Kerbosch, ``Algorithm 457: Finding all cliques of an undirected
  graph,'' \emph{Commun. ACM}, vol.~16, no.~9, p. 575–577, 1973.

\bibitem{Palla2005}
G.~Palla, I.~Derényi, I.~Farkas, and T.~Vicsek, ``Uncovering the overlapping
  community structure of complex networks in nature and society,''
  \emph{Nature}, vol. 435, pp. 814--818, 2005.

\bibitem{Bruladi}
\BIBentryALTinterwordspacing
R.~A. Brualdi, ``Matrices of zeros and ones with fixed row and column sum
  vectors,'' \emph{Linear Algebra Its Appl.}, vol.~33, pp. 159--231, 1980.
  [Online]. Available:
  \url{http://www.sciencedirect.com/science/article/pii/0024379580901056}
\BIBentrySTDinterwordspacing

\bibitem{Walkup}
D.~W. Walkup, ``Minimal interchanges of (0, 1)-matrices and disjoint circuits
  in a graph,'' \emph{Can. J. Math.}, vol.~17, p. 831–838, 1965.

\bibitem{bickson2011fault}
D.~Bickson, D.~Baron, A.~Ihler, H.~Avissar, and D.~Dolev, ``Fault
  identification via nonparametric belief propagation,'' \emph{IEEE Trans
  Signal Process.}, vol.~59, no.~6, pp. 2602--2613, 2011.

\bibitem{hanel2020boosting}
R.~Hanel and S.~Thurner, ``Boosting test-efficiency by pooled testing
  strategies for {SARS-CoV-2},'' \url{https://arxiv.org/abs/2003.09944}, Mar.
  2020.

\bibitem{Zhang2014}
J.~Zhang, L.~Chen, P.~Boufounos, and Y.~Gu, ``On the theoretical analysis of
  cross validation in compressive sensing,'' in \emph{{IEEE} Int. Conf.
  Acoust., Speech, Signal Process. (ICASSP)}, 2014, pp. 3370--3374.

\bibitem{Chicco2020}
D.~Chicco and G.~Jurman, ``The advantages of the matthews correlation
  coefficient {(MCC)} over {F1} score and accuracy in binary classification
  evaluation,'' \emph{BMC Genomics}, vol.~21, 2020.

\bibitem{everitt2002cambridge}
B.~Everitt and A.~Skrondal, \emph{The Cambridge Dictionary of Statistics},
  4th~ed.\hskip 1em plus 0.5em minus 0.4em\relax Cambridge University Press
  Cambridge, 2010.

\bibitem{morgenthaler2012advantages}
S.~Morgenthaler and R.~G. Staudte, ``Advantages of variance stabilization,''
  \emph{Scand. J. Stat.}, vol.~39, no.~4, pp. 714--728, 2012.

\bibitem{Czado-gamma}
\BIBentryALTinterwordspacing
C.~Czado, ``Lecture 8: Gamma regression.'' [Online]. Available:
  \url{https://www.groups.ma.tum.de/fileadmin/w00ccg/statistics/czado/lec8.pdf}
\BIBentrySTDinterwordspacing

\bibitem{marc2021quantifying}
A.~Marc, M.~Kerioui, F.~Blanquart, J.~Bertrand, O.~Mitj{\`a},
  M.~Corbacho-Monn{\'e}, M.~Marks, and J.~Guedj, ``Quantifying the relationship
  between {SARS-CoV-2} viral load and infectiousness,'' \emph{Elife}, Sep.
  2021.

\bibitem{Westblade2020}
L.~F. Westblade \emph{et~al.}, ``{SARS-CoV-2} viral load predicts mortality in
  patients with and without cancer who are hospitalized with {COVID-19},''
  \emph{Cancer Cell}, Nov. 2020.

\bibitem{Pujadas2020}
E.~Pujadas \emph{et~al.}, ``{SARS-CoV-2} viral load predicts {COVID-19}
  mortality,'' \emph{Lancet Respir. Med.}, vol.~8, no.~9, p. e70, Sep. 2020.

\bibitem{Kwok2021}
K.~O. Kwok \emph{et~al.}, ``Epidemic models of contact tracing: Systematic
  review of transmission studies of severe acute respiratory syndrome and
  middle east respiratory syndrome,'' \emph{Computational and Structural
  Biology Journal}, vol.~17, 2019.

\end{thebibliography}
\balance

\end{document}